\documentclass[12pt]{iopart}
\usepackage{graphicx,amsfonts,amsgen,amsbsy,amssymb,bbold,bm,cite}

\makeatletter
\def\@fnsymbol#1{^{\thefootnote}\relax}
\makeatother

\begin{document}

\topical{Linear response theories for interatomic exchange interactions}

\author{I V Solovyev}

\address{Research Center for Materials Nanoarchitectonics (MANA), National Institute for Materials Science (NIMS), 1-1 Namiki, Tsukuba, Ibaraki 305-0044, Japan}
\ead{SOLOVYEV.Igor@nims.go.jp}
\vspace{10pt}

\begin{abstract}
The linear response is a perturbation theory establishing the relationship between given physical variable and the external field inducing this variable. A well-known example of the linear response theory in magnetism is the susceptibility relating the magnetization with the magnetic field. In 1987, Liechtenstein \etal came up with the idea to formulate the problem of interatomic exchange interactions, which would describe the energy change caused by the infinitesimal rotations of spins, in terms of this susceptibility. The formulation appears to be very generic and, for isotropic systems, expresses the energy change in the form of the Heisenberg model, irrespectively on which microscopic mechanism stands behind the interaction parameters. Moreover, this approach establishes the relationship between the exchange interactions and the electronic structure obtained, for instance, in the first-principles calculations based on the density functional theory. The purpose of this review is to elaborate basic ideas of the linear response theories for the exchange interactions as well as more recent developments. The special attention is paid to the approximations underlying the original method of Liechtenstein \etal in comparison with its more recent and more rigorous extensions, the roles of the on-site Coulomb interactions and the ligand states, and calculations of antisymmetric Dzyaloshinskii-Moriya interactions, which can be performed alongside with the isotropic exchange, within one computational scheme. The abilities of the linear response theories as well as many theoretical nuances, which may arise in the analysis of interatomic exchange interactions, are illustrated on magnetic van der Walls materials Cr$X_3$ ($X$$=$ Cl, I), half-metallic ferromagnet CrO$_2$, ferromagnetic Weyl semimetal Co$_3$Sn$_2$S$_2$, and orthorhombic manganites $A$MnO$_3$ ($A$$=$ La, Ho), known for the peculiar interplay of the lattice distortion, spin, and orbital ordering.
\end{abstract}

%
\vspace{2pc}
\par \submitto{\JPCM}
%
%
%

\section{\label{sec:Intro} Introduction}
\par On many occasions our image of magnetism rests on the picture interacting spins ($\boldsymbol{e}_{i}$ and $\boldsymbol{e}_{j}$) attached to the atomic sites ($i$ and $j$). If the system is isotropic, such interactions have a form of the scalar products $\boldsymbol{e}_{i} \cdot \boldsymbol{e}_{j}$~\cite{Heisenberg,Anderson1959,RudermanKittel,Kasuya,Yosida}. The interactions are ferromagnetic (FM) if they force $\boldsymbol{e}_{i} \cdot \boldsymbol{e}_{j} > 0$ and antiferromagnetic (AFM) if $\boldsymbol{e}_{i} \cdot \boldsymbol{e}_{j} < 0$. If $i$ and $j$ are no longer connected by the spacial inversion, the spins tend to align neither ferromagnetically nor antiferromagnetically.\footnote{The collinear FM or AFM alignment is the consequence of the spacial inversion in the bond. In the former case, the spacial inversion should be amongst the symmetry operations. In the latter case, it is combined with the time reversal. Therefore, if the inversion symmetry is broken, neither FM nor AFM alignment satisfies the symmetry properties.} The corresponding interaction, which is called Dzyaloshinskii-Moriya (DM) interaction, is given by the cross product $[\boldsymbol{e}_{i} \times \boldsymbol{e}_{j}]$ and driven by the relativistic spin-orbit (SO) coupling~\cite{Dzyaloshinskii_weakF,Moriya_weakF}. Thus, it is always nice to have a transparent toy model, which would explain that certain material has a particular magnetic structure because some interactions are strong or weak, ferromagnetic or antiferromagnetic, etc. The experimental inelastic neutron scattering data are typically fitted to extract parameters of such physically meaningful model. In theory, the spin model can be constructed by averaging the energies of interatomic interactions over non-magnetic degrees of freedom~\cite{Heisenberg,Anderson1959,RudermanKittel,Kasuya,Yosida,Moriya_weakF}. In this article we will explain how the interatomic exchange interactions can be generally derived starting from the electronic structure obtained in the first-principles calculations.

\par Let us consider the simplest possible spin model with the energy
\noindent
\begin{equation}
E =  - \frac{1}{2} \sum_{i \ne j} \big( J_{ij} \boldsymbol{e}_{i} \cdot \boldsymbol{e}_{j} - \boldsymbol{d}_{ij} \cdot [\boldsymbol{e}_{i} \times \boldsymbol{e}_{j}] \big),
\label{eq:Hspin}
\end{equation}
\noindent where $J_{ij}$ is the isotropic exchange, $\boldsymbol{d}_{ij} = (d_{ij}^{x},d_{ij}^{y},d_{ij}^{z})$ is DM vector, and the spin moments $\boldsymbol{e}_{i}$ are normalized to the unity: $|\boldsymbol{e}_{i}| = 1$. Although we will be primarily interested in the behavior of isotropic interactions, it appears to be possible to consider $J_{ij}$ in the combination with $\boldsymbol{d}_{ij}$ within one computational scheme. The reason will become clear in a moment. Our goal is to find parameters of this model using the information about the electronic structure. Of course, the model \eref{eq:Hspin} is an approximation as there is no reason why the energy of a general magnetic system should have such a simple form and be described exclusively by the bilinear interactions. There are only few microscopic mechanisms, which are consistent with the form of \Eref{eq:Hspin}. These are the direct Heisenberg exchange~\cite{Heisenberg}, Anderson's superexchange~\cite{Anderson1959}, and long-range exchange interactions by Ruderman, Kittel, Kasuya, and Yosida (RKKY)~\cite{RudermanKittel,Kasuya,Yosida,Roth,BrunoChappert}. In the first example, this is the property of exchange energy, related to the antisymmetry of fermionic wave functions. In the last two examples, this is the consequence of the 2nd order perturbation theory with respect to, respectively, transfer integrals and intraatomic exchange interactions, that couples localized core spins to the outer conduction electrons.

\par Nevertheless, there is one more, very special case, where the magnetic energy can be also described by \Eref{eq:Hspin}. These are the infinitesimal rotations of spins near the equilibrium, as was realized by Liechtenstein \etal~\cite{LKG1984,LKAG1987}. Indeed, considering rotations $\boldsymbol{e}_{i} = ( \theta \cos \boldsymbol{q} \boldsymbol{R}_{i}, \theta \sin \boldsymbol{q} \boldsymbol{R}_{i}, 1-\frac{\theta^2}{2} )$ near the ground state $\boldsymbol{e}_{\rm GS} = (0,0,1)$, one can evaluate the energy change (per one unit cell) caused by interactions between the transversal ($xy$) components of spins. For the model \eref{eq:Hspin}, this energy change is given by
\noindent
\begin{equation}
\delta E_{\boldsymbol{q}} =  - \frac{1}{2} \Big( J_{\boldsymbol{q}} - i d_{\boldsymbol{q}}^{z} \Big) \theta^2,
\label{eq:Hspinq}
\end{equation}
\noindent where $X_{\boldsymbol{q}} = \sum_{j} X_{0j} \exp (-i \boldsymbol{q} \cdot \boldsymbol{R}_{j})$ is the Fourier image of $X_{ij}$. Then, the basic idea is to extract the same energy change from the electronic structure calculations, typically within spin-density functional theory (SDFT)~\cite{HK,KS,BH}, and map it on \Eref{eq:Hspinq}. This should give us the parameters of exchange interactions $J_{\boldsymbol{q}}$ and $i d_{\boldsymbol{q}}^{z}$ (or $J_{ij}$ and $d_{ij}^{z}$ after the Fourier transform to the real space). Since the rotations of $\boldsymbol{e}_{i}$ are chosen in the form of the conical spin spiral (which is compatible with the DM interactions), $J_{ij}$ can be considered in the combination with $d_{ij}^{z}$ in the one computational scheme, where the energy change is uniquely specified by $\boldsymbol{q}$~\cite{Sandratskii}. Basically, this is a perturbation theory, which can be formulated in terms of the response function (or the susceptibility).

\par One of the most attractive points of the infinitesimal rotations of spins is that the bilinear form of \Eref{eq:Hspin} remains valid irrespectively on which microscopic mechanism stands behind the interaction parameters. It can be the superexchange~\cite{Anderson1959}, RKKY~\cite{RudermanKittel,Kasuya,Yosida}, double exchange~\cite{deGennes} or any other mechanism, provided that the rotations are small. Even biquadratic exchange~\cite{Nagaev} for small $\theta$ can be reformulated in the bilinear form \eref{eq:Hspin}. Without SO coupling, the energy change caused by the infinitesimal rotations of spins can be always described by the Heisenberg model and in this sense the method is very universal.

\par Another important point of the work of Liechtenstein \etal~\cite{LKAG1987} is that they have proposed a practical scheme for calculating the exchange parameters and proved for these purposes the magnetic force theorem~\cite{Mackintosh,Heine,Oswald}, which justifies the use of the single-particle energies, obtained from the Kohn-Sham (KS) equations in SDFT~\cite{KS,BH}, for evaluating the energy change caused by the infinitesimal rotations of spins. The theorem greatly simplifies the calculations and improves the numerical accuracy.

\par The basic variable of SDFT is the magnetization density $\hat{m}$. In the ground state, $\hat{m}$ is controlled by the exchange-correlation (xc) field $\hat{b}$. Therefore, instead of rotating $\hat{m}$, Liechtenstein \etal~\cite{LKAG1987} have proposed to rotate $\hat{b}$, assuming that $\hat{m}$ will automatically rotate by the same angle. This leads to the commonly used expression for $J_{ij}$:
\noindent
\begin{equation}
J_{ij} =  \frac{1}{2\pi} {\rm Im} \int_{- \infty}^{\varepsilon_{\rm F}} d \varepsilon \, {\rm Tr}_{L} \Big\{ \hat{G}_{ij}^{\uparrow}(\varepsilon) \hat{b}_{j}^{\phantom{\uparrow}}\hat{G}_{ji}^{\downarrow}(\varepsilon) \hat{b}_{i}^{\phantom{\uparrow}} \Big\},
\label{eq:JLKAG}
\end{equation}
\noindent which is nothing but the 2nd order perturbation theory for the single-particle energy, formulated in terms of the single-particle Green's functions $\hat{G}_{ij}^{\sigma}$ with the spins $\sigma=$ $\uparrow$ or $\downarrow$ ($\varepsilon_{\rm F}$ being the Fermi energy, ${\rm Tr}_{L}$ stands for the trace over orbital indices, and ${\rm Im}$ denotes the imaginary part). Although this result was anticipated by the previous works on the RKKY interactions~\cite{Yosida,Roth}, the exchange interactions in the paramagnetic medium~\cite{Oguchi1983,Oguchi}, as well as general theories of the itinerant magnetism~\cite{SHLiu,PrangeKorenman}, the expression  \eref{eq:JLKAG} can be relatively easily combined with the first-principles electronic structure calculations for the ground state, in the framework of SDFT or its refinements. Today, it is known for almost 40 years and was successfully applied for the analysis of interatomic magnetic interactions in various substances~\cite{IS2003,Kvashnin2015,Korotin2015,Yoon2018,Matsumoto,Nomoto2020,Grytsiuk,TB2J}.

\par Nevertheless, there are also open questions. Particularly, several authors have raised doubts that the true energy change caused by the infinitesimal rotations of spins can be described by rotating only $\hat{b}$ and suggested that it should include the additional contribution steaming from the external magnetic field, which is needed to control the direction of the magnetization~\cite{Stocks,Bruno2003,Streib}. Thus, \Eref{eq:JLKAG} may be incomplete. Presumably, the most persuasive arguments were given in 2003 by Bruno~\cite{Bruno2003}, who proposed how \Eref{eq:JLKAG} should be corrected. Surprisingly, however, that even 20 years later after this publication there is no systematic analysis of the problem: \Eref{eq:JLKAG} is widely used, but little is known how good it is. The problem is complicated by rather common misunderstanding putting the equality between \Eref{eq:JLKAG} and more fundamental magnetic force theorem.

\par Another question is what is the right object to rotate? The spin model \eref{eq:Hspin} is typically formulated on the lattice. Therefore, the magnetization should be also associated with the atomic sites, in some basis of atomic-like orbitals. Then, one can define and rotate the local moments, which are scalars. Alternatively, if there are several atomic orbitals per magnetic site, one can define the magnetization matrix and rotate it. For instance, \Eref{eq:JLKAG} implies such matrix form. Nevertheless, which construction is more suitable, based on rotations of the scalar moments or the magnetization matrices, is absolutely unclear.

\par Then, what shall we do with the ligand sites, which can hardly be the source of the magnetism, but frequently carry an appreciable magnetization due to the hybridization with the magnetic transition-metal sites? The contributions of such ligand states are typically ignored, and the interactions \eref{eq:JLKAG} are computed only between the transition-metal sites without the justification. On the other hand, there are well-known Goodenough-Kanamori-Anderson (GKA) rules~\cite{Anderson1950,Goodenough1955,Goodenough1958,Kanamori1959}, which state, among others, that in certain circumstances the exchange interactions between the transition-metal sites can be controlled by the effective Stoner coupling on the intermediate ligand sites. Such coupling is typically added empirically to correct $J_{ij}$ given by \Eref{eq:JLKAG} between the transition-metal sites~\cite{Mazurenko2007,StreltsovKhomskii,JPSJ2009}. However, if the theory is general enough, it should include all such contributions automatically.

\par The aim of this article is to review some basic ideas as well as more recent developments related to the use of the linear response methods for the analysis of interatomic exchange interactions and give clear answers to all above questions. The general theory is discussed in \Sref{sec:Hspin}. Then, Sections \ref{sec:CrX3}-\ref{sec:manganites} deal with practical examples for several types of compounds, where our main goal is to explain the theoretical nuances, which may arise in various parts of calculations of the interatomic exchange interactions: the applicability of \Eref{eq:JLKAG} and its refinements, the role of on-site Coulomb correlations, the contributions of the ligand sites, the merging of correlated and uncorrelated bands, etc. Short \Sref{sec:other} outline other developments beyond the main scopes of this review. The article is summarized in \Sref{sec:Summary}. Two Appendices deal with the construction of the tight-binding (TB) Hamiltonians using the band structure obtained in first-principles calculations and the evaluation of magnetic transition temperature for the spin model in the random phase approximation (RPA).

\section{\label{sec:Hspin} Infinitesimal spin rotations and exchange interactions}

\subsection{\label{sec:Def} Basic idea, notations, and conventions}
\par In the magnetic equilibrium, the 1st derivative of the total energy with respect to a small change of the magnetization $\boldsymbol{m}(\boldsymbol{r})$ vanishes and the energy changes is described by the 2nd derivative. In a general sense, the magnetic force theorem states that not only the total energy but also its 2nd derivative with respect to the infinitesimal rotations of the magnetization is the ground-state property as it can be expressed via the eigenvalues and eigenfunctions of the ground state. This statement can be traced back to fundamentals of the quantum mechanics, where the system can be measured only via perturbations. Therefore, it is logical that the 2nd derivative can be connected to the properties of the ground state. The response function (or the susceptibility) is the useful tool, which establishes such connection by means of the perturbation theory.

\par In practical terms, we will deal mainly with SDFT~\cite{HK,KS,BH}, where the ground-state magnetization density and the total energy are described with the help of the single-particle spin-dependent KS Hamiltonian $H^{\sigma}(\boldsymbol{r})$ with some local self-consistent potential incorporating all effects of exchange and correlations~\cite{KS,BH}. This locality implies that the change of the potential in certain point $\boldsymbol{r}$ depends only on the change of magnetization in the same point $\boldsymbol{r}$. As a consequence, the energy change caused by the infinitesimal rotations of the magnetization can be presented in the form of pairwise interactions. Without SO coupling it corresponds to the isotropic bilinear Heisenberg model. This is a general property of the 2nd order perturbation theory with the local potentials.

\par From the viewpoint of analysis and interpretation, it is more convenient to adopt the TB representation, which deals with the atomically resolved properties emerging from the solution of some lattice model. Another advantage of the TB representation is the on-site Coulomb interactions, which can be easily incorporated into the model. The purpose of these interactions is to correct limitations of the local density approximation (LDA) or the generalized gradient approximation (GGA), which are derived in the limit of homogeneous electron gas and typically used to describe the effects of exchange and correlations in SDFT. Mathematically, this can be done by constructing the orthonormal basis of localized Wannier functions centered on the atomic sites~\cite{WannierRevModPhys}. The KS Hamiltonian in this basis is the matrix specified by the lattice ($i$, $j$) and orbital ($a$, $b$) indices, $\hat{H}^{\sigma} \equiv [H_{ia,jb}^{\sigma}] $, so that all other matrices can be obtained from $\hat{H}^{\sigma}$. For instance, the Green function is $\hat{G}^{\sigma}(\varepsilon) = [ \varepsilon - \hat{H}^{\sigma} ]^{-1}$ and the magnitude of the magnetization is given by $\hat{m} =  \Theta(\varepsilon_{\rm F}$$-$$\hat{H}^{\uparrow}) - \Theta(\varepsilon_{\rm F}$$-$$\hat{H}^{\downarrow}) $, in terms of the Heaviside function $\Theta$. We assume that the xc potential in this TB representation remains local in the sense that on each atomic site $i$ it depends only on the magnetization $\hat{m}_{i}$ on the same site. It is a common practice to relate the magnetic part of such potential, the so-called xc field $\hat{b}$, with the site-diagonal elements of the TB Hamiltonian: $\hat{b}_{i} = \hat{H}^{\uparrow}_{ii}-\hat{H}^{\downarrow}_{ii}$. However, such $\hat{b}$ is ill-defined because it ignores the non-local contributions steaming from the off-diagonal part of $\hat{H}^{\sigma}_{ij}$~\cite{Nomoto2020}. A more consistent definition of the \emph{local} $\hat{b}$ in terms of the response function and the ground-state magnetization will be given in \Sref{sec:sumrule}.

\par Other conventions can be formulated as follows:
\begin{itemize}
\item For periodic systems, $[\hat{H}_{ia,jb}^{\sigma}]$ can be Fourier transformed to $[H_{\mu a, \nu b}^{\sigma}(\boldsymbol{k})] $, with $\mu$ and $\nu$ denoting the atomic positions within the primitive cell. Furthermore, the analysis throughout this paper assumes the use of the periodic gauge $H_{\mu a, \nu b}^{\sigma}(\boldsymbol{k}+\boldsymbol{G}) = H_{\mu a, \nu b}^{\sigma}(\boldsymbol{k})$ for any reciprocal lattice translation $\boldsymbol{G}$~\cite{KingSmith};
\item The $n$$\times$$n$ matrix $\hat{a}_{\mu}$, specified by the $n$ orbital indices, can be viewed as the column vector $\vec{a}_{\mu}$ of the length $n^{2}$. The scalar product $\vec{a}_{\mu}^{\, \dagger} \cdot \vec{b}_{\mu}^{\phantom{\dagger}}$ is the shorthand notation for ${\rm Tr}_{L} \{ \hat{a}_{\mu}\hat{b}_{\mu} \}$ (where $\vec{a}_{\mu}^{\, \dagger}$ is the row vector corresponding to the column vector $\vec{a}_{\mu}^{\, \phantom{\dagger}}$). In the case of Cartesian vectors (such as the magnetization matrix or the magnetic field interacting with the magnetization), the notation $\vec{\boldsymbol{a}}_{\mu}^{\, \dagger} \cdot \vec{\boldsymbol{b}}_{\mu}^{\phantom{\dagger}}$ stands for the regular scalar product with the summation over the orbital indices. The notation $\vec{\boldsymbol{a}}_{\phantom{\mu}}^{\, \dagger} \cdot \vec{\boldsymbol{b}}_{\phantom{\mu}}$ implies the summation over the atomic indices as well;
\item The $n$$\times$$n$$\times$$m$$\times m$ tensor $\mathcal{A} = [\mathcal{A}_{ab,cd}]$, with first two orbitals  $(ab)$ residing on the site $\mu$ and last two orbitals $(cd)$ residing on the site $\nu$, can be viewed as the $n^{2} \times m^{2}$ matrix $\hat{\mathcal{A}}_{\mu \nu}$. The construction $\hat{\mathcal{A}}_{\mu \nu} \vec{b}_{\nu}$ implies the summation over the two orbital indices on the site ${\nu}$;
\item The 2nd derivative is a local probe and the interaction parameter depends on the point in which it is calculated. For instance, considering the simplest interaction energy $E=-J\cos \varphi$ between two spins in the bond, the 2nd derivative near FM ($\varphi = 0$) and AFM ($\varphi = \pi$) configurations of spins will be, respectively, $J$ and $-J$. Nevertheless, as it is typically done, we will additionally change the sign of interaction parameters for the antiferromagnetically coupled bonds, thus adopting the universal definition where $J>0$ and $<0$ stands for the FM and AFM interactions, respectively.
\end{itemize}

\subsection{\label{sec:SODM} Spin spirals, spin-orbit coupling, and Dzyaloshinskii-Moriya interactions}
\par The SO interaction is known to consist of the spin-diagonal, $\frac{\xi}{2}\hat{L}^{z}\hat{\sigma}^{z}$, as well as off-diagonal, $\frac{\xi}{2}(\hat{L}^{x}\hat{\sigma}^{x} + \hat{L}^{y}\hat{\sigma}^{y})$, parts, where $\xi$ is the SO coupling parameter, $\hat{\boldsymbol{L}}=(\hat{L}^{x},\hat{L}^{y},\hat{L}^{z})$ is the vector of angular momenta, and $\hat{\boldsymbol{\sigma}}=(\hat{\sigma}^{x},\hat{\sigma}^{y},\hat{\sigma}^{z})$ is that of the Pauli matrices. The antisymmetric DM interaction $\boldsymbol{d}=(d^{x},d^{y},d^{z})$ emerges in the 1st order of $\xi$ and generally one should be able to calculate all three vector projections onto $x$, $y$, and $z$ (unless they are related by the symmetry properties). Nevertheless, by proper rotations of the coordinate frame, which transform $xyz$ to $\overline{z}yx$ and $zxy$, $d^{x}$ and $d^{y}$ can be viewed as $d^{z}$ in the new coordinate frame. Therefore, we need the numerical procedure only for calculating $d^{z}$. The important point in this respect was realized by Sandratskii~\cite{Sandratskii}, who suggested that in order to calculate $d^{z}$, it is sufficient to consider only spin-diagonal part of the SO coupling. His idea was based on a simple observation that DM interactions give rise to spiral magnetic structures~\cite{Dzyaloshinskii_helix}, which can be regarded as ``eigenstates'' of the spin model \eref{eq:Hspin}. Therefore, the energies of the model \eref{eq:Hspin} can be uniquely specified by the vectors $\boldsymbol{q}$, describing propagation of the spin spiral. Then, the same should hold for the electronic model, which is used for the mapping onto the spin one, and the spin spirals should be amount possible magnetic solutions of such model. In practical terms, this means that the electronic states should obey the generalized Bloch theorem, which combines translations with the SU(2) rotations of spins in the spiral texture~\cite{Sandratskii_review}. Nevertheless, this theorem can be applied only if the spin is the good quantum number so that the Hamiltonian $\hat{H}^{\sigma}$ remains diagonal with respect to the spin indices $\sigma$. Therefore, such $\hat{H}^{\sigma}$ can include the diagonal part of the SO coupling, but not the off-diagonal one.

\par The method is suitable for the DM interactions, but not for the magnetic anisotropy, which emerges in the 2nd order of the SO coupling and typically include both diagonal and off-diagonal contributions. This is again in line with the idea of the spin-spiral approach: the DM interactions give rise to the spin spirals, while the magnetic anisotropy acts against them, by deforming the spin spirals and locking them to the crystallographic lattice~\cite{Koehler,PRB2011}. The alternative to the spin-spiral technique is to work in the real space separately for each magnetic bond~\cite{Kvashnin2020,Ebert,Mahfouzi,Ebert2023}. Such methods, which have certain limitations, will be briefly considered in~\Sref{sec:brealspace}.

\subsection{\label{sec:echange} General expression for the energy change}
\par
\begin{figure}[t]
\begin{center}
\includegraphics[height=6cm]{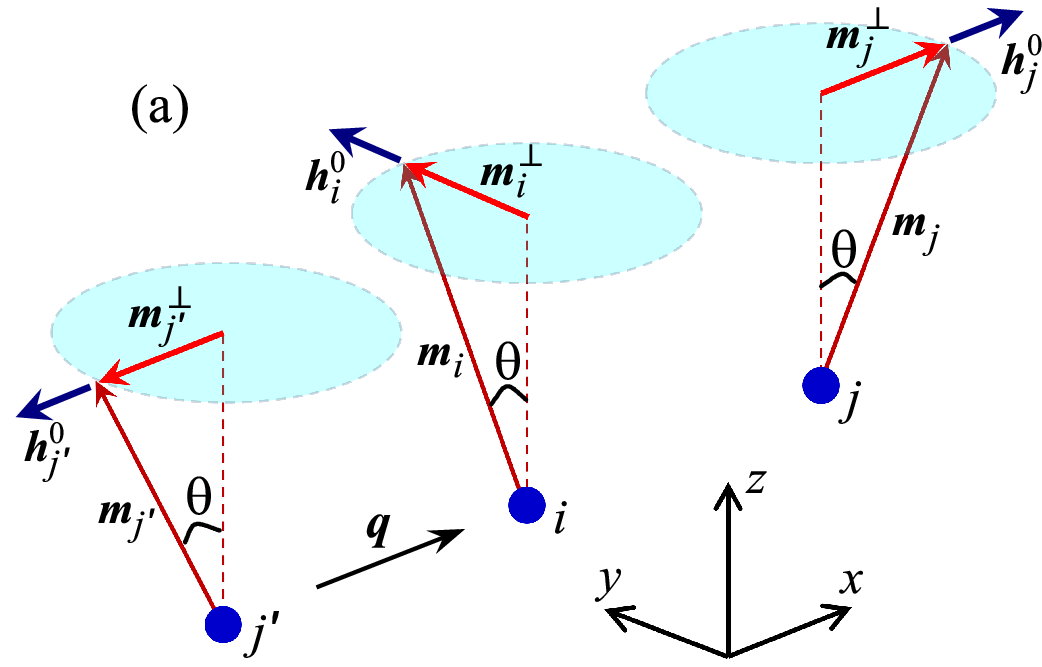} \hspace{0.5cm} \includegraphics[height=6cm]{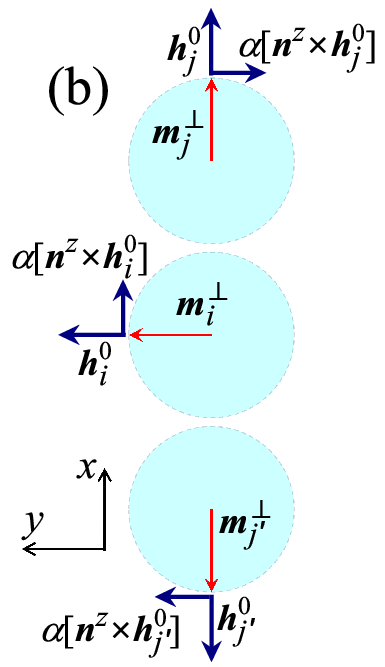}
\end{center}
\caption{(a) Conical spin spiral, which is assumed in calculations of interatomic exchange interactions: $\boldsymbol{q}$ is the propagation vector, $\boldsymbol{h}^{0}$ is the constraining field inducing the transversal magnetization $\boldsymbol{m}_{i}^{\perp}$ without the spin-orbit coupling, and $\boldsymbol{m}_{i}$ is the rotated magnetization on the site $i$. (b) Configuration of constraining field in the $xy$ plane: $\boldsymbol{h}^{0}$ is required to induce given transversal magnetization $\boldsymbol{m}^{\perp}$, while the additional perpendicular field, $\alpha \boldsymbol{n}^{z} \times \vec{\boldsymbol{h}}^{0}$, is required to compensate the additional rotation of $\boldsymbol{m}^{\perp}$ caused by the DM interaction $d^{z}$.}
\label{fig.cartoon}
\end{figure}

 As was already pointed out before, our basic idea is to ``excite'' the spin spiral, rotating the ground-state magnetization $\hat{\boldsymbol{m}}_{\rm GS} = (0,0,\hat{m})$ as
\noindent
\begin{equation}
\hat{\boldsymbol{m}}_{\boldsymbol{q},i} = \Big( \theta \cos \boldsymbol{q} \boldsymbol{R}_{i}, \theta \sin \boldsymbol{q} \boldsymbol{R}_{i}, 1-\frac{\theta^2}{2} \Big) \, \hat{m},
\label{eq:mq}
\end{equation}
\noindent (see \Fref{fig.cartoon}) and evaluate the interactions between the transversal ``fluctuations'' of the magnetization $\hat{\boldsymbol{m}}_{\boldsymbol{q},i}^{\perp} = ( \cos \boldsymbol{q} \boldsymbol{R}_{i}, \sin \boldsymbol{q} \boldsymbol{R}_{i}, 0 ) \, \theta \hat{m}$ for small $\theta$. In order to induce such $\hat{\boldsymbol{m}}_{\boldsymbol{q},i}^{\perp}$, we have to apply the external field
\noindent
\begin{equation}
\hat{\boldsymbol{h}}_{\boldsymbol{q},i} = \Big( \cos (\boldsymbol{q} \boldsymbol{R}_{i} + \alpha_{\boldsymbol{q}} ), \sin (\boldsymbol{q} \boldsymbol{R}_{i} + \alpha_{\boldsymbol{q}} ), 0 \Big) \hat{h}
\label{eq:extfield}
\end{equation}
\noindent in the direction perpendicular to the ground state magnetization. If the SO coupling is included, $\hat{\boldsymbol{h}}_{\boldsymbol{q},i}$ is not necessarily parallel to $\hat{\boldsymbol{m}}_{\boldsymbol{q},i}^{\perp}$ as the latter can experience the effect of the DM interaction $d^{z}$, which tend to additionally rotate the magnetization in the $xy$ plane. Therefore, the phases $\alpha_{\boldsymbol{q}}$ are needed to compensate the effect of DM interactions (see \Fref{fig.cartoon}). For small $\alpha_{\boldsymbol{q}}$, $\hat{\boldsymbol{h}}_{\boldsymbol{q},i}$ can be written as
\noindent
\begin{equation}
\hat{\boldsymbol{h}}_{\boldsymbol{q},i} \approx  \hat{\boldsymbol{h}}_{\boldsymbol{q},i}^{0} + \alpha_{\boldsymbol{q}} \boldsymbol{n}^{z} \times \hat{\boldsymbol{h}}_{\boldsymbol{q},i}^{0},
\label{eq:extfield1}
\end{equation}
\noindent where $\hat{\boldsymbol{h}}_{\boldsymbol{q},i}^{0} = \left( \cos \boldsymbol{q} \boldsymbol{R}_{i} , \sin \boldsymbol{q} \boldsymbol{R}_{i}, 0 \right) \hat{h}$ and $\boldsymbol{n}^{z} = (0,0,1)$.

\par The corresponding energy can be evaluated in the framework of constrained SDFT~\cite{Bruno2003,Dederichs} as:
\noindent
\begin{equation}
{\cal E}[\vec{\boldsymbol{m}}] = {\cal T}[\vec{\boldsymbol{m}}] + {\cal E}_{\rm xc}[\vec{\boldsymbol{m}}] + \frac{1}{2} \vec{\boldsymbol{h}}_{\boldsymbol{q}}^{\dagger} \cdot \left( \vec{\boldsymbol{m}} - \vec{\boldsymbol{m}}_{\boldsymbol{q}}^{\perp} \right),
\label{eq:tenergy}
\end{equation}
\noindent where ${\cal T}$ and ${\cal E}_{\rm xc}$ are, respectively, the kinetic and exchange-correlation (xc) energies, while the last term is the penalty term controlling the size of the transversal magnetization $\vec{\boldsymbol{m}}_{\boldsymbol{q}}^{\perp}$. For simplicity, we drop here all irrelevant dependencies of ${\cal E}$ on the charge density.
\noindent

\par Then, the kinetic energy can be expressed as the sum of the occupied KS single-particle energies, ${\cal E}_{\rm sp}$, calculated for the external field $\vec{\boldsymbol{h}}_{\boldsymbol{q}}$ and the xc field
\noindent
\begin{displaymath}
\vec{\boldsymbol{b}}_{\boldsymbol{q}} = 2  \frac{\partial {\cal E}_{\rm xc}[\vec{\boldsymbol{m}}]}{\partial \vec{\boldsymbol{m}}} \Big\vert_{\vec{\boldsymbol{m}} = \vec{\boldsymbol{m}}_{\boldsymbol{q}}},
\end{displaymath}
\noindent minus the interaction energy of $\vec{\boldsymbol{m}}_{\boldsymbol{q}}$ with $\vec{\boldsymbol{h}}_{\boldsymbol{q}}$ and $\vec{\boldsymbol{b}}_{\boldsymbol{q}}$~\cite{KS,BH}, yielding
\begin{equation}
{\cal E}[\vec{\boldsymbol{m}}_{\boldsymbol{q}}] = {\cal E}_{\rm sp} ( \vec{\boldsymbol{h}}_{\boldsymbol{q}} + \vec{\boldsymbol{b}}_{\boldsymbol{q}} ) - \frac{1}{2} (\vec{\boldsymbol{h}}_{\boldsymbol{q}} + \vec{\boldsymbol{b}}_{\boldsymbol{q}})^{\dagger} \cdot \vec{\boldsymbol{m}}_{\boldsymbol{q}} + {\cal E}_{\rm xc}[\vec{\boldsymbol{m}}_{\boldsymbol{q}}].
\label{eq:tenergy2}
\end{equation}

\par The important property of the xc energy in this respect, being the consequence of fundamental gauge invariance of the density functional theory~\cite{Vignale1987}, is that rotations of the spin magnetization $\hat{\boldsymbol{m}}_{\rm GS} \rightarrow \hat{\boldsymbol{m}}_{\boldsymbol{q},i}$ do not change ${\cal E}_{\rm xc}$: ${\cal E}_{\rm xc}[ \hat{\boldsymbol{m}}_{\boldsymbol{q},i}] = {\cal E}_{\rm xc}[ \hat{\boldsymbol{m}}_{\rm GS} ]$~\cite{Vignale1988,PRB1998,Katsnelson_DM}. This is a general property of SDFT, which becomes especially transparent in the local spin-density approximation (LSDA), based on the picture of homogeneous electron gas. In this case, ${\cal E}_{\rm xc}$ in each point $\boldsymbol{r}$ depends only on the magnitude of the magnetization, ${\cal E}_{\rm xc}^{\rm LSDA} \equiv {\cal E}_{\rm xc}^{\rm LSDA}[|\boldsymbol{m}(\boldsymbol{r})|]$~\cite{Kuebler,EichGross}, and therefore does not changes under rotations of $\boldsymbol{m}(\boldsymbol{r})$. Since ${\cal E}_{\rm xc}[ \hat{\boldsymbol{m}}_{\boldsymbol{q},i}] = {\cal E}_{\rm xc}[ \hat{\boldsymbol{m}}_{\rm GS} ]$, the rotation of the magnetization will rotate the xc field by the same angle:
\noindent
\begin{equation}
\hat{\boldsymbol{b}}_{\boldsymbol{q},i} = \Big( \theta \cos \boldsymbol{q} \boldsymbol{R}_{i}, \theta \sin \boldsymbol{q} \boldsymbol{R}_{i}, 1-\frac{\theta^2}{2} \Big) \, \hat{b}.
\label{eq:chtenergy}
\end{equation}
\noindent Therefore, $\vec{\boldsymbol{b}}_{\boldsymbol{q}}^{\dagger} \cdot \vec{\boldsymbol{m}}_{\boldsymbol{q}}$ does not change either and \Eref{eq:tenergy2} will lead to the following energy change:
\noindent
\begin{displaymath}
\delta{\cal E}_{\boldsymbol{q}} = \delta{\cal E}_{\rm sp} ( \vec{\boldsymbol{h}}_{\boldsymbol{q}} + \vec{\boldsymbol{b}}_{\boldsymbol{q}} ) - \frac{1}{2} \vec{\boldsymbol{h}}_{\boldsymbol{q}}^{\dagger} \cdot \vec{\boldsymbol{m}}_{\boldsymbol{q}}.
\end{displaymath}
\noindent
Then, $\delta {\cal E}_{\rm sp}$ can be evaluated by treating $\vec{\boldsymbol{h}}_{\boldsymbol{q}} + \vec{\boldsymbol{b}}_{\boldsymbol{q}}$ as a perturbation, to the 2nd order in $\vec{\boldsymbol{h}}_{\boldsymbol{q}}^{\phantom{\perp}} + \vec{\boldsymbol{b}}_{\boldsymbol{q}}^{\perp}$ and the 1st order in the longitudinal change of the xc field, $-\frac{1}{2} \hat{b} \theta^2$. The details are elaborated in ref.~\cite{PRB2021}, leading to the simple but general expression:
\noindent
\begin{equation}
\delta {\cal E}_{\boldsymbol{q}} = -\frac{1}{4} \, \vec{\boldsymbol{h}}_{\boldsymbol{q}}^{\,0 \dagger} \cdot  \vec{\boldsymbol{m}}_{\boldsymbol{q}}^{\perp }.
\label{eq:techange}
\end{equation}
\noindent In fact, this result is well anticipated. On the one hand, $\delta {\cal E}_{\boldsymbol{q}}$ should be proportional to $\vec{\boldsymbol{h}}_{\boldsymbol{q}}$ as there would be no energy change without the external field. On the other hand, there only possible interaction of $\vec{\boldsymbol{h}}_{\boldsymbol{q}}$ with the constrained magnetization $\vec{\boldsymbol{m}}_{\boldsymbol{q}}^{\perp }$ is the scalar product given by \Eref{eq:techange}. It may look incomplete because $\delta {\cal E}_{\boldsymbol{q}}$ does not seem to know anything about $\alpha_{\boldsymbol{q}}$ and the DM interactions. Nevertheless, all necessary information is in \Eref{eq:techange} and in \Sref{sec:exexact} we will show how it should be used to derive practical expressions for the isotropic exchange \emph{and} DM interactions.

\par In addition to the rotations given by \eref{eq:mq}, the magnetization can experience the longitudinal change, which is caused by these rotations. It will affect $\hat{m}$, resulting in an additional change of each of the terms in \Eref{eq:tenergy2}. Nevertheless, these contributions can be shown to cancel out in the lowest order of $\theta$~\cite{LKAG1987,PRB1998}.

\subsection{\label{sec:rtensor} Response tensor}
\par The response theory is basically the perturbation theory relating the small change of the potential $\vec{v}$ with the induced density $\vec{n}$: $\vec{n} = \hat{\mathcal{R}} \vec{v}$. Spin-dependent $\vec{v}$ can be generally specified by four elements:
\noindent
\begin{displaymath}
\vec{v} = \left(
\begin{array}{cc}
\vec{v}^{\, \uparrow   \uparrow} & \vec{v}^{\, \uparrow   \downarrow} \\
\vec{v}^{\, \downarrow \uparrow} & \vec{v}^{\, \downarrow \downarrow}
\end{array}
\right).
\end{displaymath}
\noindent Then, each $\vec{v}^{\, \sigma   \sigma'}$ induces the corresponding change $\vec{n}^{\, \sigma   \sigma'}$:
\begin{equation}
\vec{n}^{\, \sigma   \sigma'} = \hat{\mathcal{R}}^{ \sigma   \sigma'} \vec{v}^{\, \sigma   \sigma'},
\label{eq:dresponse}
\end{equation}
\noindent where the rank-4 tensor $\hat{\mathcal{R}}^{ \sigma   \sigma'}$ can be found in terms of the 1st-order perturbation theory for the wave functions~\cite{PRB2014}. In our case, the perturbation is $\vec{\boldsymbol{h}}_{\boldsymbol{q}} + \vec{\boldsymbol{b}}_{\boldsymbol{q}}^{\perp}$ and our goal is to evaluate $\vec{\boldsymbol{m}}_{\boldsymbol{q}}^{\perp}$. Then, it is convenient to use the local coordinate frame where $\hat{\boldsymbol{h}}_{i} = ( 1, \alpha_{\boldsymbol{q}}, 0 ) \hat{h}^{0} $, which is obtained by rotating $\hat{\boldsymbol{h}}_{\boldsymbol{q},i}$ about $z$ by the angles $-$$\boldsymbol{q} \boldsymbol{R}_{i}$, and employ the generalized Bloch theorem, combining lattice translations with the SU(2) rotations of spins~\cite{Sandratskii_review}. This will lead to the additional shift of the $\boldsymbol{k}$-mesh for the states with $\sigma = \uparrow$ relative to those with $\sigma = \downarrow$. Moreover, since $\hat{\boldsymbol{h}}_{i} = ( 1, \alpha_{\boldsymbol{q}}, 0 )\hat{h}^{0} $ corresponds to
\begin{displaymath}
\vec{v}_{h} = \frac{1}{2} \left(
\begin{array}{cc}
0 & 1-i\alpha_{\boldsymbol{q}} \\
1+i\alpha_{\boldsymbol{q}} & 0
\end{array}
\right)\vec{h}^{0},
\end{displaymath}
\noindent we have to consider only $\hat{\mathcal{R}}^{ \uparrow \downarrow}$ and $\hat{\mathcal{R}}^{ \downarrow \uparrow}$. Then, the perturbation theory yields
\noindent
\begin{equation}
{\cal R}_{ab,cd}^{\uparrow \downarrow} (\boldsymbol{q}) = \sum_{ml \boldsymbol{k}} \frac{f_{m \boldsymbol{k}}^{\uparrow} - f_{l \boldsymbol{k}+\boldsymbol{q}}^{\downarrow}}{\varepsilon_{m \boldsymbol{k}}^{\uparrow} - \varepsilon_{l \boldsymbol{k}+\boldsymbol{q}}^{\downarrow}} (C_{m \boldsymbol{k}}^{a \, \uparrow})^{*}C_{l \boldsymbol{k}+\boldsymbol{q}}^{b \, \downarrow} (C_{l \boldsymbol{k}+\boldsymbol{q}}^{c \, \downarrow})^{*}C_{m \boldsymbol{k}}^{d \, \uparrow},
\label{eq:gresponse}
\end{equation}
\noindent where $\varepsilon_{m \boldsymbol{k}}^{\sigma}$ and $| C_{l \boldsymbol{k}}^{\sigma} \rangle = [ C_{l \boldsymbol{k}}^{a\sigma}]$ are, respectively, the eigenvalues and  eigenvectors of $\hat{H}^{\sigma}$, and $f_{m \boldsymbol{k}}^{\sigma} \equiv \Theta(\varepsilon_{\rm F} - \varepsilon_{m \boldsymbol{k}}^{\sigma})$ is the Fermi distribution function. The orbital indices in each of the pairs $ab$ and $cd$ belong to the same atomic sites in the unit cell. Furthermore, ${\cal R}^{\downarrow \uparrow}$ can be obtained from ${\cal R}^{\uparrow \downarrow}$ using the property
\noindent
\begin{equation}
{\cal R}_{ab,cd}^{\uparrow \downarrow} (\boldsymbol{q}) = \big[ {\cal R}_{ba,dc}^{\downarrow \uparrow} (-\boldsymbol{q}) \big]^{*}.
\label{eq:Rproperty}
\end{equation}
\noindent If $\hat{H}^{\sigma}$ remains invariant under the time reversal (e.g., without SO interaction), \Eref{eq:Rproperty} is reduced to ${\cal R}_{ab,cd}^{\uparrow \downarrow} (\boldsymbol{q}) = {\cal R}_{ba,dc}^{\downarrow \uparrow} (\boldsymbol{q})$.\footnote{Without SO interaction, ${\cal R}_{ab,cd}^{\uparrow \downarrow}$ and ${\cal R}_{ab,cd}^{\downarrow \uparrow}$ are considered only in the combination with the \emph{symmetric} matrices $\hat{m}$ and $\hat{b}$. Therefore, one can write ${\cal R}_{ab,cd}^{\uparrow \downarrow} = {\cal R}_{ab,cd}^{\downarrow \uparrow}$.}

\par ${\cal R}_{ab,cd}^{\uparrow \downarrow} (\boldsymbol{q})$ can be also related to the Green function $\hat{G}^{\sigma}(\varepsilon,\boldsymbol{k}) = [ \varepsilon - \hat{H}^{\sigma}(\boldsymbol{k}) ]^{-1}$ as~\cite{KL2004}
\noindent
\begin{equation}
\mathcal{R}^{\uparrow \downarrow}_{ab , cd} (\boldsymbol{q}) = -\frac{1}{\pi} \sum_{\boldsymbol{k}}^{\rm BZ} {\rm Im} \int_{- \infty}^{\varepsilon_F} d \varepsilon \left\{ G_{da}^{\uparrow}(\varepsilon,\boldsymbol{k}) G_{bc}^{\downarrow}(\varepsilon,\boldsymbol{k}+\boldsymbol{q}) \right\}
\label{eq:rabcd}
\end{equation}
\noindent with the summation running over the 1st Brillouin zone (BZ).

\par Then, using the definition \eref{eq:dresponse}, one can find:
\noindent
\begin{displaymath}
\vec{n}^{\, \uparrow \downarrow} = \frac{1}{2} \hat{\mathcal{R}}^{\uparrow \downarrow}_{\boldsymbol{q}}  \left( \vec{h}^{0} + \vec{b}^{\perp} -i\alpha_{\boldsymbol{q}} \vec{h}^{0} \right)
\end{displaymath}
\noindent and
\noindent
\begin{displaymath}
\vec{n}^{\, \downarrow \uparrow} = \frac{1}{2} \hat{\mathcal{R}}^{\downarrow \uparrow}_{\boldsymbol{q}}  \left( \vec{h}^{0} + \vec{b}^{\perp} +i\alpha_{\boldsymbol{q}} \vec{h}^{0} \right),
\end{displaymath}
\noindent where $\hat{\mathcal{R}}^{\sigma \sigma'}_{\boldsymbol{q}} \equiv \hat{\mathcal{R}}^{\sigma \sigma'}({\boldsymbol{q}})$. In the local coordinate frame, these $\vec{n}^{\, \uparrow \downarrow}$ and $\vec{n}^{\, \downarrow \uparrow}$ should give us the magnetization $\vec{m}^{\perp} = \vec{n}^{\, \uparrow \downarrow} + \vec{n}^{\, \downarrow \uparrow}$ along $x$:
\noindent
\begin{equation}
\vec{m}^{\perp} = \hat{\mathcal{R}}^{+}_{\boldsymbol{q}} (\vec{h}^{0} + \vec{b}^{\perp}) - i \alpha_{\boldsymbol{q}}  \hat{\mathcal{R}}^{-}_{\boldsymbol{q}}  \vec{h}^{0},
\label{eq:mx}
\end{equation}
\noindent where $\hat{\mathcal{R}}^{\pm}_{\boldsymbol{q}} = \frac{1}{2} \left( \hat{\mathcal{R}}^{\uparrow \downarrow}_{\boldsymbol{q}} \pm \hat{\mathcal{R}}^{\downarrow \uparrow}_{\boldsymbol{q}} \right)$. Another equation,
\begin{equation}
i \hat{\mathcal{R}}^{-}_{\boldsymbol{q}} (\vec{h}^{0} + \vec{b}^{\perp})+ \alpha_{\boldsymbol{q}}  \hat{\mathcal{R}}^{+}_{\boldsymbol{q}} \vec{h}^{0} = 0,
\label{eq:my}
\end{equation}
\noindent requires that the perpendicular to it magnetization along $y$, $i (\vec{n}^{\, \uparrow \downarrow}$$-$$\vec{n}^{\, \downarrow \uparrow} )$, should vanish (so as the $y$ component of the xc field) according to our constraint conditions. These are the equations for $\vec{h}^{0}$ and $\alpha_{\boldsymbol{q}}$ for given $\vec{m}^{\perp} = \theta \vec{m}$. Their meaning is very straightforward. For instance, in \Eref{eq:my}, the isotropic part of the magnetization $\alpha_{\boldsymbol{q}}  \hat{\mathcal{R}}^{+}_{\boldsymbol{q}} \vec{h}^{0}$, which is induced by $\alpha_{\boldsymbol{q}}  \vec{h}^{0}$ along $y$, is compensated by the one, which is induced due to the DM interaction by the field $\vec{h}^{0}$$+$$\vec{b}^{\perp}$ acting in the perpendicular direction $x$. The same is with \Eref{eq:mx}, where $\vec{m}^{\perp}$ has two components: the isotropic one, induced by $\vec{h}^{0}$$+$$\vec{b}^{\perp}$ along $x$, and the one caused by the DM interaction, transferring the effect of the magnetic field $\alpha_{\boldsymbol{q}}  \vec{h}^{0}$, applied along $y$, to the magnetization along $x$. This explains how one can naturally separate the contributions of the isotropic and DM interactions in \Eref{eq:techange}.

\subsection{\label{sec:exexact} Exchange interactions}
\par The next step is the mapping of the total energy change \eref{eq:techange} onto the spin model:
\noindent
\begin{equation}
\delta {\cal E}_{\boldsymbol{q}} = - \frac{1}{2} \sum_{\mu \nu} \left( J_{\boldsymbol{q}, \, \mu \nu} - id_{\boldsymbol{q}, \, \mu \nu}^{z} \right) \theta_{\mu} \theta_{\nu},
\label{eq:eqmapping}
\end{equation}
\noindent where we explicitly consider the possibility of having several magnetic sublattices. In the local coordinate frame, \Eref{eq:techange}, can be rearranged as
\noindent
\begin{displaymath}
\delta {\cal E}_{\boldsymbol{q}} = -\frac{1}{4} \, (\vec{h}^{\,0} + \vec{b}^{\perp })^{\dagger} \cdot  \vec{m}^{\perp } + \frac{1}{4}\vec{b}^{\perp \dagger} \cdot  \vec{m}^{\perp },
\end{displaymath}
\noindent where we have added and subtracted the xc field $\vec{b}^{\perp }$. Our strategy is to start with the expression for $\vec{m}^{\perp}$ without the SO coupling, which is given by the 1st term in \Eref{eq:mx}, and then consider the corrections arising in the 1st order of the SO coupling, which are given by the 2nd term.\footnote{Since we neglect spin-off-diagonal elements of the SO coupling (see \Sref{sec:SODM}), higher-order corrections are meaningless.} Then, noting that
\begin{equation}
\vec{h}^{0} + \vec{b}^{\perp}  = \hat{\mathcal{Q}}^{+}_{\boldsymbol{q}} \vec{m}^{\perp},
\label{eq:bx}
\end{equation}
\noindent where $\hat{\mathcal{Q}}^{\sigma \sigma'}_{\boldsymbol{q}} = \left[ \hat{\mathcal{R}}^{\sigma \sigma'}_{\boldsymbol{q}} \right]^{-1}$, $\hat{\mathcal{Q}}^{\pm}_{\boldsymbol{q}} = \frac{1}{2} \left( \hat{\mathcal{Q}}^{\uparrow \downarrow}_{\boldsymbol{q}} \pm \hat{\mathcal{Q}}^{\downarrow \uparrow}_{\boldsymbol{q}} \right)$, and without SO coupling $\hat{\mathcal{Q}}^{+}_{\boldsymbol{q}} = [\hat{\mathcal{R}}^{+}_{\boldsymbol{q}}]^{-1}$, one immediately finds the following expression for the isotropic exchange interactions:
\noindent
\begin{equation}
J_{\boldsymbol{q}, \, \mu \nu} = \frac{1}{2} \left( \vec{m}^{\, \dagger}_{\mu} \cdot \hat{\mathcal{Q}}^{+}_{\boldsymbol{q}, \, \mu \nu} \vec{m}^{\, \phantom{\dagger}}_{\nu} - \vec{b}^{\, \dagger}_{\mu} \cdot \vec{m}^{\, \phantom{\dagger}}_{\mu} \delta_{\mu \nu} \right).
\label{eq:jq}
\end{equation}

\par Considering the 2nd term in \Eref{eq:mx}, the construction $i (\vec{h}^{\,0}$$+$$\vec{b}^{\perp})^{\dagger} \cdot \hat{\mathcal{R}}^{-}_{\boldsymbol{q}} \alpha_{\boldsymbol{q}}  \vec{h}^{0}$ describes the interaction between $x$ and $y$ components of the magnetic field caused by the DM interactions. Corresponding interaction parameter should satisfy the condition
\noindent
\begin{equation}
d_{\boldsymbol{q}, \, \mu \nu}^{z} \, \theta_{\mu} \theta_{\nu} = \frac{i}{2} (\vec{h}^{\,0}_{\mu}+\vec{b}^{\perp}_{\mu})^{\dagger} \cdot \hat{\mathcal{R}}^{-}_{\boldsymbol{q}, \, \mu \nu} (\vec{h}^{\,0}_{\nu}+\vec{b}^{\perp}_{\nu}),
\label{eq:dzqb}
\end{equation}
\noindent where we had to ``rescale'' $y$ components of the magnetic field, $\alpha_{\boldsymbol{q}}  \vec{h}^{\, 0}_{\mu} \to \vec{h}^{\,0}_{\mu}+\vec{b}^{\perp}_{\mu}$, in order to specify $x$ and $y$ components of the transversal magnetization by the same set of parameters $\theta_{\mu}$ and $\theta_{\nu}$. Then, using \Eref{eq:bx} and noting that to the 1st order in the SO coupling $\hat{\mathcal{Q}}^{+} \hat{\mathcal{R}}^{-} \hat{\mathcal{Q}}^{+} = -\hat{\mathcal{Q}}^{-}$, one can find that
\noindent
\begin{equation}
d_{\boldsymbol{q}, \, \mu \nu}^{z} = -\frac{i}{2} \vec{m}^{\, \dagger}_{\mu} \cdot \hat{\mathcal{Q}}^{-}_{\boldsymbol{q}, \, \mu \nu} \vec{m}^{\, \phantom{\dagger}}_{\nu}.
\label{eq:dzq}
\end{equation}
\noindent This expression was obtained in ref.~\cite{PRB2023} basically heuristically, by the analogy with isotropic interactions and similar expression formulated in terms of the xc fields, which will be considered in \Sref{sec:rbxc}. Here, we have provided a more rigorous proof of \Eref{eq:dzq}. The real space parameters can obtained by the Fourier transform of $J_{\boldsymbol{q}, \, \mu \nu}$ and $d_{\boldsymbol{q}, \, \mu \nu}^{z}$.

\par Thus, the exchange interactions are proportional to the \emph{inverse} response function. For the isotropic exchange, this is basically the result of Bruno~\cite{Bruno2003}. For the Hubbard model, similar relationship has been established by Szczech \etal~\cite{Szczech}.

\par For practical purposes, it may be more convenient to calculate
\noindent
\begin{equation}
X_{\boldsymbol{q}, \, \mu \nu} = \frac{1}{2} \left( \vec{m}^{\, \dagger}_{\mu} \cdot \hat{\mathcal{Q}}^{\uparrow \downarrow}_{\boldsymbol{q}, \, \mu \nu} \vec{m}^{\, \phantom{\dagger}}_{\nu} - \vec{b}^{\, \dagger}_{\mu} \cdot \vec{m}^{\, \phantom{\dagger}}_{\mu} \delta_{\mu \nu} \right),
\label{eq:xq}
\end{equation}
\noindent in terms of only $\hat{\mathcal{Q}}^{\uparrow \downarrow}_{\boldsymbol{q}}$, and then relate it with $J_{\boldsymbol{q}, \, \mu \nu}$ and $d_{\boldsymbol{q}, \, \mu \nu}^{z}$ using the property \eref{eq:Rproperty}, which yields
\noindent
\begin{equation}
J_{\boldsymbol{q}, \, \mu \nu} = \frac{1}{2} \left( X_{\boldsymbol{q}, \, \mu \nu} + X_{-\boldsymbol{q}, \, \mu \nu}^{*} \right)
\label{eq:jq1}
\end{equation}
\noindent and
\noindent
\begin{equation}
d_{\boldsymbol{q}, \, \mu \nu}^{z} = -\frac{i}{2} \left( X_{\boldsymbol{q}, \, \mu \nu} - X_{-\boldsymbol{q}, \, \mu \nu}^{*} \right).
\label{eq:dzq1}
\end{equation}
\noindent Thus, $J_{\boldsymbol{q}, \, \mu \nu}$ is related to the average energy of spin spirals propagating in $\boldsymbol{q}$ and $-\boldsymbol{q}$, while $d_{\boldsymbol{q}, \, \mu \nu}^{z}$ is related to the energy difference~\cite{Sandratskii}.

\subsection{\label{sec:sumrule} Sum rule and local exchange-correlation field}
\par The sum rule is obtained from the identity
\noindent
\begin{displaymath}
\left[ \hat{G}^{\downarrow}(\varepsilon,\boldsymbol{k}) \right]^{-1} - \left[ \hat{G}^{\uparrow}(\varepsilon,\boldsymbol{k}) \right]^{-1} = \hat{b},
\end{displaymath}
which can be further rearranged as
\noindent
\begin{displaymath}
\hat{G}^{\uparrow}(\varepsilon,\boldsymbol{k}) - \hat{G}^{\downarrow}(\varepsilon,\boldsymbol{k}) = \hat{G}^{\downarrow}(\varepsilon,\boldsymbol{k}) \, \hat{b} \, \hat{G}^{\uparrow}(\varepsilon,\boldsymbol{k})
                                                                                                  = \hat{G}^{\uparrow}(\varepsilon,\boldsymbol{k}) \, \hat{b} \, \hat{G}^{\downarrow}(\varepsilon,\boldsymbol{k}),
\end{displaymath}
\noindent where $\hat{b} = \hat{H}^{\uparrow}(\boldsymbol{k})$$-$$\hat{H}^{\downarrow}(\boldsymbol{k})$ is assumed to be local (i.e., site-diagonal and not depending on $\boldsymbol{k}$). Then, integrating over $\varepsilon$ and $\boldsymbol{k}$, and using the definition \eref{eq:rabcd} for the response tensor, one can find:
\noindent
\begin{equation}
\vec{m} = \hat{\mathcal{R}}^{\uparrow \downarrow}_{0} \vec{b}.
\label{eq:sumrule}
\end{equation}
\noindent This sum rule has very straightforward meaning: $\boldsymbol{q} = 0$ corresponds to the uniform rotation of the ground-state magnetization, where all spins are rotated in the same direction by the same angle. Therefore, the transversal magnetization is described by the same xc field $\vec{b}$ as in the ground state (without any constraining fields).

\par Nevertheless, in the TB representation, such xc field is not necessary local. For instance, in LSDA, the splitting $\hat{H}^{\uparrow}(\boldsymbol{k})$$-$$\hat{H}^{\downarrow}(\boldsymbol{k})$ can have interatomic matrix elements and depend on $\boldsymbol{k}$. In such a situation, it can be important to reenforce the sum rule, by defining new \emph{local} xc field as $\vec{b} = \hat{\mathcal{Q}}^{\uparrow \downarrow}_{0} \vec{m}$, which would yield the given ground-state magnetization $\vec{m}$. For instance, this is a simple and transparent alternative to the kernel polynomial method, which was recently proposed to deal with nonlocal matrix elements of the xc field~\cite{Nomoto2020}. In fact, if the xc field is nonlocal, the total energy change for the infinitesimal rotations of spins is no longer representable in the form of pairwise interactions.

\subsection{\label{sec:object} Right object to rotate: magnetization matrices versus local magnetic moments}
\par So far, we did not properly specify the spin object which should be rotated on the magnetic sites in order to obtain the total energy change \eref{eq:techange}. All above discussions implied that it is the magnetization matrix $\hat{m}_{\mu}$, while the spin model is typically formulated in terms of the magnetic moments $M_{\mu} = {\rm Tr}_{L} \{ \hat{m}_{\mu} \}$. Undoubtedly, the rotation of $\hat{m}_{\mu}$, as a whole, by the angle $\theta_{\mu}$ will rotate $M_{\mu}$ by the same angle. However, is this choice unique? Are there other perturbations of $\hat{m}_{\mu}$, resulting in the same rotations of $M_{\mu}$ but preferably at lower energy cost? Here, we will follow the discussion in ref.~\cite{PRB2021}. Nevertheless, we would like to note that somewhat similar ideas can be found in the work of Antropov \etal~\cite{Antropov2006}.

\par Indeed, for the Hermitian matrix $\hat{m}_{\mu}$, one can always choose the diagonal representation $\hat{m}_{\mu} = {\rm diag} ( \, \dots, \, m^{a}_{\mu}, \, \dots )$ with respect to the orbital indices. In principle, each orbital $a$ in such representation can be rotated by its own angle $\theta_{\mu}^{a}$. Then, the transversal magnetization in the local coordinate frame, where it is parallel to $x$, will be $\hat{m}_{\mu}^{\perp} = {\rm diag} ( \, \dots, \, \theta_{\mu}^{a} m^{a}_{\mu}, \, \dots )$. Nevertheless, these angles are subjected to the additional constraint because ${\rm Tr}_{L} \{ \hat{m}_{\mu}^{\perp} \}$ should be equal to $\theta_{\mu} M_{\mu}$. Importantly, this condition is softer than the rotation of $\hat{m}_{\mu}$ as a whole, where all orbitals are rotated by the same $\theta_{\mu}^{a} = \theta_{\mu}$. Therefore, it is reasonable to expect that the energy change will be smaller, so as the exchange parameters.

\par Mathematically, we have to minimize the energy change \eref{eq:techange} with the additional condition $\sum_{a} \left( \theta^{a}_{\mu} - \theta_{\mu}^{\phantom{a}} \right) m^{a}_{\mu} =0$ on each site $\mu$:
\noindent
\begin{equation}
\delta {\cal E} = - \frac{1}{4} \sum_{\mu a} \left\{ \theta^{a}_{\mu} m^{a}_{\mu} h^{0 a}_{\mu} - \left( \theta^{a}_{\mu} - \theta_{\mu}^{\phantom{a}} \right) m^{a}_{\mu} \lambda_{\mu} \right\},
\label{eq:ctheta}
\end{equation}
\noindent where $h^{0 a}_{\mu}$ are the constraining fields acting on $m^{a}_{\mu}$ and $\lambda_{\mu}$ are the Lagrange multipliers. Then, minimizing $\delta {\cal E}$ with respect to $\theta^{a}_{\mu}$, it is straightforward to find that $h^{0 a}_{\mu} = \lambda_{\mu}$. Thus, in order to rotate the spin moments at the minimal energy cost, one have to apply the \emph{scalar} field $h_{\mu}$, i.e. the same for all orbitals $a$. Moreover, in this case it is convenient to use the ``spherically averaged'' version of the linear response, where $\hat{m}_{\mu}$ is replaced by $M_{\mu}$, $\hat{b}_{\mu}$ is replaced by $B_{\mu} = \frac{1}{n}{\rm Tr}_{L} \{ \hat{b}_{\mu} \}$, and ${\cal R}_{ab,cd}^{\uparrow \downarrow}(\boldsymbol{q})$ is replaced by
\noindent
\begin{displaymath}
\mathbb{R}^{\uparrow \downarrow}_{\mu \nu}(\boldsymbol{q}) = \sum_{a \in \mu, c \in \nu} {\cal R}_{aa,cc}^{\uparrow \downarrow} (\boldsymbol{q}).
\end{displaymath}
\noindent The corresponding exchange interaction parameters will be given by
\noindent
\begin{equation}
J_{\boldsymbol{q}, \, \mu \nu} = \frac{1}{2} \left( M_{\mu} \mathbb{Q}^{+}_{\boldsymbol{q}, \, \mu \nu} M_{\nu} - B_{\mu} M_{\mu} \delta_{\mu \nu} \right)
\label{eq:jqs}
\end{equation}
\noindent and
\noindent
\begin{equation}
d_{\boldsymbol{q}, \, \mu \nu}^{z} = -\frac{i}{2} M_{\mu} \mathbb{Q}^{-}_{\boldsymbol{q}, \, \mu \nu} M_{\nu},
\label{eq:dzqs}
\end{equation}
\noindent where $\hat{\mathbb{Q}}^{\pm}_{\boldsymbol{q}} = \frac{1}{2} \left( \hat{\mathbb{Q}}^{\uparrow \downarrow}_{\boldsymbol{q}} \pm \hat{\mathbb{Q}}^{\downarrow \uparrow}_{\boldsymbol{q}} \right)$, $\hat{\mathbb{Q}}^{\sigma \sigma'}_{\boldsymbol{q}} = \left[ \hat{\mathbb{R}}^{\sigma \sigma'}_{\boldsymbol{q}} \right]^{-1}$, and $\hat{\mathbb{R}}^{\sigma \sigma'}_{\boldsymbol{q}} \equiv \hat{\mathbb{R}}^{\sigma \sigma'}({\boldsymbol{q}})$ is the matrix specified by the atomic indices in the unit cell.

\par In comparison with Equations \eref{eq:jq} and \eref{eq:dzq}, based on rotations of the magnetization matrix, Equations \eref{eq:jqs} and \eref{eq:dzqs} are expected to be more suitable for the analysis of low-energy excitations, of course, provided that the latter can be described by the spin model \eref{eq:Hspin}. In the following, these two methods will be denoted as $\hat{m}$ and $M$, after the basic variable describing the infinitesimal rotations of spins.

\subsection{\label{sec:rbxc} Rotations of the exchange-correlation field as an alternative perturbation}
\par In this section, we consider the original formulation of the linear response theory, as it was proposed by Liechtenstein \etal~\cite{LKAG1987}, which is frequently called the magnetic force theorem~\cite{Bruno2003}. However, there are two important points about the work of Liechtenstein \etal~\cite{LKAG1987}, which should be distinguished from each other~\cite{Antropov2004}:
\noindent
\begin{itemize}
\item
The general claim that, in SDFT, the energy change caused by the infinitesimal rotations of spins can be related to the KS eigenstates in the ground state is certainly correct and should not be revised. This is what is actually called the ``magnetic force theorem'' stating that the interaction parameters are the ground state properties and can be found by knowing the electronic structure in the ground state;
\item
Nevertheless, the practical expression \eref{eq:JLKAG}, which was derived by Liechtenstein \etal~\cite{LKAG1987} for the exchange interactions, relies on additional approximations and, in principle, can be improved.
\end{itemize}

\par The starting assumption of Liechtenstein \etal~\cite{LKAG1987} is that since the rotation of the magnetization results in the rotation of the xc field by the same angle (\Sref{sec:echange}), it is logical to treat the change of the xc field, $\delta \hat{\boldsymbol{b}}_{\boldsymbol{q}} = \hat{\boldsymbol{b}}_{\boldsymbol{q}} - \hat{\boldsymbol{b}}_{\rm GS}$, as a perturbation without the constraining field. Then, we have to consider only $\delta {\cal E}_{\rm sp}$ in \Eref{eq:chtenergy} caused by this $\delta \hat{\boldsymbol{b}}_{\boldsymbol{q}}$. Furthermore, the transversal magnetization, $\vec{\boldsymbol{m}}_{\boldsymbol{q}}^{\perp '}$, which is induced by $\delta \hat{\boldsymbol{b}}_{\boldsymbol{q}}$ will generally differ from $\vec{\boldsymbol{m}}_{\boldsymbol{q}}^{\perp}$ because, without the constraining field, $\vec{\boldsymbol{m}}_{\boldsymbol{q}}$ will tend to relax towards the ground state~\cite{Bruno2003,Stocks,Streib}. The DM interactions, if any, will tend to additionally rotate $\vec{\boldsymbol{m}}_{\boldsymbol{q}}^{\perp '}$ relative to $\vec{\boldsymbol{b}}_{\boldsymbol{q}}^{\perp}$: $\vec{\boldsymbol{m}}_{\boldsymbol{q}}^{\perp '} \approx  \vec{\boldsymbol{m}}_{\boldsymbol{q}}^{\perp 0 \, '} + \beta_{\boldsymbol{q}} \boldsymbol{n}^{z} \times \vec{\boldsymbol{m}}_{\boldsymbol{q}}^{\perp 0 \, '}$. Thus, instead of \Eref{eq:techange}, this method relies on (in the local coordinate frame)
\noindent
\begin{equation}
\delta {\cal E}_{\boldsymbol{q}} \approx \frac{1}{4} \, \vec{b}^{\perp \dagger} \cdot  \vec{m}^{\perp '}
- \frac{1}{4} \, \vec{b}^{\, \dagger} \cdot \vec{m} \, \theta^2,
\label{eq:spchangeL}
\end{equation}
\noindent arising from the single-particle energies for the perturbations caused by the transversal and longitudinal parts of the xc field. Again, the important point here is that $\vec{m}^{\perp '}$ deviates from $\vec{m}^{\perp}$. Otherwise, the right-hand side of \Eref{eq:spchangeL} would identically be equal to zero, as was discussed in \Sref{sec:echange}. Then, using the definitions $\vec{m}^{\perp 0 \, '} = \hat{\mathcal{R}}^{+}_{\boldsymbol{q}} \vec{b}^{\perp}$ and $\beta_{\boldsymbol{q}} \vec{m}^{\perp 0 \, '} = i\hat{\mathcal{R}}^{-}_{\boldsymbol{q}} \vec{b}^{\perp}$, one can find that
\noindent
\begin{equation}
J_{\boldsymbol{q}, \, \mu \nu} = -\frac{1}{2} \left( \vec{b}^{\, \dagger}_{\mu} \cdot \hat{\mathcal{R}}^{+}_{\boldsymbol{q}, \, \mu \nu} \vec{b}^{\, \phantom{\dagger}}_{\nu} - \vec{b}^{\, \dagger}_{\mu} \cdot \vec{m}^{\, \phantom{\dagger}}_{\mu} \delta_{\mu \nu} \right),
\label{eq:jqL}
\end{equation}
\noindent and
\noindent
\begin{equation}
d_{\boldsymbol{q}, \, \mu \nu}^{z} = \frac{i}{2} \vec{b}^{\, \dagger}_{\mu} \cdot \hat{\mathcal{R}}^{-}_{\boldsymbol{q}, \, \mu \nu} \vec{b}^{\, \phantom{\dagger}}_{\nu}.
\label{eq:dzqL}
\end{equation}
\noindent Alternatively, $d_{\boldsymbol{q}, \, \mu \nu}^{z}$ can be obtained from \Eref{eq:dzqb} for $\vec{h}^{\,0}_{\mu} = 0$ and $\vec{b}^{\perp}_{\mu} = \vec{b}_{\mu} \theta_{\mu}$. Substituting \Eref{eq:rabcd} into \Eref{eq:jqL} and Fourier transforming it to the real space, one obtains the well-known \Eref{eq:JLKAG}. Nevertheless, these are the \emph{approximate} expressions, which can be formally obtained from the exact ones, Equations \eref{eq:jqL} and \eref{eq:dzqL}, replacing $\vec{m}$ by $\vec{b}$ and $\hat{\mathcal{Q}}$ by $\hat{\mathcal{R}}$, with the additional minus sign. In the following, we will refer to this method as ``method $\hat{b}$'' or ``approximate method $\hat{b}$''.

\par In principle, one can also introduce the ``spherically averaged'' version of this method (the so-called method $B$) replacing $\vec{b}_{\nu}$ by $B_{\nu}$ and $\hat{\mathcal{R}}$ by $\mathbb{R}$~\cite{Bruno2003}, though it is rarely used. Without the SO coupling, $\hat{\mathbb{R}}^{+} = \hat{\mathbb{R}}^{\uparrow \downarrow}$ and corresponding exchange interactions $\hat{J}_{\boldsymbol{q}}^{B} \equiv [ J_{\boldsymbol{q}, \, \mu \nu}^{B} ]$ can be written as $\hat{J}_{\boldsymbol{q}}^{B} = - \frac{1}{2} \hat{B} ( \, \hat{\mathbb{R}}_{\boldsymbol{q}}^{\uparrow \downarrow} + \hat{\cal I}^{-1} \, ) \hat{B}$, where $\hat{B} = {\rm diag} ( \, \dots, \, B_{\nu}, \, \dots )$ and $\hat{\cal I} = {\rm diag} ( \, \dots, \, -B_{\nu}/M_{\nu}, \, \dots )$ are the diagonal matrices of, respectively, exchange splittings and effective Stoner parameters. By adapting the same matrix form for the ``exact'' interactions \eref{eq:jqs},  $\hat{J}_{\boldsymbol{q}} = \frac{1}{2} \hat{M} ( \, [\hat{\mathbb{R}}_{\boldsymbol{q}}^{\uparrow \downarrow} ]^{-1} + \hat{\cal I} \, ) \hat{M}$ with $\hat{M} = {\rm diag} ( \, \dots, \, M_{\nu}, \, \dots )$, one can find the following expression, connecting $\hat{J}_{\boldsymbol{q}}^{\phantom{B}}$ with $\hat{J}_{\boldsymbol{q}}^{B}$~\cite{Bruno2003}:
\noindent
\begin{equation}
\hat{J}_{\boldsymbol{q}}^{\phantom{B}} = \hat{J}_{\boldsymbol{q}}^{B} \left( 1 - 2 \hat{B}^{-1} \hat{M}^{-1} \hat{J}_{\boldsymbol{q}}^{B} \right)^{-1}.
\label{eq:eLconnect}
\end{equation}
\noindent Thus, $\hat{J}_{\boldsymbol{q}}^{\phantom{B}}$ can be indeed replaced by $\hat{J}_{\boldsymbol{q}}^{B}$ at least in two cases: (i) the long wavelength limit $\boldsymbol{q} \to 0$ and (ii) the strong-coupling limit $\hat{B} \to \infty$. Therefore, the spin-wave stiffness in the limit $\boldsymbol{q} \to 0$ is expected to be the same in both methods. Nevertheless, this statement should not be exaggerated because \Eref{eq:eLconnect} holds only in the spherical case, where the xc field and the magnetization on each magnetic site are given by the scalar parameters $B_{\nu}$ and $M_{\nu}$. In the matrix case, the simple relationship \eref{eq:eLconnect} is no longer valid~\cite{PRB2021}. That is why even the spin-wave stiffness in the methods $\hat{b}$ and $M$ can be different. Furthermore, we will see that there is indeed a number examples, where the approximate method $\hat{b}$ fails to reproduce the correct magnetic ground state, while the method $M$ dramatically improves the description.

\par Of course, it is reasonable to ask what are the right objects to rotate in this case: whether they should be the whole matrices $\hat{b}_{\nu}$ or only the spherical parts of these matrices $B_{\nu}$? If in the case of the magnetization, the answer can be found by minimizing the energy change \eref{eq:techange} (see \Sref{sec:object}), \Eref{eq:techange} is not applicable for rotations of the xc field. Therefore, the answer is open. However, historically most of the applications deal with the rotations of the matrices $\hat{b}_{\nu}$.

\par Considering the strong-coupling limit in \Eref{eq:jqL}~\cite{PRL99,PRB99}, one can derive all known expressions for the double exchange $J_{ij} \sim \langle \hat{H}_{ij} \rangle$~\cite{deGennes}, superexchange $J_{ij} \sim -\langle \hat{H}_{ij}^{2} \rangle/U$~\cite{Anderson1959}, superexchange with the interatomic Coulomb repulsion $J_{ij} \sim -\langle \hat{H}_{ij}^{2} \rangle/(U-V)$~\cite{extendedH}, etc., where $U$ and $V$ is the on-site and intersite Coulomb repulsion, respectively, $\hat{H}_{ij}$ are the transfer integrals (see \ref{sec:Hubbard}), which are typically associated with the matrix elements of the KS Hamiltonian in LDA (GGA), and $\langle \dots \rangle$ denotes the expectation value in the ground state. The strong-coupling limit for the DM interaction \eref{eq:dzqL} results in the spin-current model~\cite{Katsnelson_DM,Kikuchi}, which can be viewed as the relativistic counterpart of the double exchange mechanism~\cite{PRB2023}. The expression for RKKY interactions can be also derived starting from \Eref{eq:jqL}, but using slightly different philosophy~\cite{BrunoChappert}. In this case, $\vec{b}$ is the field created by localized core spins and acting on outer conduction electrons. Without $\vec{b}$, the conduction bands are non-magnetic (and the tensor $\hat{\mathcal{R}}$ is evaluated in this non-magnetic state~\cite{BrunoChappert}).

\subsection{\label{sec:sw} Relationship to the spin-wave spectra}
\par In the previous sections, we have considered how the spin model can be generally derived from the electronic one using the concept of infinitesimal rotations of spins. The parameters of such spin model are expressed in terms of the static spin susceptibility (or the response function). On the other hand, the spin-wave dispersion, $\omega_{\boldsymbol{q}}$, which is the experimentally measurable quantity, can be derived in the framework of RPA from the poles of the dynamic spin susceptibility~\cite{Cooke,Callaway,Savrasov}. However, this $\omega_{\boldsymbol{q}}$ does not necessary coincide with the one of the spin model with the parameters derived from the static spin susceptibility~\cite{KL2004,Szczech}. In this respect, Katsnelson and Lichtenstein~\cite{KL2004} have argued that although the method proposed by Bruno~\cite{Bruno2003} is more consistent with the static response formulation, the method $\hat{b}$ should more suitable for the analysis of the spin-wave spectra. Here, we will briefly consider this problem. For simplicity, we assume that there is only one magnetic site in the unit cell and drop all matrix notations. Then, in the spherical case, the dynamic response function is given by $\tilde{\mathbb{R}}_{\boldsymbol{q}}^{\uparrow \downarrow}(\omega) = \mathbb{R}_{\boldsymbol{q}}^{\uparrow \downarrow}(\omega)\Big[ 1 + {\cal I} \mathbb{R}_{\boldsymbol{q}}^{\uparrow \downarrow}(\omega)\Big]^{-1}$, where $\mathbb{R}_{\boldsymbol{q}}^{\uparrow \downarrow}(\omega)$ is obtained from \Eref{eq:gresponse} replacing in the denominator $(\varepsilon_{m \boldsymbol{k}}^{\uparrow}$$-$$\varepsilon_{l \boldsymbol{k}+\boldsymbol{q}}^{\downarrow})$ by $(\omega$$+$$\varepsilon_{m \boldsymbol{k}}^{\uparrow}$$-$$\varepsilon_{l \boldsymbol{k}+\boldsymbol{q}}^{\downarrow})$. Therefore, one has to solve the equation
\noindent
\begin{equation}
1+ {\cal I} \mathbb{R}_{\boldsymbol{q}}^{\uparrow \downarrow}(\omega_{\boldsymbol{q}})=0,
\label{eq:sw}
\end{equation}
\noindent which can be equivalently rearranged as: $2M^{-1}J_{\boldsymbol{q}}(\omega_{\boldsymbol{q}})=0$, where $J_{\boldsymbol{q}}(\omega)$ is given by \Eref{eq:jqs} with $\mathbb{Q}^{\uparrow \downarrow}_{\boldsymbol{q}}(\omega)$ instead of $\mathbb{Q}^{\uparrow \downarrow}_{\boldsymbol{q}} = \mathbb{Q}^{+}_{\boldsymbol{q}}$. Thus, the problem is that the spin-wave energies are given by the zeros of $J_{\boldsymbol{q}}(\omega)$ and do not necessary coincide with $\omega_{\boldsymbol{q}}^{M} = 2M^{-1}J_{\boldsymbol{q}}(0)$, expected from the solution of the spin model.

\par In the limit $\omega \ll \varepsilon_{l \boldsymbol{k}+\boldsymbol{q}}^{\downarrow}$$-$$\varepsilon_{m \boldsymbol{k}}^{\uparrow}$ (which takes place, for instance, for insulating and half-metallic materials, where the occupied and unoccupied states with opposite projections of spins are separated by an energy gap), one can use the linearization $\mathbb{R}_{\boldsymbol{q}}^{\uparrow \downarrow}(\omega) \approx \mathbb{R}_{\boldsymbol{q}}^{\uparrow \downarrow}(0) + \omega \dot{\mathbb{R}}_{\boldsymbol{q}}^{\uparrow \downarrow}(0)$, where $\dot{\mathbb{R}}_{\boldsymbol{q}}^{\uparrow \downarrow}(0) = \frac{\partial}{\partial \omega} \mathbb{R}_{\boldsymbol{q}}^{\uparrow \downarrow}(\omega)\vert_{\omega=0}$, and find the following expression: $\omega_{\boldsymbol{q}} = \omega_{\boldsymbol{q}}^{B} {\cal I}^{-1} [\dot{\mathbb{R}}_{\boldsymbol{q}}^{\uparrow \downarrow}(0)]^{-1}B^{-1}$, where $\omega_{\boldsymbol{q}}^{B} = 2M^{-1}J^B_{\boldsymbol{q}}(0)$ is the spin-wave dispersion calculated with the parameters of the scheme $B$. This example clearly shows that $\omega_{\boldsymbol{q}}^{B}$ should be additionally renormalized, though this renormalization is generally different from the one given by \Eref{eq:eLconnect}, connecting the parameters of the methods $M$ and $B$. $\dot{\mathbb{R}}_{\boldsymbol{q}}^{\uparrow \downarrow}(0)$ depends on the details of the electronic structure. In practical terms, it can be calculated replacing $(\varepsilon_{m \boldsymbol{k}}^{\uparrow}$$-$$\varepsilon_{l \boldsymbol{k}+\boldsymbol{q}}^{\downarrow})$ by $-(\varepsilon_{m \boldsymbol{k}}^{\uparrow}$$-$$\varepsilon_{l \boldsymbol{k}+\boldsymbol{q}}^{\downarrow})^2$ in \Eref{eq:gresponse}. Then, in certain circumstances, the method $M$ can be a good starting point for the analysis of the spin-wave dispersion. For instance, if the $\uparrow$-spin ($\downarrow$-spin) states are fully occupied (empty) and $B$ is large compare to the band dispersion, it is straightforward to obtain that $\dot{\mathbb{R}}_{\boldsymbol{q}}^{\uparrow \downarrow}(0) \approx -B^{-1} \mathbb{R}_{\boldsymbol{q}}^{\uparrow \downarrow}(0)$ and $\omega_{\boldsymbol{q}} \approx \omega_{\boldsymbol{q}}^{M}$.

\par Thus, it would be fair to conclude that the analysis of the spin-wave dispersion requires the additional renormalization of the parameters derived from the static spin susceptibility~\cite{Szczech,KL2004}. This conclusion applies to \emph{all} methods ($M$, $B$, and $\hat{b}$). Therefore, this is an open question which method serves better for the analysis of the spin-wave dispersion. It is certainly true that, in LSDA, the method $\hat{b}$ better reproduces the experimental spin-wave dispersion in the canonical case of bcc-Fe and fcc-Ni~\cite{PRB2021,KL2004}. However, this conclusion does not seem to be general and for other materials the comparison can be less favorable.

\par Finally, we note that \Eref{eq:sw} can be further rearranged as $\mathbb{R}_{\boldsymbol{q}}^{\uparrow \downarrow}(0)(B$$+$$h_{\omega}) = M$, where $h_{\omega} = [\mathbb{R}_{\boldsymbol{q}}^{\uparrow \downarrow}(0)]^{-1} [\mathbb{R}_{\boldsymbol{q}}^{\uparrow \downarrow}(\omega) - \mathbb{R}_{\boldsymbol{q}}^{\uparrow \downarrow}(0)] B \approx \omega [\mathbb{R}_{\boldsymbol{q}}^{\uparrow \downarrow}(0)]^{-1}\dot{\mathbb{R}}_{\boldsymbol{q}}^{\uparrow \downarrow}(0) B$ has a meaning of the constraining field, which is needed to correct the effect of the xc field $B$ in order to reproduce the ground-state magnetic moment for an arbitrary $\boldsymbol{q}$. In this sense, there is an analogy between the search of the poles of the dynamic susceptibility and the constrained SDFT considered in \Sref{sec:echange}.

\subsection{\label{sec:downf} Elimination of the ligand spins}
\par By knowing $J_{\boldsymbol{q}, \, \mu \nu}$ and $d_{\boldsymbol{q}, \, \mu \nu}^{z}$, one can, in principle, calculate isotropic and DM interactions operating between all sites in the unit cell. Nevertheless, the magnetization at these sites may have completely different origin. For instance, the transition-metal (${\rm T}$) sites in many oxide materials participate as a source of the magnetism, being primarily responsible for the spontaneous time-reversal symmetry breaking, while the oxygen or any other ligand (${\rm L}$) sites behave as ``magnetic slaves'': although they can host an appreciable portion of the magnetization, it is solely induced by hybridization with the ${\rm T}$ sites and strictly follow the change of the magnetization on the ${\rm T}$ sites. The corresponding energy change for each $\boldsymbol{q}$ can be schematically expressed as
\noindent
\begin{equation}
\delta {\cal E} = - \frac{1}{2} \left( \vec{\theta}_{\rm T}^{\, \dagger} \cdot \hat{X}_{\rm TT}^{\phantom{T}} \vec{\theta}_{\rm T}^{\phantom{\dagger}} + \vec{\theta}_{\rm T}^{\, \dagger} \cdot \hat{X}_{\rm TL}^{\phantom{T}} \vec{\theta}_{\rm L}^{\phantom{\dagger}} + \vec{\theta}_{\rm L}^{\, \dagger} \cdot \hat{X}_{\rm LT}^{\phantom{T}} \vec{\theta}_{\rm T}^{\phantom{\dagger}} + \vec{\theta}_{\rm L}^{\, \dagger} \cdot \hat{X}_{\rm LL}^{\phantom{T}} \vec{\theta}_{\rm L}^{\phantom{\dagger}} \right),
\label{eq:eLT}
\end{equation}
\noindent where $\hat{X}_{\rm AB}^{\phantom{T}}$ is the matrix specified by the atomic sites of the types ${\rm A}$ or ${\rm B}$, $\vec{\theta}_{\rm A}^{\phantom{\dagger}}$ are the polar angles specifying the rotations of magnetic moments of the type ${\rm A}$ in the form of the column vector, and $\vec{\theta}_{\rm A}^{\, \dagger}$ is the corresponding to it row vector. Then, one can try to eliminate the ${\rm L}$ degrees of freedom by transferring their effect into the interaction parameters between the ${\rm T}$ sites. This can be done by employing the ideas of adiabatic spin dynamics~\cite{AntropovSD,HalilovSD} and assuming that the ${\rm T}$ spins are sufficiently slow so that the ${\rm L}$ spins have sufficient time to adjust each change in the system of the ${\rm T}$ spins. Mathematically, this means that for each $\vec{\theta}_{\rm T}^{\phantom{\dagger}}$, $\vec{\theta}_{\rm L}^{\phantom{\dagger}}$ can be found from the condition $\frac{\partial}{\partial \vec{\theta}_{\rm L}^{\, \dagger}} \delta {\cal E} = 0$, yielding
\noindent
\begin{equation}
\vec{\theta}_{\rm L} = - \left[ \hat{X}_{\rm LL} \right]^{-1} \hat{X}_{\rm LT} \vec{\theta}_{\rm T}
\label{eq:LfromT}
\end{equation}
\noindent and $\delta {\cal E} = - \frac{1}{2} \vec{\theta}_{\rm T}^{\, \dagger} \cdot \hat{\tilde{X}}_{\rm TT}^{\phantom{T}} \vec{\theta}_{\rm T}^{\phantom{\dagger}} $ with
\noindent
\begin{equation}
\hat{\tilde{X}}_{\rm TT} = \hat{X}_{\rm TT} -  \hat{X}_{\rm TL} \left[ \hat{X}_{\rm LL} \right]^{-1} \hat{X}_{\rm LT}.
\label{eq:jTT}
\end{equation}
\noindent The corresponding parameters of isotropic and DM interactions can be obtained from $\hat{\tilde{X}}_{\rm TT}$ using Equations \eref{eq:jq1} and \eref{eq:dzq1}.

\par In the method $M$, the matrix inversion in \Eref{eq:jTT} can be combined with the one of the response matrix $\hat{\mathbb{Q}}^{\uparrow \downarrow} = \left[ \hat{\mathbb{R}}^{\uparrow \downarrow} \right]^{-1}$ to obtain the following expression for $\hat{\tilde{X}}_{\rm TT}$:
\noindent
\begin{equation}
\hat{\tilde{X}}_{\rm TT} = \hat{\tilde{X}}_{\rm TT}^{0} + \Delta \hat{\tilde{X}}_{\rm TT}^{\phantom{0}},
\label{eq:jTT1}
\end{equation}
\noindent where
\noindent
\begin{equation}
\hat{\tilde{X}}_{\rm TT}^{0} = \hat{M}_{\rm T}^{\phantom{\uparrow}} \left[ \hat{\mathbb{R}}^{\uparrow \downarrow}_{\rm TT} \right]^{-1} \hat{M}_{\rm T}^{\phantom{\uparrow}},
\label{eq:jTT0}
\end{equation}
\noindent and
\noindent
\begin{equation}
\Delta \hat{\tilde{X}}_{\rm TT}^{\phantom{0}} = \hat{M}_{\rm T}^{\phantom{\uparrow}} \left[ \hat{\mathbb{R}}^{\uparrow \downarrow}_{\rm TT} \right]^{-1} \hat{\mathbb{R}}^{\uparrow \downarrow}_{\rm TL} \hat{\mathbb{Q}}^{\uparrow \downarrow}_{\rm LL} \left( \hat{\mathbb{Q}}^{\uparrow \downarrow}_{\rm LL} + \hat{\cal I}_{\rm L}^{\phantom{\uparrow}} \right)^{-1} \hat{\cal I}_{\rm L}^{\phantom{\uparrow}} \hat{\mathbb{R}}^{\uparrow \downarrow}_{\rm LT} \left[ \hat{\mathbb{R}}^{\uparrow \downarrow}_{\rm TT} \right]^{-1}  \hat{M}_{\rm T}^{\phantom{\uparrow}}.
\label{eq:djTT}
\end{equation}
\noindent Here, $\hat{M}_{\rm T}$ is the diagonal matrix of magnetic moments on the sites ${\rm T}$, and $\hat{\cal I}_{\rm L}$ is the diagonal matrix of effective Stoner parameters on the sites ${\rm L}$. In this expression, the explicit dependence of $\hat{\tilde{X}}_{\rm TT}$ on $\hat{\cal I}_{\rm L}$ is incorporated into $\Delta \hat{\tilde{X}}_{\rm TT}^{\phantom{0}}$, while $\hat{\tilde{X}}_{\rm TT}^{0}$ formally does not depend on $\hat{\cal I}_{\rm L}$. The parameters $\hat{\cal I}_{\rm L}$ can play a very important role in the theory of exchange interactions. For instance, according to GKA rules, they largely contribute to the FM coupling in the systems, where the ${\rm T}$-${\rm L}$-${\rm T}$ bond angle is close to $90^{\circ}$~\cite{Anderson1950,Goodenough1955,Goodenough1958,Kanamori1959}. In the linear response theories, these effects are incorporated into $\Delta \hat{\tilde{X}}_{\rm TT}$~\cite{PRB2022}.

\par Furthermore, the representation \eref{eq:jTT1} allows us to improve the numerical accuracy. Since in most of the systems the ligand band is filled, the matrix elements of $\hat{\mathbb{R}}^{\uparrow \downarrow}$ associated with the ${\rm L}$ sites are typically small.  Therefore, when we invert $\hat{\mathbb{R}}^{\uparrow \downarrow}$, we have to deal with very large numbers even for the ${\rm T}$ sublattice. This is the reason wgy the bare interactions $\hat{X}_{\rm TT}^{\phantom{\uparrow}} \sim \hat{\mathbb{Q}}^{\uparrow \downarrow}_{\rm TT}$ are typically large and strongly compensated by the second term in \Eref{eq:jTT}~\cite{PRB2021,PRB2022}. Similar situation occurs in the scheme $\hat{m}$. On the contrary, the calculation of $\hat{\tilde{X}}_{\rm TT}^{0}$ and $\Delta \hat{\tilde{X}}_{\rm TT}$ using Equations \eref{eq:jTT0} and \eref{eq:djTT} involves the inversion of only the ${\rm T}$ block of $\hat{\mathbb{R}}^{\uparrow \downarrow}$. This procedure is numerically much more stable rather than the inversion of the whole matrix $\hat{\mathbb{R}}^{\uparrow \downarrow}$ in \Eref{eq:jTT}.

\par The idea of downfolding somewhat similar to ours was previously considered by Mryasov \etal in order to eliminate the $5d$ states of the heavy Pt atoms and explain the unusual temperature dependence of the magnetic anisotropy energy in the ordered FePt alloy~\cite{Mryasov}. A simplified approach in the framework of the scheme $\hat{b}$, which did not take into account the effects of ligand-ligand interactions and $\hat{\cal I}_{\rm L}$, was also considered by Logemann~\etal\cite{Logemann}.

\section{\label{sec:CrX3} Chromium trihalides}
\par
\begin{figure}[t]
\begin{center}
\includegraphics[width=6.0cm]{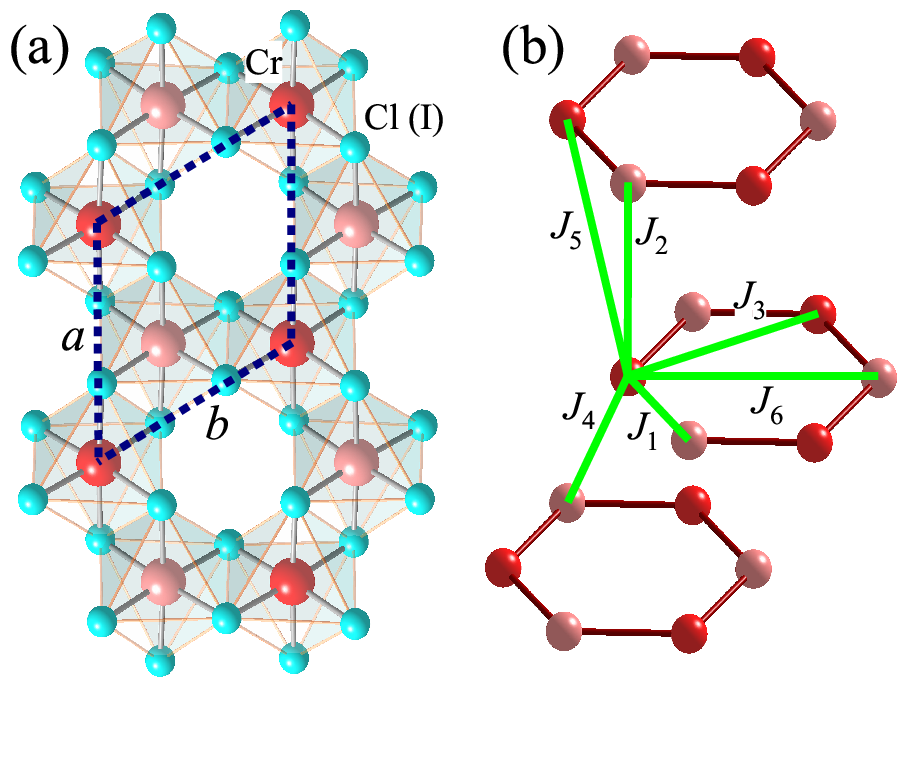}\hspace*{0.0cm}
\includegraphics[width=4.5cm]{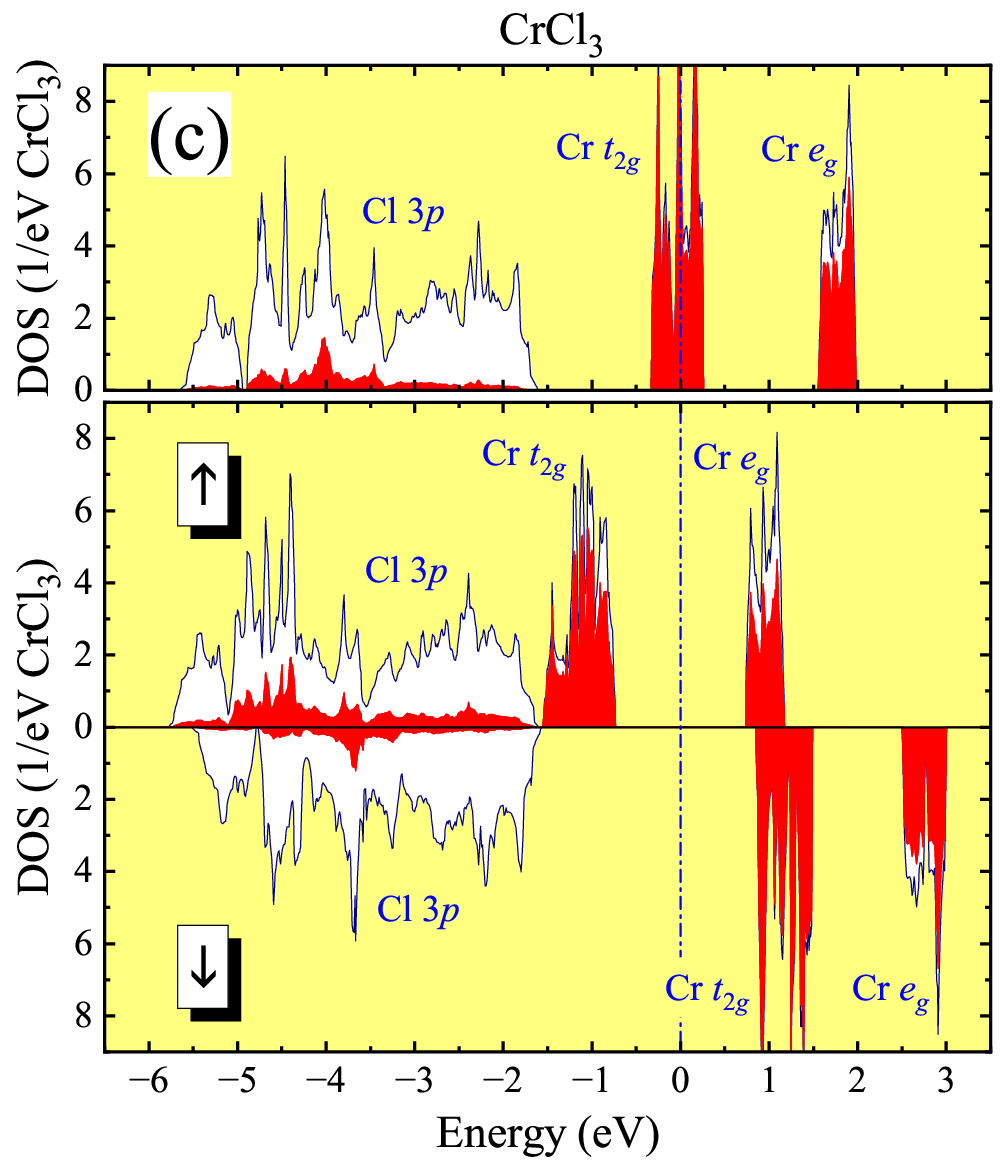}
\includegraphics[width=4.5cm]{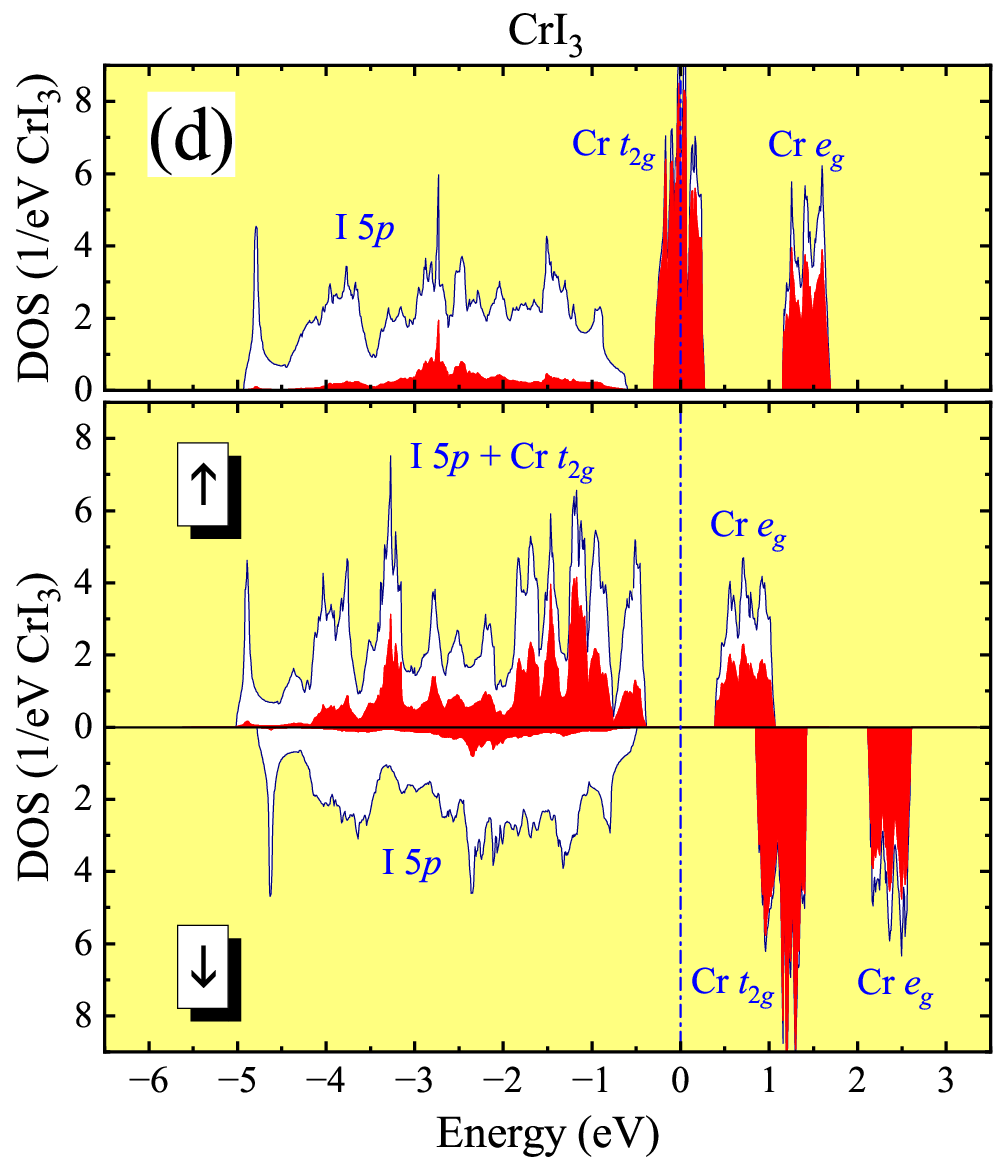}
\end{center}
\caption{(a) Top view on the CrCl$_3$ (CrI$_3$) layer. The hexagonal unit cell is denoted by the broken line. (b) Stacking of adjacent honeycomb layers with the notation of main exchange interactions. Two Cr atoms in the primitive rhombohedral unit cell are denoted by different colors. (c),(d) Densities of states (DOS) of CrCl$_3$ and CrI$_3$ in LDA (top) and LSDA for the ferromagnetic state (bottom). Shaded areas show partial contributions of the Cr $3d$ states. The Fermi level is at zero energy (the middle of the band gap in the insulating phase).}
\label{fig.CrX3_basic}
\end{figure}

 In order to illustrate abilities of considered linear response techniques, we start with the detailed analysis of exchange interactions in chromium trihalides Cr$X_3$ ($X$$=$ Cl and I). These van der Walls compounds crystallize in the rhombohedral $R\overline{3}$ structure, which is built from the honeycomb layers as shown in \Fref{fig.CrX3_basic}a,b~\cite{CrCl3str,CrI3str}. The interactions between the layers are weak, but not negligible. For instance, sizable exchange interactions spread up to 6th coordination sphere, in and between the layers, as shown in \Fref{fig.CrX3_basic}b.
\noindent

\par CrI$_3$ is the ferromagnet with the Curie temperature $T_{\rm C} = 61$ K~\cite{CrI3str,McGuire,ChenPRX}, while CrCl$_3$ is a antiferromagnet with the N\'eel temperature $T_{\rm N} = 14.1$ K~\cite{Chen2022,Schneeloch2022}. The AFM transition in CrCl$_3$ is followed by another transition to a pseudo-FM phase with $T_{\rm C} \sim 17$ K. In both cases, the magnetic moments tend to order ferromagnetically within the honeycomb layers. Below $T_{\rm N}$, the interlayer coupling is weakly AFM, while in the temperature interval $T_{\rm N} < T < T_{\rm C}$ the magnetic behavior of CrCl$_3$ is explained by the interlayer disorder~\cite{Schneeloch2022}.

\par CrI$_3$ is viewed as the prominent two-dimensional ferromagnet~\cite{CrI3_Huang_Nature}, where one of the key ingredients is the strong SO coupling steaming from the heavy I atoms~\cite{Lado2017}, which is mainly responsible for the exchange anisotropy and emergence of the long-range FM order at relatively high $T_{\rm C}$ (i.e., contrary to what would be expected from the Mermin-Wagner theorem in the isotropic case~\cite{MerminWagner}). Besides that, CrCl$_3$ and CrI$_3$ are regarded as the testbed materials for studying fundamental aspects related to the origin of the ferromagnetism. Namely, why are these materials ferromagnetic? The popular answer is that since the Cr-$X$-Cr angle is close to $90^{\circ}$, the interaction is expected to be ferromagnetic due to intraatomic exchange coupling ${\cal I}_{X}$ on the ligand sites, as prescribed by the GKA rules~\cite{Anderson1950,Goodenough1955,Goodenough1958,Kanamori1959}. However, the GKA rule for the $90^{\circ}$ exchange is not very conclusive, because typically there are several competing mechanisms supporting either ferromagnetism or antiferromagnetism~\cite{Chaloupka}. In fact, Kanamori himself admitted that there are several exceptions from this rule in the $90^{\circ}$ case~\cite{Kanamori1959}. Moreover, below we will see that, under certain circumstances, ${\cal I}_{X}$ can easily become negative and act against the ferromagnetism.

\par Although CrI$_3$ is more useful practically, CrCl$_3$ is interesting from the explanatory point of view. Even in LSDA, the electronic structure of CrCl$_3$ consists of well separated Cr $t_{2g}$, Cr $e_{g}$, and Cl $3p$ bands, as explained in \Fref{fig.CrX3_basic}c.\footnote{In the octahedral environment, the Cr $3d$ states are split into triply-degenerate $t_{2g}$ levels and doubly-degenerate $e_{g}$ levels. In many transition-metal oxides or related materials, the strong crystal-field splitting, $10Dq$, is caused by the hybridization between $3d$ and ligand $p$ states~\cite{Kanamori1957}.} Therefore, it is possible to study separately the contributions of each of these bands to the exchange interactions by constructing proper TB models in the basis of Wannier functions~\cite{review2008,WannierRevModPhys}. Besides that, one can also consider correlated models, both for CrCl$_3$ and CrI$_3$, which explicitly consider the on-site Coulomb interactions. This procedure is briefly explained in \ref{sec:Hubbard}.
The numerical calculations are performed on the basis of the linear muffin-tin orbital (LMTO) method in the atomic-spheres approximation~\cite{LMTO1,LMTO2}, using the mesh of the $10$$\times$$10$$\times$$10$ points, both for the $\boldsymbol{k}$- and $\boldsymbol{q}$-integration.

\par First, we will review the behavior of interatomic exchange interactions depending on each new ingredient added to the model as well as the type of infinitesimal spin rotations (whether the rotated object is $\hat{b}$, $\hat{m}$, or $M$). Having estimated the parameters of interatomic exchange interactions, one can find the magnetic ground state and evaluate the magnetic transition temperature as explained in \ref{sec:RPA}. Then, the detailed comparison with the experimental data will be given in \Sref{sec:CrX3_exp}.

\subsection{\label{sec:CrX3_3o} 3-orbital $t_{2g}$ model in LSDA}
\par The simplest model for CrCl$_3$ is the half-filled $t_{2g}$ model. It contains only occupied $\uparrow$-spin $t_{2g}$ band and unoccupied $\downarrow$-spin $t_{2g}$ band. Since all basis functions of this model are associated with the Cr states, there are no additional complications coming from the ligand states. Even in LSDA, the exchange splitting $B_{\rm Cr}$ between the $\uparrow$- and $\downarrow$-spin $t_{2g}$ bands is large compared to the bandwidth. Thus, the system should be in the strong-coupling limit. The corresponding parameters of exchange interactions, calculated by rotating $\hat{b}$, $\hat{m}$, or $M$ are summarized in \Tref{tab.CrCl3_lsda_3o}.
\noindent
\begin{table}[ht]
\caption{\label{tab.CrCl3_lsda_3o} Isotropic exchange interactions in CrCl$_3$ (in meV): results of the $t_{2g}$ model in LSDA. $\hat{b}$, $\hat{m}$, and $M$ stand for the methods based on the infinitesimal rotations of, respectively, the xc field, the magnetization matrix, and the local spin moments. The notations of $J_{k}$ are explained in Fig.~\protect\ref{fig.CrX3_basic}.}
\begin{indented}
\lineup
\item[]\begin{tabular}{@{}ccccccc}
\br
method    & $J_{1}$ & $J_{2}$ & $J_{3}$ & $J_{4}$ & $J_{5}$ & $J_{6}$ \cr
\mr
$\hat{b}$ & $-6.44$ & $-0.50$ & $-0.69$ & $-0.17$ & $-0.18$ & $-0.63$ \cr
$\hat{m}$ & $-6.32$ & $-0.49$ & $-0.68$ & $-0.17$ & $-0.18$ & $-0.62$ \cr
$M$       & $-6.30$ & $-0.49$ & $-0.69$ & $-0.17$ & $-0.18$ & $-0.63$ \cr
\br
\end{tabular}
\end{indented}
\end{table}

\par In this case we obtain very consistent description as all three methods provide very similar sets of parameters. This is not surprising: the approximate scheme $\hat{b}$ is justified in the strong coupling limit. Moreover, three $t_{2g}$ levels are nearly degenerate. Therefore, the asphericity of $\hat{m}$ is small and corresponding parameters are practically identical to the ones obtained in the $M$ scheme. However, all interactions are antiferromagnetic, which is quite expected for the half filling~\cite{Anderson1959}, but totally contradicts to the experimental situation.

\subsection{\label{sec:CrX3_5o} 5-orbital Cr $3d$ model}
\par The next important question is whether the ferromagnetism of Cr$X_{3}$ can be explained without the ligand states, in the model including both $t_{2g}$ \emph{and} $e_{g}$ bands. In LSDA, the $\uparrow$- and $\downarrow$-spin bands are subjected to the exchange splitting $B_{\rm Cr}$. At the first sign, there is only a small addition in comparison with the $t_{2g}$ model -- the unoccupied $e_{g}$ bands. However, it changes the story dramatically. First, the crystal-field splitting between $t_{2g}$ and $e_{g}$ bands is comparable with $B_{\rm Cr}$. Moreover, since only the transitions between the occupied $\uparrow$-bands and unoccupied $\downarrow$-bands contribute to $\hat{\mathcal{R}}^{\uparrow \downarrow}$, such exchange interactions do not know anything about the existence of the $\uparrow$-spin $e_{g}$ band. Thus, although $B_{\rm Cr}$ is large (as in the $t_{2g}$ model), the behavior of the exchange interactions is also controlled by other details of the electronic structure and the system is no longer in the strong coupling limit. Furthermore, the matrix $\hat{m}$ in the basis of all five $3d$ orbitals acquires additional degrees of freedom besides $M$, which is only the spherical part of $\hat{m}$.

\par All these tendencies are reflected in the behavior of interatomic exchange interactions (\Tref{tab.CrCl3_lsda_5o}). First, the inclusion of the $e_{g}$ band into the model gives rise to the FM interactions. These interactions prevail in the nearest-neighbor (nn) bonds and considerably weaken the AFM interactions in other neighbors, in agreement with the experimental situation. Then, the schemes $\hat{b}$, $\hat{m}$, and $M$ provide quite a different description. Compared to the methods $\hat{m}$ and $M$, the approximate method $\hat{b}$ underestimates the FM interaction $J_{1}$. Moreover, the method $\hat{m}$ (in comparison with $M$) has a tendency to overestimate the in-plane interactions $J_{1}$, $J_{3}$, and $J_{6}$, as expected from the analysis in \Sref{sec:object}.
\noindent
\begin{table}[ht]
\caption{\label{tab.CrCl3_lsda_5o} Isotropic exchange interactions in CrCl$_3$ (in meV): results of the ${\rm Cr} \, 3d$ model in LSDA. $\hat{b}$, $\hat{m}$, and $M$ stand for the methods based on the infinitesimal rotations of, respectively, the xc field, the magnetization matrix, and the local spin moments. The xc field was either computed from the on-site spin splitting of the TB Hamiltonian (H) or obtained from the sum rule (sr). $T_{\rm S}$ is corresponding spin transition temperature in RPA (in K). The method $\hat{m}$ yields the ferromagnetic ground state, while the methods $\hat{b}$ and $M$ result in an incommensurate spin-spiral structure propagating both in and between the planes. The notations of $J_{k}$ are explained in \Fref{fig.CrX3_basic}b.}
\begin{indented}
\lineup
\item[]\begin{tabular}{@{}cccccccc}
\br
method         &  $J_{1}$           & $J_{2}$ & $J_{3}$ & $J_{4}$            & $J_{5}$           & $J_{6}$ & $T_{\rm S}$ \cr
\mr
$\hat{b}$ (H)  &  $\phantom{1}2.39$ & $-0.27$ & $-0.45$ & $-0.03$            & $-0.04$           & $-0.50$ & $8$         \cr
$\hat{b}$ (sr) &  $\phantom{1}2.27$ & $-0.16$ & $-0.37$ & $\phantom{-}0.01$  & $0$               & $-0.46$ & $5$         \cr
$\hat{m}$      &  $11.94$           & $-0.01$ & $-0.69$ & $\phantom{-}0.08$  & $\phantom{-}0.09$ & $-0.69$ & $30$        \cr
$M$            &  $\phantom{1}4.68$ & $-0.21$ & $-0.40$ & $0$                & $-0.01$           & $-0.48$ & $5$         \cr
\br
\end{tabular}
\end{indented}
\end{table}

\par Another important question is how to define the xc field in the TB model. The common practice is to use $\hat{b} = \hat{H}^{\uparrow}$$-$$\hat{H}^{\downarrow}$ and take only site-diagonal (local) part of this splitting. However, in LSDA, the off-diagonal elements of $\hat{H}$ can also depend on spin, giving rise to non-local contributions to the xc field~\cite{Nomoto2020}. Then, the use of the site-diagonal part alone will violate the sum rule because $\hat{\mathcal{R}}_{0}^{\uparrow \downarrow} \vec{b}$ with such $\vec{b}$ will no longer reproduce the ground-state magnetization $\vec{m}$ (see \Sref{sec:sumrule}). Another possibility is to reenforce the sum rule by defining the \emph{local} xc field $\vec{b} = \hat{\mathcal{Q}}^{\uparrow \downarrow}_{0}\vec{m}$, which would reproduce the ground state magnetization $\vec{m}$. In the 3-orbital $t_{2g}$ model, this results only in minor change of the exchange interactions. However, starting from the 5-orbital model, the difference becomes more significant (see Table~\ref{tab.CrCl3_lsda_5o}).

\par
\begin{figure}[t]
\begin{center}
\includegraphics[width=4.5cm]{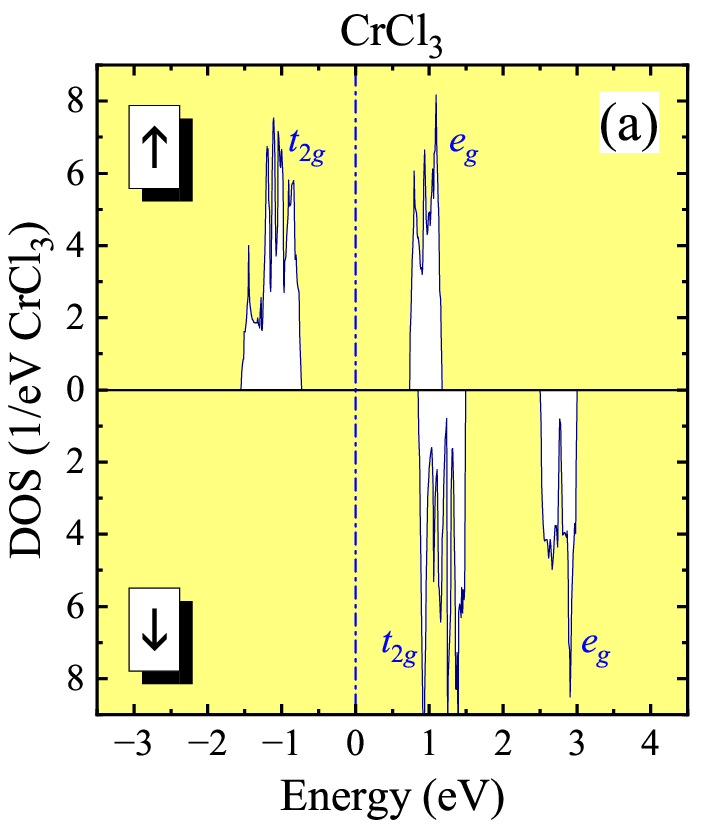}
\includegraphics[width=4.5cm]{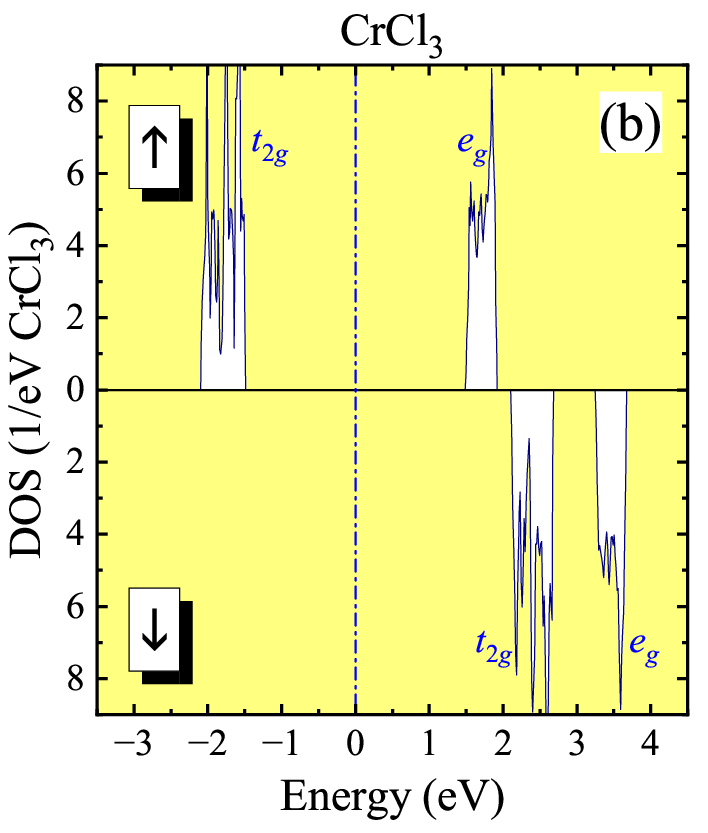}
\includegraphics[width=4.5cm]{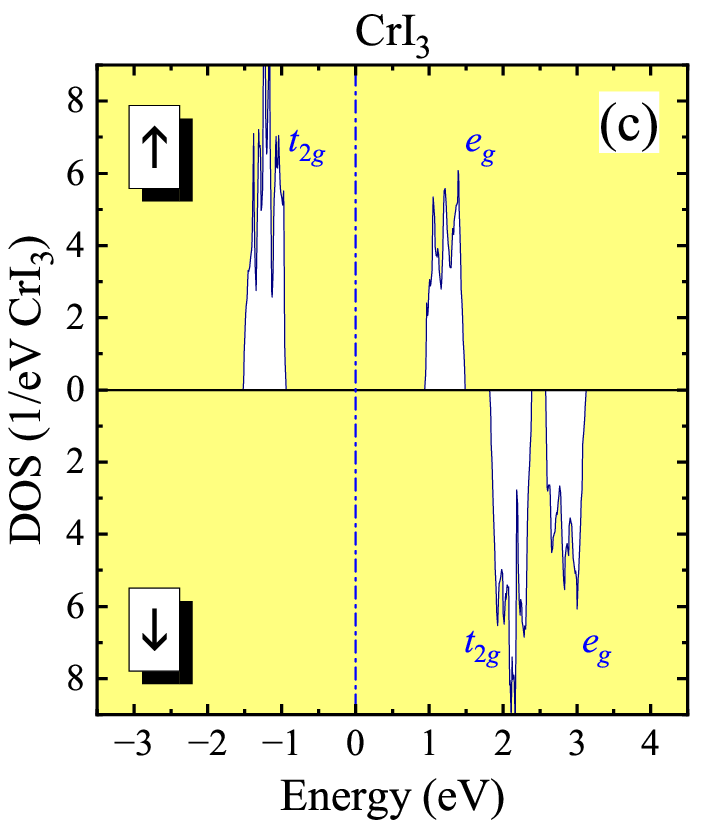}
\end{center}
\caption{Densities of states (DOS) for the FM state of 5-orbital Cr $3d$ model in (a) LSDA for CrCl$_3$, (b) Hartree-Fock approximation for CrCl$_3$, and (c) Hartree-Fock approximation for CrI$_3$. The zero energy is in the middle of the band gap.}
\label{fig.trihalides_5o}
\end{figure}

 Similar analysis can be performed for the correlated model in the Hartree-Fock approximation (see \ref{sec:Hubbard}). The non-interacting electron part of the model Hamiltonian in this case is evaluated in LDA. The corresponding electronic structure is shown in \Fref{fig.CrX3_basic}c,d: since in CrCl$_3$ and CrI$_3$ the Cr $3d$ bands are well separated from the ligand ones, such model can be easily constructed for both compounds. The screened Coulomb interactions are evaluated within constrained random-phase approximation (cRPA)~\cite{cRPA}, as explained in ref.~\cite{review2008}. The obtained (averaged) parameters of on-site Coulomb repulsion $U$, intraatomic exchange interaction $J$, and nonsphericity $B$ (see \ref{sec:Hubbard} for explanations) are $U$$=$ $1.79$ ($1.15$), $J$$=$ $0.85$ ($0.78$), and $B$$=$ $0.09$ ($0.07$) eV for CrCl$_3$ (CrI$_3$). The screened $U$ is not particularly large. Moreover, the screening is more efficient in CrI$_3$ due to proximity of the I $5d$ and Cr $t_{2g}$ bands~\cite{review2008}. The corresponding densities of states are shown in \Fref{fig.trihalides_5o}. As expected~\cite{AZA,PRB94}, the Coulomb $U$ further increases the band gap in comparison with LSDA. The band gap is smaller in CrI$_3$ because $U$ is smaller.
\noindent

\par Corresponding parameters of exchange interactions are summarized in Tables~\ref{tab.CrCl3_H_5o} and \ref{tab.CrI3_H_5o} for CrCl$_3$ and CrI$_3$, respectively. The approximate method $\hat{b}$ systematically underestimates the FM interactions. For instance, all interactions in CrCl$_3$ are antiferromagnetic, which clearly contradicts to the experimental situation. In CrI$_3$, only $J_{1}$ is weakly ferromagnetic, which is not enough to stabilize the FM ground state~\cite{PRB2019}. The method $M$ systematically improves the situation: the FM interactions clearly prevail and the FM ground state is stabilized both in CrCl$_3$ and CrI$_3$.\footnote{The situation with the AFM interlayer coupling in the case of CrCl$_3$ is rather fragile: the interaction $J_{2}$ is indeed weakly antiferromagnetic. However, in the $M$ method, it is counterbalanced by two weakly FM interaction $J_{4}$ and $J_{5}$.} The corresponding Curie temperature, evaluated in RPA~\cite{PRM2019} (see \ref{sec:RPA} for details), is also in reasonable agreement with the experimental data. The method $\hat{m}$ has a tendency to overestimate the interactions $J_{1}$, $J_{3}$, and $J_{6}$, making the FM ground state in CrI$_3$ unstable.
\noindent
\begin{table}[ht]
\caption{\label{tab.CrCl3_H_5o} Isotropic exchange interactions in CrCl$_3$ (in meV): results of the correlated Cr $3d$ model in the Hartree-Fock approximation. $\hat{b}$, $\hat{m}$, and $M$ stand for the methods based on the infinitesimal rotations of, respectively, the xc field, the magnetization matrix, and the local spin moments. $T_{\rm S}$ is the corresponding spin transition temperature in RPA (in K). The methods $\hat{m}$ and $M$ yield the ferromagnetic ground state, while the method $\hat{b}$ results in an incommensurate spin-spiral state. The notations of $J_{k}$ are explained in Fig.~\protect\ref{fig.CrX3_basic}b.}
\begin{indented}
\lineup
\item[]\begin{tabular}{@{}cccccccc}
\br
method    &  $J_{1}$           & $J_{2}$ & $J_{3}$ & $J_{4}$            & $J_{5}$           & $J_{6}$ & $T_{\rm S}$ \cr
\mr
$\hat{b}$ &  $-1.69$           & $-0.13$ & $-0.28$ & $-0.02$            & $-0.03$           & $-0.18$ & $6$         \cr
$\hat{m}$ &  $\phantom{-}6.66$ & $-0.06$ & $-0.58$ & $\phantom{-}0.02$  & $\phantom{-}0.02$ & $-0.35$ & $11$        \cr
$M$       &  $\phantom{-}3.76$ & $-0.02$ & $-0.13$ & $\phantom{-}0.05$  & $\phantom{-}0.05$ & $-0.14$ & $27$        \cr
\br
\end{tabular}
\end{indented}
\end{table}
\noindent
\begin{table}[ht]
\caption{\label{tab.CrI3_H_5o} The same as \Tref{tab.CrCl3_H_5o} but for CrI$_3$. The method $M$ yields the ferromagnetic ground state, while the methods $\hat{b}$ and $\hat{m}$ result in an incommensurate spin-spiral state. (A) denotes the same parameters calculated in the antiferromagnetic state.}
\begin{indented}
\lineup
\item[]\begin{tabular}{@{}cccccccc}
\br
method    &  $J_{1}$ & $J_{2}$ & $J_{3}$ & $J_{4}$ & $J_{5}$ & $J_{6}$ & $T_{\rm S}$ \cr
\mr
$\hat{b}$ &  $0.76$  & $-0.30$ & $-0.39$ & $0.04$  & $0$     & $-0.46$ & $10$        \cr
$\hat{m}$ &  $5.87$  & $-0.10$ & $-0.21$ & $0.33$  & $0.16$  & $-0.71$ & $14$        \cr
$M$       &  $4.58$  & $-0.08$ & $-0.06$ & $0.37$  & $0.25$  & $-0.37$ & $51$        \cr
$M$ (A)   &  $4.52$  & $-0.07$ & $-0.06$ & $0.36$  & $0.25$  & $-0.41$ & $49$        \cr
\br
\end{tabular}
\end{indented}
\end{table}

\par Thus, the minimal model, which can capture the FM character of the exchange interactions in the chromium trihalides is the 5-orbital model. Formally, the ferromagnetism can be obtained without invoking the ligand states. Nevertheless, it is crucial to consider the unoccupied $e_g$ bands. Furthermore, it is crucially important to use the exact technique, formulated in terms of the \emph{inverse} response function. The approximate method $\hat{b}$, which is linear in $\hat{\mathcal{R}}$, can lead to an incorrect magnetic ground state (see Tables~\ref{tab.CrCl3_H_5o} and \ref{tab.CrI3_H_5o}).

\par Generally, the interatomic exchange interactions, defined via infinitesimal rotations of spins, can depend on the magnetic state. In a number of cases, this dependence can be very strong, reflecting the dependence of the electronic structure on the magnetic arrangement~\cite{PRL1999,PRB2001}. Considering how $J_{k}$ in Cr$X_3$ change depending on the method used for their calculations (for instance, $M$ versus $\hat{b}$), it is reasonable to expect that the system is no longer in the strong-coupling limit and, therefore, these parameters can also depend on the magnetic state. However, this appears to be not the case. For instance, the exchange parameters calculated in the AFM state of CrI$_3$ are practically identical to the ones in the FM state (\Tref{tab.CrI3_H_5o}). This observation is very important and means, for example, that the parameters derived in the ordered FM state can be also used to evaluate the spin transition temperature, $T_{\rm S}$. The main reason why the $\hat{b}$ and $M$ methods yield different $J_{k}$ is related to the fact that the crystal-field splitting, $10Dq$, is comparable with the intraatomic exchange splitting. However, this $10Dq$ does not depend on the magnetic state, and therefore does not contribute to the magnetic-state dependence of $J_{k}$.

\subsection{\label{sec:CrX3_L} Role of the ligand states}
\par Now we turn to the analysis of the most general model, which explicitly includes the contributions of the Cr $3d$ as well we the ligand Cl $3p$ or I $5p$ states.

\subsubsection{\label{sec:CrX3_L_LSDA} Cr $3d$ and $X$ $p$ bands in LSDA}
\par
\begin{figure}[t]
\begin{center}
\includegraphics[width=6cm]{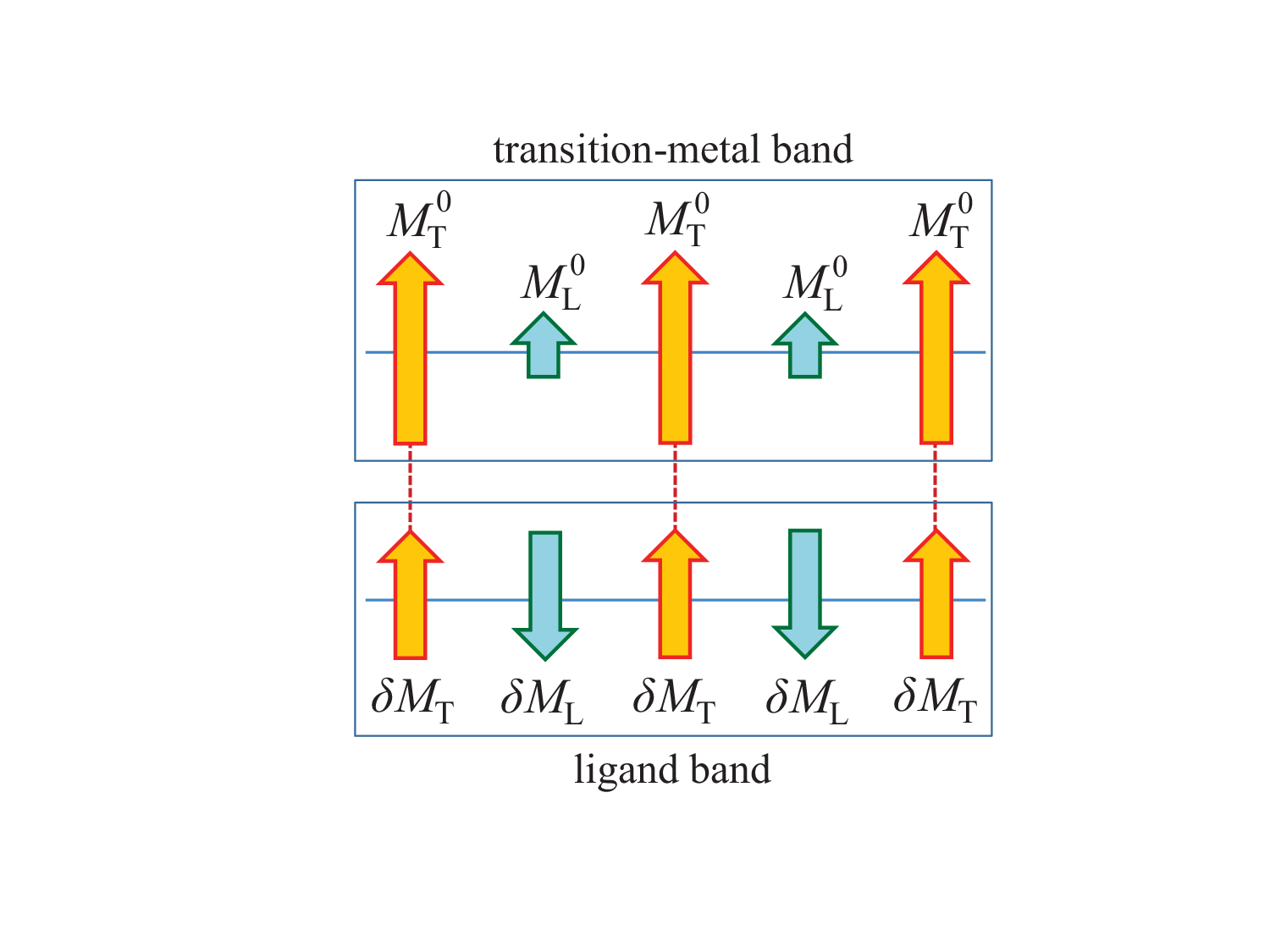}
\end{center}
\caption{Schematic view on the distribution of magnetic moments in polarized transition-metal (T) and ligand (L) bands: $M_{\rm T}^{0}$ and $M_{\rm L}^{0}$ denote the spin moments on the T and L sites in the T bands, and $\delta M_{\rm T}$ and $\delta M_{\rm L}$ denote those in the L band. The total moments are $M_{\rm T(L)} = M_{\rm T(L)}^{0}$$+$$\delta M_{\rm T(L)}^{\phantom{0}}$. In the T band, the hybridization between T and L sites induces $M_{\rm L}^{0}$ in the same direction as $M_{\rm T}^{0}$. Then, $\delta M_{\rm T}$ in the L band emerges as the joint effect of hybridization and intraatomic exchange interactions (denoted by dashed lines). Since the L bands is fully occupied, $\delta M_{\rm L} = -$$\delta M_{\rm T}$, resulting in the \emph{antipolarization} of $M_{\rm L}^{0}$ and $\delta M_{\rm L}^{\phantom{0}}$ (and also $M_{\rm L}$ and $M_{\rm T}$).}
\label{fig.Lpol}
\end{figure}

 We start with the analysis of general TB model, including both Cr $3d$ and $X$ $p$ bands, in LSDA. First, we note that the Cr $3d$ and ligand $p$ states are \emph{antipolarized}. In this case $M_{\rm Cr}$ exceeds the nominal value of $3$ $\mu_{\rm B}$ (where $\mu_{\rm B}$ denotes the Bohr magneton). Nevertheless, it is compensated by the negative $M_{X}$ on the ligand atoms, so that the total moment, $M_{\rm tot} =  M_{\rm Cr} + 3 M_{X}$, is integer and equal to $3$ $\mu_{\rm B}$. This antipolarization is a joint effect of the spin splitting on the Cr atoms and the hybridization between the occupied ligand states and unoccupied Cr $e_g$ states. Since the $\uparrow$-spin Cr $e_g$ states are closer to the ligand band, the hybridization is stronger, resulting in the stronger admixture of the $\uparrow$-spin Cr $e_g$ states into the occupied ligand band (and transfer of the $\uparrow$-spin ligand states into the unoccupied Cr $e_g$ band). Therefore, when the ligand band is added to the model, we have $M_{\rm Cr}>3$ $\mu_{\rm B}$ but $M_{X}<0$, as schematically illustrated in \Fref{fig.Lpol}.
\noindent
\noindent For instance, the LSDA yields $M_{\rm Cr}$$=3.14$ ($3.37$) $\mu_{\rm B}$ and $M_{X}$$=-0.05$ ($-$$0.12$) $\mu_{\rm B}$ for CrCl$_3$ (CrI$_3$). As expected, the effect is stronger in CrI$_3$ where the Cr $e_g$ states are closer to the I $5p$ band (see \Fref{fig.CrX3_basic}d). Moreover, the I $5p$ states are more extended in comparison with the Cl $3p$ ones, resulting in stronger hybridization.

\par The new aspect of calculations of the exchange interactions is that now they can also depend on the strength of the Stoner coupling $\mathcal{I}_{X}$, which contributes to the second term of \Eref{eq:jTT1}. Therefore, the first problem we have to solve is how to properly define $\mathcal{I}$. In fact, there are several possible definitions:
\noindent
\begin{itemize}
\item[(I)] $\mathcal{I}_{\mu} = -{\rm Tr}_{L} \{ \hat{b}_{\mu}  \hat{m}_{\mu} \}/(M_{\mu})^{2}$, where $\hat{b}_{\mu}$ is associated with the intraatomic spin splitting of $\hat{H}^{\sigma}$ and $\hat{m}_{\mu}$ is the ground-state magnetization;
\item[(II)] $\mathcal{I}_{\mu} = -B_{\mu}/M_{\mu}$, where $B_{\mu} = \frac{1}{n} {\rm Tr}_{L} \{ \hat{b}_{\mu} \}$ and $M_{\mu} = {\rm Tr}_{L} \{ \hat{m}_{\mu} \}$, which is nothing but the spherically averaged version of (I);
\item[(III)] the same as (I), but taking $\hat{b}_{\mu}$ from the sum rule \eref{eq:sumrule};
\item[(IV)] the same as (II), but taking $B_{\mu}$ from the sum rule $\vec{M} = \hat\mathbb{R}^{\uparrow \downarrow} \vec{B}$;
\item[(V)] the same as (I), but taking $\hat{m}_{\mu}$ and $M_{\mu}$ from the sum rule (and defining $\hat{b}_{\mu}$ as the intraatomic spin splitting of $\hat{H}^{\sigma}$);
\item[(VI)] the same as (II), but taking $M_{\mu}$ from the sum rule.
\end{itemize}
\noindent The results are summarized in \Tref{tab.CrX3_Stoner}.
\noindent
\begin{table}
\caption{\label{tab.CrX3_Stoner} The effective Stoner parameters in CrCl$_3$ and CrI$_3$ (in eV), obtained using six possible definitions (as explained in the text).}
\begin{indented}
\lineup
\item[]\begin{tabular}{@{}ccccc}
\br
           & \centre{2}{CrCl$_3$}                             & \centre{2}{CrI$_3$}                            \cr
\ns
           & \crule{2}                                        & \crule{2}                                      \cr
definition &  $\mathcal{I}_{\rm Cr}$ & $\mathcal{I}_{\rm Cl}$ & $\mathcal{I}_{\rm Cr}$ & $\mathcal{I}_{\rm I}$ \cr
\mr
I          &  $0.82$                 & $-3.08$                & $0.76$                 & $-0.55$               \cr
II         &  $0.81$                 & $-0.75$                & $0.75$                 & $-0.13$               \cr
III        &  $0.83$                 & $-0.28$                & $0.77$                 & $\phantom{-}0.31$     \cr
IV         &  $0.83$                 & $-3.32$                & $0.77$                 & $-0.46$               \cr
V          &  $0.83$                 & $-2.50$                & $0.77$                 & $-0.51$               \cr
VI         &  $0.83$                 & $-0.64$                & $0.77$                 & $-0.13$               \cr
\br
\end{tabular}
\end{indented}
\end{table}
\noindent One can see that $\mathcal{I}_{\rm Cr}$ only weakly depends on the definition (thought it is different in CrCl$_3$ and CrI$_3$). However, we do not need this parameter in our calculations. On the other hand, $\mathcal{I}_{X}$ is very sensitive to the definition~\cite{PRB2022}. Apparently, one can discard the definitions V and VI as they yield incorrect $M_{\rm tot}$ violating the fundamental property $M_{\rm tot} =3\mu_{\rm B}$.\footnote{$M_{\rm tot}$$= 2.93$ ($2.93$) and $2.91$ ($2.93$) $\mu_{\rm B}$ for CrCl$_3$ (CrI$_3$) in the schemes V and VI, respectively.} Furthermore, it makes sense to enforce the sum rule by using the definitions III and IV instead of I and II. Finally, the definitions II and IV are more appropriate for the spherically averaged method $M$, while the definitions I and III should be used in the combination with the matrix form of the methods $\hat{m}$ and $\hat{b}$.

\par While $\mathcal{I}_{\rm Cr}$ in Cr$X_3$ is close to atomic values ($\sim$$0.7$ eV) and can be interpreted as the intraatomic exchange integral responsible for Hund's first rule~\cite{Gunnarsson,Janak,Brooks}, $\mathcal{I}_{X}$ is definitely not. In most of the cases $\mathcal{I}_{X}$ is negative (except scheme III for CrI$_3$). This is another manifestation of the fact that the moments $M_{X}$ are merely induced by the hybridization with the Cr $3d$ states, which acts against $B_{X}$. According to \Eref{eq:djTT}, the negative $\mathcal{I}_{X}$ will tend to \emph{decrease} the FM coupling, contrary to rather common believe that the ferromagnetism in Cr$X_3$ is driven by the Hund's rule coupling on the ligand sites, as suggested by the phenomenological GKA rules\cite{Anderson1950,Goodenough1955,Goodenough1958,Kanamori1959}.

\par The parameters of exchange interactions are summarized in Tables~\ref{tab.CrCl3_lsda_L} and \ref{tab.CrI3_lsda_L}, for CrCl$_3$ and CrI$_3$, respectively. We note the following. In comparison with the 5-orbital model (\Sref{sec:CrX3_5o}), the parameters $J_{k}$ becomes mostly ferromagnetic, except antiferromagnetic $J_{6}$. The exchange interactions are sensitive to $\mathcal{I}_{X}$. This dependence is illustrated only for the method $M$, but very similar behavior is observed also for the methods $\hat{b}$ and $\hat{m}$. The use of the $M$ method in the combination with $\mathcal{I}_{X}$ obtained in the spherically averaged scheme IV substantially improves the agreement with the experimental data for $T_{\rm C}$ (but not for the spin-wave dispersion, which will be considered in~\Sref{sec:CrX3_exp}). The exchange parameters obtained in the method $\hat{m}$ are unrealistically large~\cite{PRB2021} and not shown here. Furthermore, in the earlier work~\cite{PRB2021}, the approximate method $\hat{b}$ was concluded to provide systematically worse description, especially when considers contributions of the ligand states. However, the situation crucially depends to two factors: (i) the proper choice for $\mathcal{I}_{X}$; (ii) the proper definition of the xc field $\hat{b}$. For instance, abilities of this method can be substantially improved by enforcing the sum rule in the definition of $\hat{b}$ and $\mathcal{I}_{X}$, and taking into account the aspherical contributions to $\mathcal{I}_{X}$, following the definition III.
\noindent
\begin{table}
\caption{\label{tab.CrCl3_lsda_L} Isotropic exchange interactions in CrCl$_3$ (in meV): results of the ${\rm Cr} \, 3d$+${\rm Cl} \, 3p$  model in LSDA. $\hat{b}$ and $M$ stand for the methods based on the infinitesimal rotations of, respectively, the xc field and the local spin moments. The xc field was either associated with the on-site spin splitting of the TB Hamiltonian (H) or obtained from the sum rule (sr). $\mathcal{I}_{\rm Cl}$ is the value of the effective Stoner parameter used in the calculations (the results of the methods III and IV from Table~\protect\ref{tab.CrX3_Stoner}, in eV). In the ${\rm L}0$ method, the xc field on the ligand sites is set to be zero, while that on the Cr sites is set to reproduce the total magnetization $M_{\rm tot} =3\mu_{\rm B}$ (further details are given in the text). $T_{\rm C}$ is corresponding Curie temperature in RPA (in K). The notations of $J_{k}$ are explained in~\Fref{fig.CrX3_basic}b.}
\begin{indented}
\lineup
\item[]\begin{tabular}{@{}ccccccccc}
\br
method         & $\mathcal{I}_{\rm Cl}$ & $J_{1}$           & $J_{2}$ & $J_{3}$ & $J_{4}$  & $J_{5}$  & $J_{6}$ & $T_{\rm C}$ \cr
\mr
$\hat{b}$ (H)  &       $-0.28$          &           $-0.51$ &  $0.26$ &  $0.34$ &  $0.16$  & $0.15$   & $-0.31$ & $21$        \cr
$\hat{b}$ (sr) &       $-0.28$          & $\phantom{-}4.21$ &  $0.28$ &  $0.26$ &  $0.13$  & $0.13$   & $-0.34$ & $49$        \cr
$M$            &       $-0.28$          & $\phantom{-}4.36$ &  $0.28$ &  $0.25$ &  $0.16$  & $0.16$   & $-0.32$ & $52$        \cr
$M$            &       $-3.32$          & $\phantom{-}0.64$ &  $0.25$ &  $0.53$ &  $0.08$  & $0.10$   & $-0.42$ & $20$        \cr
${\rm L}0$     &       $0$              & $\phantom{-}4.73$ &  $0.29$ &  $0.25$ &  $0.17$  & $0.17$   & $-0.32$ & $56$        \cr
\br
\end{tabular}
\end{indented}
\end{table}
\noindent
\begin{table}
\caption{\label{tab.CrI3_lsda_L} Isotropic exchange interactions in CrI$_3$ (in meV): results of the ${\rm Cr} \, 3d$+${\rm I} \, 5p$  model in LSDA (see \Tref{tab.CrCl3_lsda_L} for the notations).}
\begin{indented}
\lineup
\item[]\begin{tabular}{@{}ccccccccc}
\br
method         & $\mathcal{I}_{\rm I}$ & $J_{1}$  & $J_{2}$ & $J_{3}$ & $J_{4}$  & $J_{5}$  & $J_{6}$ & $T_{\rm C}$     \cr
\mr
$\hat{b}$ (H)  &   $\phantom{-}0.31$   & $3.93$   &  $0.77$ &  $0.75$ &  $0.87$  & $0.58$   & $-0.44$ & $\phantom{1}97$ \cr
$\hat{b}$ (sr) &   $\phantom{-}0.31$   & $3.40$   &  $0.76$ &  $0.71$ &  $0.85$  & $0.56$   & $-0.46$ & $\phantom{1}88$ \cr
$M$            &   $\phantom{-}0.31$   & $4.29$   &  $0.87$ &  $0.89$ &  $1.03$  & $0.67$   & $-0.41$ & $113$           \cr
$M$            &       $-0.46$         & $1.32$   &  $0.75$ &  $0.77$ &  $0.82$  & $0.54$   & $-0.51$ & $\phantom{1}62$ \cr
${\rm L}0$     &       $0$             & $3.21$   &  $0.84$ &  $0.86$ &  $0.96$  & $0.63$   & $-0.46$ & $\phantom{1}96$ \cr
\br
\end{tabular}
\end{indented}
\end{table}

\par Recently, Ke and Katsnelson~\cite{KeKatsnelson} reported the exchange parameters for CrI$_3$, which were also derived using the inverse response function (i.e. similar to our method $M$). In SDFT, they have found (using our notations, in meV): $J_{1} = 6.58$, $J_{2} = 0.38$, $J_{3} = 1.14$, $J_{4} = 0.62$, $J_{5} = 0.94$, and $J_{6} = -0.14$. These parameters are in reasonable agreement with our data, perhaps except $J_{1}$, which is systematically smaller in our case. Nevertheless, our experience shows that in order to obtain reasonable parameters $J_{k}$ in CrI$_3$ it is absolutely crucial to consider the contributions of the ligand states (using the downfolding technique described in \Sref{sec:downf}). The bare exchange interactions between the Cr $3d$ states are strongly antiferromagnetic and fails to reproduce the correct FM ground state~\cite{PRB2021}. Unfortunately, it is not clear how this problem was tackled by Ke and Katsnelson~\cite{KeKatsnelson}. Furthermore, our parameters are sensitive to the choice of $\mathcal{I}_{\rm I}$, as seen in \Tref{tab.CrI3_lsda_L}.

\par The sensitivity of the exchange interactions to the parameter $\mathcal{I}_{X}$ can be regarded as the weak point of the downfolding technique. Nevertheless, one can propose an alternative option, which can be viewed as an extension of the method $M$ for the FM insulators and half-metals. In the future, we will call it the ${\rm L}0$ scheme. The crucial observation is that $M_{X}$ is induced by the hybridization between the Cr $3d$ and ligand $p$ states (which is totally in line with the general idea of the downfolding technique, considered in \Sref{sec:downf}). Then, it is reasonable to enforce $B_{X} =0$ and find the remaining parameter of the xc field, $B_{\rm Cr}$, from the equation $\vec{M} = \hat\mathbb{R}^{\uparrow \downarrow} \vec{B}$. This $B_{\rm Cr}$ induces the magnetic moments on the Cr sites as well as the ligand sites. Then, the parameter $B_{\rm Cr}$ can be chosen to reproduce $M_{\rm tot} =3$ $\mu_{\rm B}$. For CrCl$_3$ (CrI$_3$) in LSDA, such procedure yields $M_{\rm Cr}$$=3.18$ ($3.40$) $\mu_{\rm B}$ and $M_{X}$$=-$$0.06$ ($-$$0.13$) $\mu_{\rm B}$, which are pretty close to the values of spin magnetic moments in the LSDA ground state. The good aspect of this approximation is that $\mathcal{I}_{X}=0$ and the exchange interactions are solely determined by the 1st term of \Eref{eq:jTT1} (but with the redefined $M_{\rm Cr}$ for $B_{X} =0$). The corresponding values of $J_{k}$ and $T_{\rm C}$ are also listed in Tables~\ref{tab.CrCl3_lsda_L} and \ref{tab.CrI3_lsda_L}. In comparison with the $M$ scheme, these parameters somewhat overestimate $T_{\rm C}$ but otherwise fall within the range of typical estimates for $J_{k}$. We will continue to use this ${\rm L}0$ scheme in the future. It is especially good for semiquantitative estimates, which would illustrate the basic trend in the behavior of interatomic exchange interactions. The 2nd term in \Eref{eq:jTT1} also vanishes if $\hat{\mathbb{Q}}^{\uparrow \downarrow}_{\rm LL} = 0$, meaning that there is no ligand-ligand interactions. In this sense, there is certain similarity with the downfolding method proposed by Logemann \etal~\cite{Logemann}, but reformulated for the scheme $M$.

\subsubsection{\label{sec:CrX3_XinCr3d} Ligand states in correlated Cr $3d$ band}
\par In this section, we try to extend the correlated 5-orbital model, by expanding its Wannier basis $W$ into the pseudo-atomic Wannier basis, $A$, of the more general model, containing Cr $3d$ as well as the ligand $p$ bands. Namely, we still consider only ten Cr $3d$ bands (for each spin), but transform $| C \rangle$ in \Eref{eq:gresponse} from the basis $W$ to the basis $A$: $| C^{W} \rangle \to | C^{A} \rangle = \hat{T}| C^{W} \rangle$, where $\hat{T}$ is the transformation matrix. Our intension here is to consider explicitly the $X$ $p$ states and all the contributions of these states to the exchange interactions in the 5-orbital model. The magnetic moments in the $A$ basis are $M_{\rm Cr} =2.63$ ($2.56$) $\mu_{\rm B}$ and $M_{X} =0.12$ ($0.15$) $\mu_{\rm B}$ for CrCl$_3$ (CrI$_3$). In the 5-orbital model, the only occupied $\uparrow$-spin $t_{2g}$ band provides $3$ electrons, which are distributed between one Cr and three ligand atoms. Therefore, the Cr and ligand moments in the $t_{2g}$ band are ferromagnetically coupled (see also \Fref{fig.Lpol}), while in order to obtain the opposite polarization of these states, it is essential to consider a more general model, which would explicitly include the ligand $p$ band.

\par Without the ligand bands, the inversion of the matrix $\hat{\mathbb{R}}^{\uparrow \downarrow}$ in the $A$ basis becomes unstable as it contains small matrix elements associated with the ligand sites. While for calculations of the exchange interactions for the given $\mathcal{I}_{X}$ such inversion can be largely avoided using Equations (\ref{eq:jTT0}) and (\ref{eq:djTT}), the inverse matrix is still needed to evaluate $\mathcal{I}_{X}$ using the sum rule. Nevertheless, one can still try to estimate $J_{k}$ using the method ${\rm L}0$, where it is sufficient to invert only the subblock of $\hat{\mathbb{R}}^{\uparrow \downarrow}$ in the subspace of the Cr sites. It yields the magnetic moments: $M_{\rm Cr} =2.64$ ($2.59$) $\mu_{\rm B}$ and $M_{X} =0.12$ ($0.14$) $\mu_{\rm B}$ for CrCl$_3$ (CrI$_3$), which are close to the ones reported above.

\par The corresponding exchange interactions and $T_{\rm C}$ for the 5-orbital model reformulated in the pseudo-atomic basis, are clearly overestimated (see \Tref{tab.CrX35o_L0}, the rows denoted 5o).
\noindent
\begin{table}[ht]
\caption{\label{tab.CrX35o_L0} Exchange interactions in CrCl$_3$ and CrI$_3$ (in meV) obtained in the scheme ${\rm L}0$ for the 5-orbital model with the ligand states (5o) and after merging this model with the ligand $p$ bands (5o+$p$). $T_{\rm C}$ is the corresponding Curie temperature in RPA (in K). The notations of $J_{k}$ are explained in~\Fref{fig.CrX3_basic}b.}
\begin{indented}
\lineup
\item[]\begin{tabular}{@{}lccccccc}
\br
     system                  & $J_{1}$            & $J_{2}$ & $J_{3}$ & $J_{4}$  & $J_{5}$  & $J_{6}$            & $T_{\rm C}$   \cr
\mr
CrCl$_3$ (5o)                & $14.27$            &  $0.85$ &  $1.65$ &  $0.42$  & $0.45$   & $\phantom{-}0.21$  & $242$         \cr
CrCl$_3$ (5o+$p$)            & $10.81$            &  $0.28$ &  $0.35$ &  $0.18$  & $0.15$   & $-0.27$            & $102$         \cr
CrI$_3$\phantom{C} (5o)      & $14.03$            &  $2.37$ &  $3.63$ &  $1.89$  & $1.59$   & $\phantom{-}0.76$  & $484$         \cr
CrI$_3$\phantom{C} (5o+$p$)  & $\phantom{1}6.87$  &  $0.72$ &  $0.91$ &  $1.14$  & $0.63$   & $-0.31$            & $119$         \cr
\br
\end{tabular}
\end{indented}
\end{table}
\noindent However, this is quite in line with general understanding. The interactions mediated by the tails of the Wannier functions spreading to the ligand sites and transferring there the FM magnetization can be viewed as an analog of the direct Heisenberg exchange~\cite{Heisenberg}. In order to evaluate these direct exchange contributions numerically, one typically performs the 6-dimensional integration in the real space~\cite{Mazurenko2007,Ku2002,Danis2016}. Nevertheless, the downfolding method suggests how these contributions can be naturally evaluated within SDFT. The bare direct exchange integrals are typically large and need to be additionally scaled, in the spirit of cRPA, to account for the screening caused by other bands~\cite{Mazurenko2007,Danis2016}. The same situation is here: the attempt to include the ligand states into 5-orbital model only worsen the description of exchange interactions. The discrepancy can be resolved by considering more general model, which would explicitly include the contributions of the ligand $p$ band.

\subsection{\label{sec:CrX3_L_5o} Merging correlated and ligand bands}
\par
\begin{figure}[t]
\begin{center}
\includegraphics[width=6.0cm]{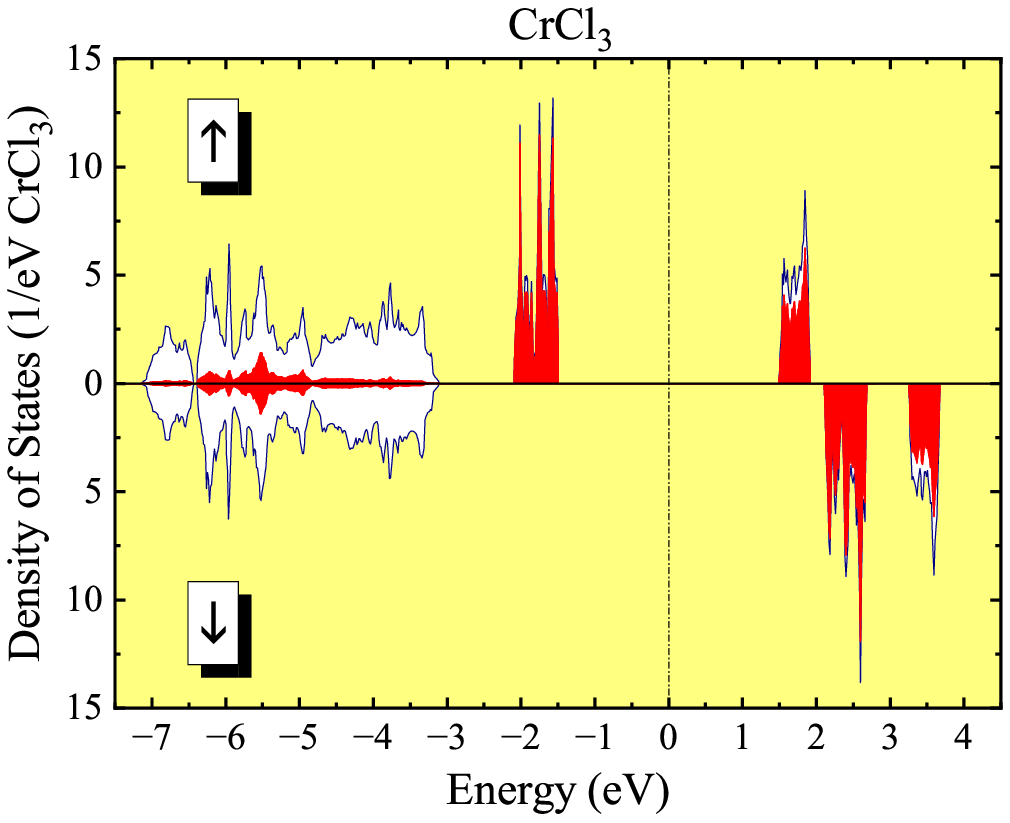}
\includegraphics[width=6.0cm]{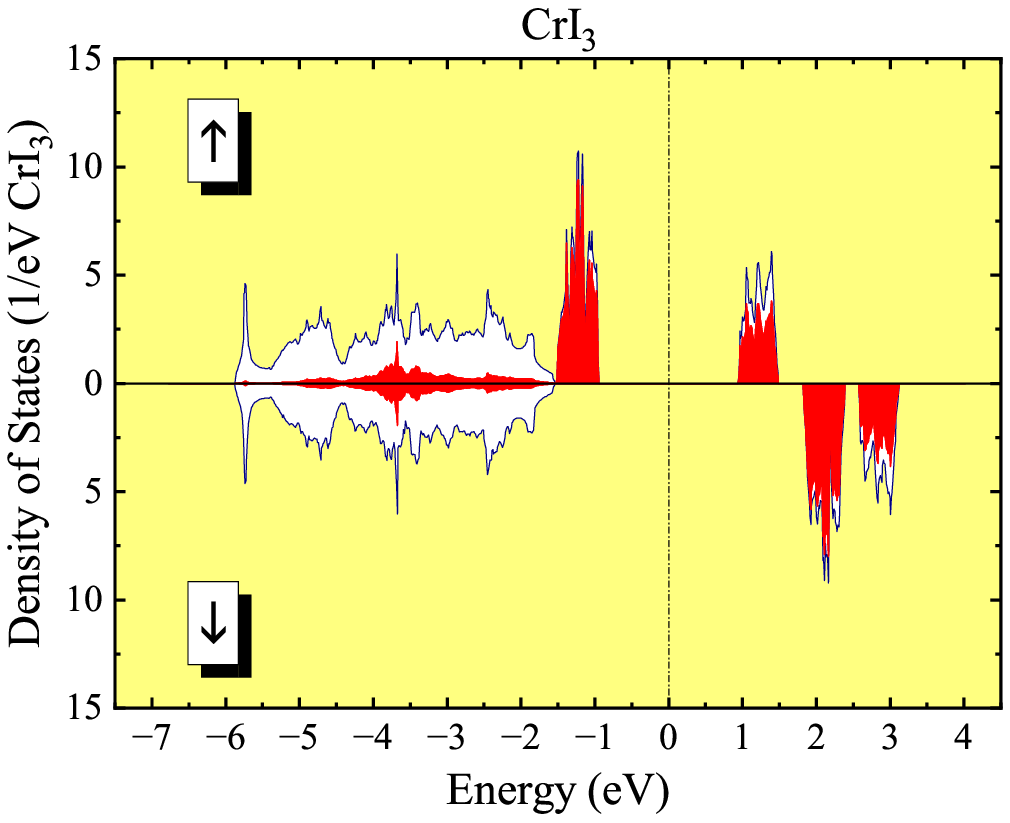}
\end{center}
\caption{Electronic structure obtained by merging the correlated Cr $3d$ bands, treated in the model Hartree-Fock approximation, and uncorrelated ligand $p$ bands in LDA for CrCl$_3$ (left) and CrI$_3$ (right). The zero energy is in the middle of the band gap.}
\label{fig.CrX3_merging}
\end{figure}

 The next important step is to merge the correlated Cr $3d$ bands with the ligand bands in the framework of the Cr $3d$ $+$ $X$ $p$ model. The correlated 5-orbital model is formulated in the Wannier basis for the Cr $3d$ bands in LDA. After solving this model in the Hartree-Fock approximation, we want to replace the original Cr $3d$ bands in the all-electron LDA band structure by these correlated bands in the pseudo-atomic $A$ basis. Since such basis states of the 5-orbital model and the ligand $p$ bands are orthogonal to other states, such merging is not unique and an arbitrary energy shift of the Cr $3d$ and $X$ $p$ bands relative to each other will not change the magnetization (but will change the exchange interactions). Similar problem arises in the LDA$+$$U$ method~\cite{AZA}, where the relative position of the transition-metal $3d$ and ligand $p$ states is controlled by the empirical double-counting term~\cite{PRB94,PRB1998}. Therefore, we have to make some empirical conjecture about this merging and we choose it from the ``charge neutrality'' condition, by requesting the Fermi energy of the correlated model to coincide with the one in LDA. If the system is insulating, Fermi energy is chosen in the middle of the gap. The electronic structure after such merging is shown in Fig.~\ref{fig.CrX3_merging}.
\noindent

\par We would like to emphasize that this electronic structure is rather artificial and would require different sets of the xc fields for the Cr $3d$ and $X$ $p$ bands (zero for the $X$ $p$ band, but finite for the Cr $3d$ one), which is hardly useful for our purposes. Therefore, in order to evaluate the exchange interactions, we again employ the ${\rm L}0$ technique. It yields the following magnetic moments: $M_{\rm Cr} =3.35$ ($3.49$) $\mu_{\rm B}$ and $M_{X} =-0.12$ ($-0.16$) $\mu_{\rm B}$ for CrCl$_3$ (CrI$_3$). Thus, after adding the $X$ $p$ band, the ${\rm L}0$ method nicely capture the antipolarization of the Cr $3d$ and $X$ $p$ states, which is again stronger in CrI$_3$. However, it should be understood that this effect is ``mimicked'' by the specific choice of the xc fields, while from the viewpoint of electronic structure itself, the $X$ $p$ band remains unpolarized and the Cr $3d$ band is polarized as in \Sref{sec:CrX3_XinCr3d}. In the other words, in the ${\rm L}0$ method, the imperfection of the electronic structure is corrected by the physical choice of the xc fields (though the electronic structure itself was obtained \emph{not} for these fields).

\par The corresponding exchange parameters are listed in \Tref{tab.CrX35o_L0} (and denoted as 5o$+$$p$). We note a systematic improvement in comparison with the 5-orbital model with the ligand states (denoted as 5o): all exchange interactions become smaller so as $T_{\rm C}$. However, this improvement is only partial as these parameters are still too large and clearly overestimate the tendency towards the ferromagnetism. This demonstrates the complexity of the problem: the ${\rm L}0$ method is only as the starting point of the self-consistent solution, which should take into account the feedback of correlated Cr $3d$ bands onto the ligand bands via the xc field and results in the magnetic polarization of the ligand bands. However, such field will inevitably mix up the Cr $3d$ and $X$ $p$ bands. After that, one has to redefine the Wannier basis for the correlated model and recalculate the model parameters. An alternative solution is LDA$+$$U$~\cite{AZA,PRB94}, which considers both transition-metal and ligand states, and takes into account the polarization of the ligand bands. Today there are many applications of this method for calculating the interatomic exchange interactions in various transition-metal oxides and other strongly correlated systems~\cite{Kvashnin2015,Korotin2015,Yoon2018}. However, it should be understood that the LDA$+$$U$ approach is supplemented with additional approximations of purely empirical character, such as the form of the double counting as well as the choice of the Coulomb interaction parameters and the correlated subspace for the solution of the Hubbard-type model~\cite{PRB1998}.

\subsection{\label{sec:CrX3_DM} Dzyaloshinskii-Moriya interactions}
\par In this section, we briefly discuss the behavior of DM interactions in CrI$_3$, which are driven by the strong SO coupling of the heavy I atoms (so as other magnetic interactions of the relativistic origin~\cite{Lado2017}). The DM interactions in CrCl$_3$ are considerably weaker~\cite{PRB2023,Kvashnin2020} and not significant~\cite{Chen2022}. In addition to $d^{z}$, two other components of the DM vector, $d^{x}$ and $d^{y}$, can be computed by rotating the coordinate frame and applying the same \Eref{eq:dzqs} (or similar Equations for the $\hat{b}$ or $\hat{m}$ rotations). The symmetry of CrI$_3$ is such that two Cr sublattices (shown by different colors in \Fref{fig.CrX3_basic}a,b) are transformed to each other by the spacial inversion. Therefore, all intersublattice interactions will identically vanish~\cite{Dzyaloshinskii_weakF,Moriya_weakF}. On the other hand, the interactions within the sublattices can exist and for each vector $\boldsymbol{R}$ connecting two Cr sites, the interactions in the sublattices $I$ and $II$ are related as: $\boldsymbol{d}^{II}(\boldsymbol{R}) = - \boldsymbol{d}^{I}(\boldsymbol{R})$. The strongest interaction is $\boldsymbol{d}_{3}$, which occurs in the 2nd coordination sphere of the honeycomb plane (together with the isotropic interaction $J_{3}$). In total, there are six bonds $n$ connecting the central atom with the atoms in the 2nd coordination sphere. The corresponding DM vectors are $\boldsymbol{d}_{3,n} = \Big( d^{xy}\cos ( \phi_{n}+\psi ), d^{xy}\sin ( \phi_{n}+ \psi ), (-1)^{n}d^{z} \Big)$, where $\phi_{n} = (2n$$+$$1)\frac{\pi}{6}$ is the azimuthal angle specifying the direction of the bond $n$  in the $xy$ plane (relative to the bond along $\boldsymbol{a}$ in \Fref{fig.CrX3_basic}a, corresponding to $n=1$) and $\psi$ specifies the direction of the DM vector in the plane relative to this bond. For the $R\overline{3}$ symmetry, the parameters $d^{xy}$, $d^{z}$, and $\psi$ do not depend on $n$. They are listed in \Tref{tab.CrX3DM}.
\noindent
\begin{table}
\caption{\label{tab.CrX3DM} Parameters of Dzyaloshinskii-Moriya interactions in CrI$_3$ ($d^{xy}$ and $d^{z}$ are in meV and $\psi$ is in degrees) in correlated 5-orbital model (5o) and LSDA for the all-electron ${\rm Cr} \, 3d$+${\rm I} \, 5p$ model.}
\begin{indented}
\lineup
\item[]\begin{tabular}{@{}ccccc}
\br
method    & model      & $|d^{xy}|$ & $\psi$ & $|d^{z}|$    \cr
\mr
$\hat{b}$ & 5o         & $0.01$        & $39$   &  $0.22$    \cr
$M$       & 5o         & $0.02$        & $19$   &  $0.20$    \cr
$\hat{b}$ & LSDA       & $0.08$        & $-2$   &  $0.25$    \cr
$M$       & LSDA       & $0.08$        & $0$    &  $0.28$    \cr
\br
\end{tabular}
\end{indented}
\end{table}

\par We note that the schemes $\hat{b}$ and $M$ provide very consistent description for the DM interactions. The small discrepancy for $\psi$ in the correlated 5-orbital model is probably due to the fact that the angle $\psi$ is ill-defined when $d^{xy}$ is small. There is also surprisingly good agreement between results of the correlated 5-orbital model and the Cr $3d$ $+$ I $5p$ model in LSDA. However, such agreement is probably fortuitous because the models are very different.\footnote{The correlated 5-orbital model includes orbital polarization caused by on-site Coulomb interactions~\cite{PRB2014}. On the other hand, the ${\rm Cr} \, 3d$+${\rm I} \, 5p$ model in LSDA includes the additional contributions steaming from the magnetically polarized I $5p$ band. Apparently, these two effects are comparable with each other, which explain a good agreement in \Tref{tab.CrX3DM}.} The DM interactions in CrCl$_3$ are considerably weaker ($|d^{z}| \sim 0.02$ meV~\cite{PRB2023}). However, this is to be expected and the order of magnitude difference of the DM interactions in CrCl$_3$ and CrI$_3$ well correlates with the strength of the SO coupling on the ligand sites, which differs by the same order of magnitude: $\xi_{\rm Cl}=95$ meV versus $\xi_{\rm I}=881$ meV.

\subsection{\label{sec:CrX3_exp} Comparison with experimental data}
The experimental parameters of exchange interactions, derived from the inelastic neutron scattering data~\cite{ChenPRX,Chen2022,Schneeloch2022}, are summarized in \Tref{tab.CrXexp}.\footnote{In order to be consistent with our definition of the spin model, \Eref{eq:Hspin}, the experimental parameters have been multiplied by $S^{2}=(3/2)^{2}$.} One of the interesting features of the experimental spin-wave dispersion in CrI$_3$ is a $\sim 4$ meV gap at the Dirac (K) point, indicating at the existence of large DM interaction $d^{z}$ between 2nd neighbors in the honeycomb plane~\cite{ChenPRX}. Quite expectably, no such feature was observed in CrCl$_3$~\cite{Chen2022,Schneeloch2022}, where this DM interaction is small. For the isotropic interactions, the agreement between theoretical and experimental data depends on the model and approximations employed for treating the Coulomb correlations and the ligand states, which we will discuss below. The theoretical DM interaction in CrI$_3$ appears to be underestimated by factor two, both in the correlated 5-orbital model and LSDA for the  ${\rm Cr} \, 3d$+${\rm I} \, 5p$ model.
\noindent
\begin{table}[ht]
\caption{\label{tab.CrXexp} Experimental parameters of exchange interactions in CrCl$_3$ and CrI$_3$.}
\begin{indented}
\lineup
\item[]\begin{tabular}{@{}ccccccc}
\br
compound  & reference             & $J_{1}$ & $J_{2}$            & $J_{3}$ & $J_{6}$ & $d^{z}$     \cr
\mr
CrCl$_3$  & \cite{Chen2022}       & $2.14$  & $-0.02$            &  $0.05$ & $-0.11$ & $0.03$      \cr
CrCl$_3$  & \cite{Schneeloch2022} & $2.12$  &                    &  $0.08$ & $-0.16$ &             \cr
CrI$_3$   & \cite{ChenPRX}        & $4.52$  & $\phantom{-}1.33$  &  $0.36$ & $-0.18$ & $0.50$      \cr
\br
\end{tabular}
\end{indented}
\end{table}

\par
\begin{figure}[t]
\begin{center}
\includegraphics[width=6.0cm]{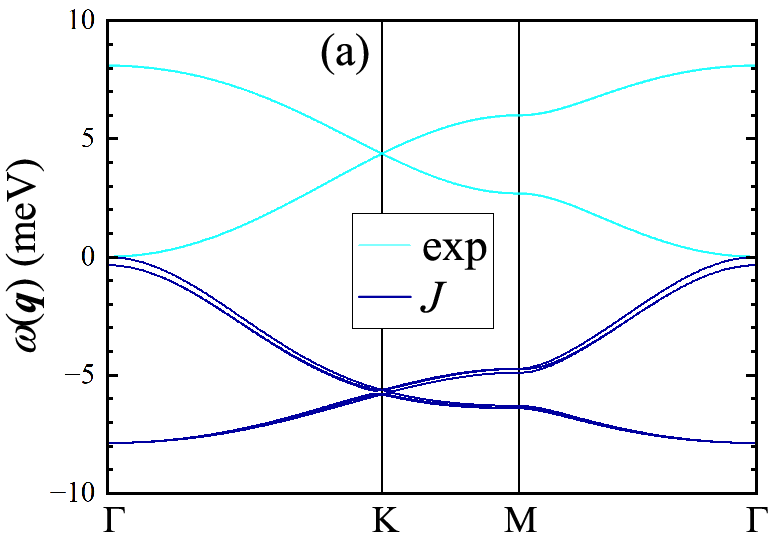} \includegraphics[width=6.0cm]{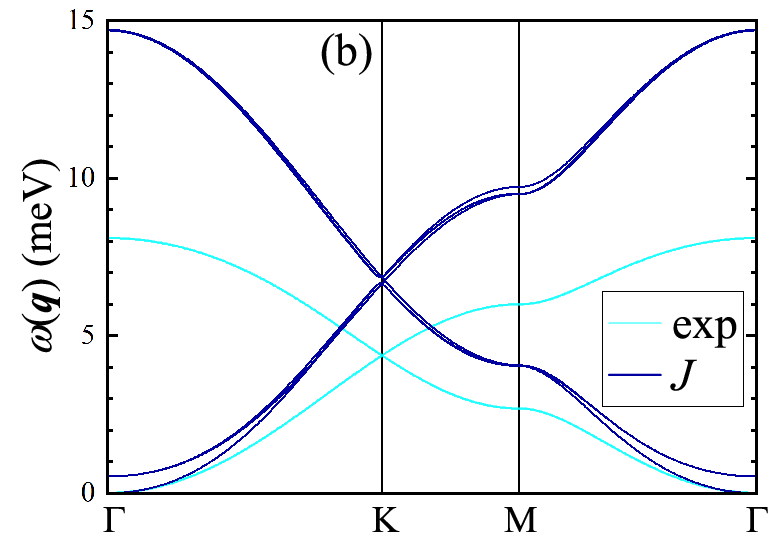} \\
\includegraphics[width=6.0cm]{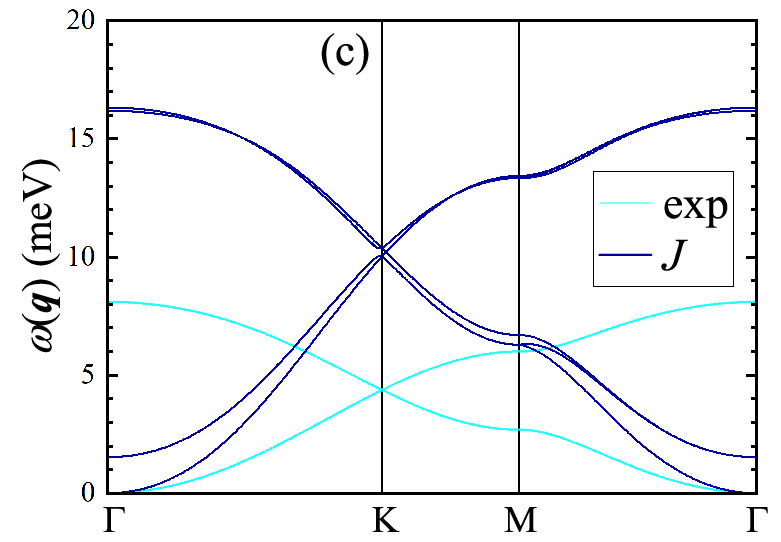} \includegraphics[width=6.0cm]{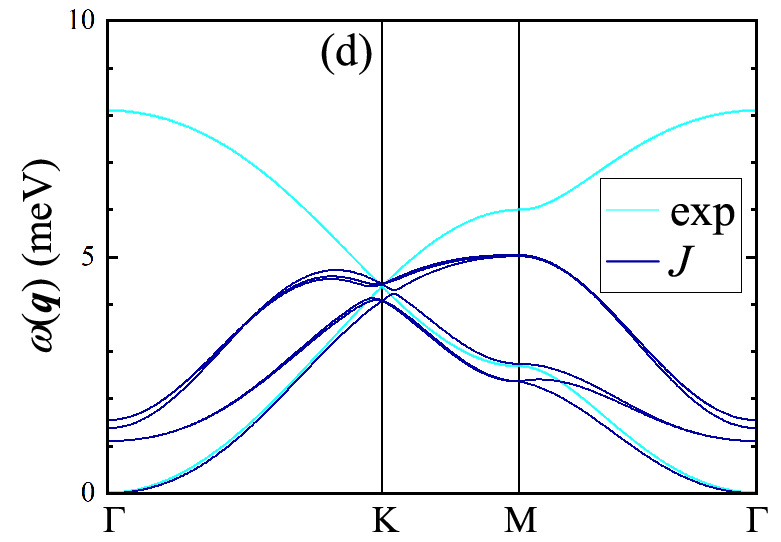}
\end{center}
\caption{Theoretical and experimental spin-wave dispersion for CrCl$_3$: Results of correlated 5-orbital model with the isotropic parameters ($J$) obtained in the schemes $\hat{b}$ (a) and $M$ (b); Results of all-electron ${\rm Cr} \, 3d$+${\rm Cl} \, 3p$ model in LSDA with the parameters obtained in the schemes $\hat{b}$ (c) and $M$ (d) (data rows 2 and 4 in \Tref{tab.CrCl3_lsda_L}). The calculations are performed for the hexagonal cell, where six branches of $\omega(\boldsymbol{q})$ correspond to six magnetic Cr sublattices.}
\label{fig.CrCl3_SW}
\end{figure}

\begin{figure}[t]
\begin{center}
\includegraphics[width=6.0cm]{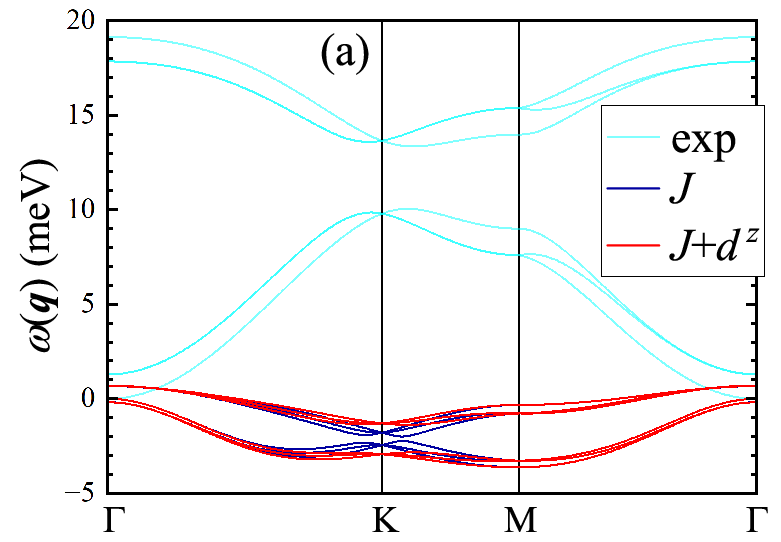} \includegraphics[width=6.0cm]{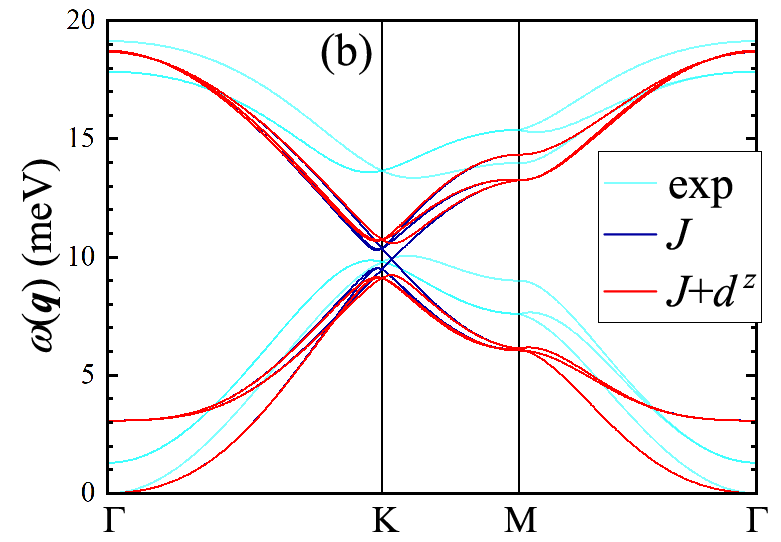} \\
\includegraphics[width=6.0cm]{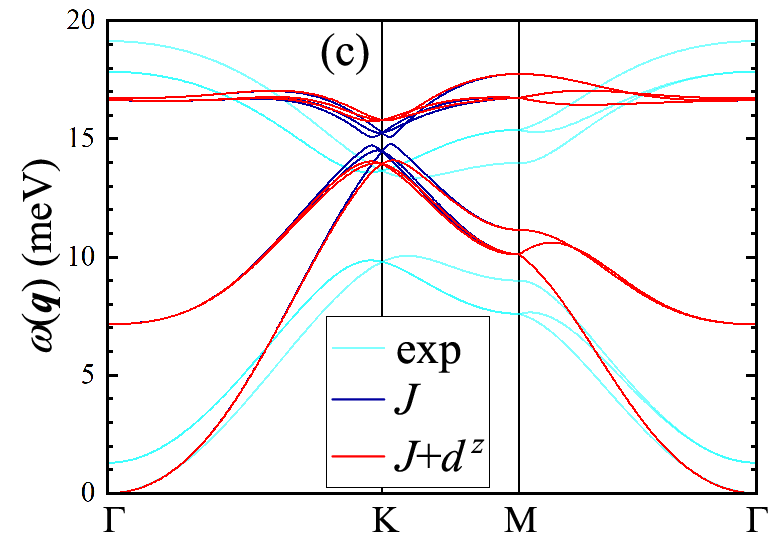} \includegraphics[width=6.0cm]{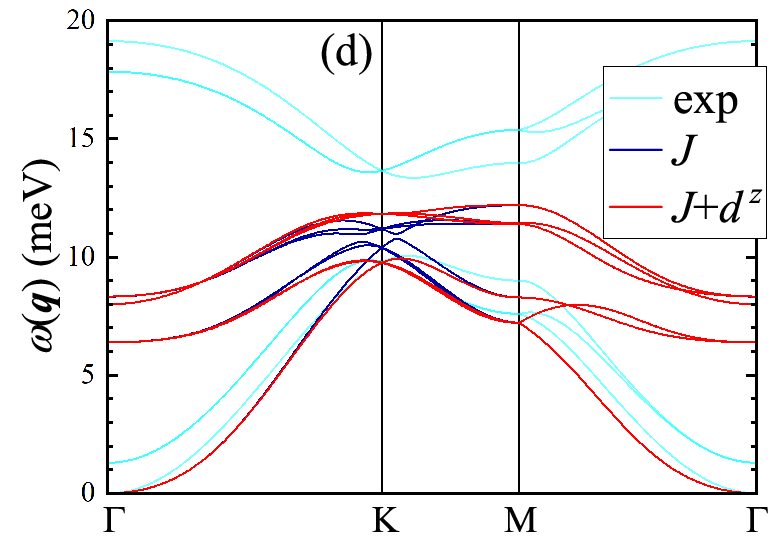}
\end{center}
\caption{Theoretical and experimental spin-wave dispersion for CrI$_3$: Results of correlated 5-orbital model with the isotropic only ($J$) as well as both isotropic and DM ($J$$+$$d^{z}$) parameters obtained in the schemes $\hat{b}$ (a) and $M$ (b); Results of all-electron Cr ${\rm Cr} \, 3d$+${\rm I} \, 5p$ model in LSDA with the parameters obtained in the schemes $\hat{b}$ (c) and $M$ (d) (data rows 2 and 4 in \Tref{tab.CrI3_lsda_L}). The calculations are performed for the hexagonal cell, where six branches of $\omega(\boldsymbol{q})$ correspond to six magnetic Cr sublattices.}
\label{fig.CrI3_SW}
\end{figure}

 The theoretical spin-wave dispersion in CrCl$_3$ and CrI$_3$, calculated for some representative sets of parameters, is plotted in Figures~\ref{fig.CrCl3_SW} and \ref{fig.CrI3_SW}, respectively, in comparison with the experimental data.
\noindent
\noindent
\noindent First, let us consider results of correlated 5-orbital model (shown in the panels a and b of these Figures), where there are no additional complications caused by the ligand states. Within this 5-orbital model, the approximate scheme $\hat{b}$ severely underestimates the FM interactions, making the FM ground state unstable both in CrCl$_3$ and CrI$_3$. The situation is dramatically improved when we use the ``exact'' scheme $M$: it strengthens the FM interactions and stabilizes the FM ground state in both the compounds. In CrCl$_3$, the dispersion is overestimated by factor two, while in CrI$_3$ we obtain an overall fair agreement with the experimental data, except for the band splitting in the point K, which is underestimated by factor two. Nevertheless, the situation changes significantly when we explicitly consider the contributions of the ligand states in the framework of the all-electron ${\rm Cr} \, 3d$+$X \, p$ model in LSDA (panels c and d). Now, the FM state becomes stable already in the scheme $\hat{b}$. The spin-wave dispersion in CrCl$_3$ remains overestimated by factor two. The agreement with the experimental data is better for CrI$_3$, again except for the band splitting in the point K. When we turn to the $M$ scheme, which is expected to be more accurate, the theoretical spin-wave dispersion in CrCl$_3$ is substantially reduced, so that the lowest experimental band is reproduced pretty well. However, the behavior of theoretical upper bands is not satisfactory. This is related to the fact that the theoretical $J_{1}=0.64$ meV (\Tref{tab.CrCl3_lsda_L}) is underestimated by factor three. A similar situation occurs in CrI$_3$, where the theoretical $J_{1}=1.32$ meV (\Tref{tab.CrI3_lsda_L}) is underestimated by the same amount, which worsens the agreement with the experimental data. The problem is partially related to our choice of $\mathcal{I}_{X}$, which appears to be more negative for the scheme $M$ due to the additional spherical approximation, as discussed in \Sref{sec:CrX3_L_LSDA}. For instance, if one uses the same $\mathcal{I}_{X}$ as in the scheme $\hat{b}$ (or the ${\rm L}0$ scheme, where we do not need any $\mathcal{I}_{X}$), one could get a better agreement with the experimental data for the upper bands (but not for the dispersion in the entire energy region).

\par Regarding the value of $d^{z}$ and the band gap, first we note that there is an alternative mechanism of opening the band gap, related to long-range isotropic exchange interactions beyond the 6th coordination sphere~\cite{KeKatsnelson}. These interactions were taken into account also in our calculations.\footnote{We have considered all the interactions within the coordination sphere of about $16$ ($20$) \AA~in CrCl$_3$ (CrI$_3$). \Fref{fig.CrX3_basic}b shows only representative parameters up to 6th coordination sphere.} However, their effect is not particularly strong. Therefore, the band gap is mainly controlled by $d^{z}$ and the discrepancy with the experimental data is caused by the underestimation of this interaction in the theoretical calculations, as was also reported by Kvashnin \etal~\cite{Kvashnin2020} and Olsen~\cite{Olsen}. In this respect, we note that by explicitly considering the contributions of the ligand states in the correlated 5-orbital model (\Sref{sec:CrX3_XinCr3d}) and using the ${\rm L}0$ scheme for the evaluation of the exchange parameters, we can easily get $|d^{z}|$ as large as $1.10$ meV, which exceeds the value obtained in the regular correlated 5-orbital model by factor five and the experimental value by factor two~\cite{ChenPRX}. Such enhancement is caused by the additional contribution of the basis functions, which explicitly takes into account the effect of the heavy I atoms. Nevertheless, when we try to combine this effect with another contribution stemming from the I $5p$ band itself (employing for these purposes the merging of the Cr $3d$ and I $5p$ bands, as discussed in \Sref{sec:CrX3_L_5o}), again in the framework of the ${\rm L}0$ scheme, the DM interaction is reduced till $|d^{z}|=0.16$ meV. Such strong reduction is apparently caused by the cancelation of contributions in the Cr $3d$ and I $5p$ bands. Since the I $5p$ states in the Cr $3d$ and I $5p$ bands are \emph{antipolarized}, the fact of the cancelation itself is not surprising. However, such analysis clearly demonstrates the fragility of the situation, where the value of $d^{z}$ appears to depend on the delicate balance of two large contributions arising from the Cr $3d$ and I $5p$ bands. The merging of these two bands considered in \Sref{sec:CrX3_L_5o} was probably too crude to explain the experimental situation.

\subsection{\label{sec:CrX3_summary} Brief summary}
\par To summarize this section, we would like to stress again the main points:
\begin{itemize}
\item
The minimal model, which captures the magnetic properties of Cr$X_{3}$, is the 5-orbital model, constructed for the magnetic Cr $3d$ bands near the Fermi level. The main advantage of this model is the simplicity: in this case there is simply no contributions associated with the ligand states and all calculations of the exchange interactions become pretty straightforward. Nevertheless, it is absolutely essential to use for these purposes the ``exact'' approach dealing with rotations of magnetic moments $M$. The approximate scheme, dealing with rotations of the xc fields $\hat{b}$, severely underestimate the FM interactions and fails to reproduce the FM ground state. The scheme $M$ improves the situation tremendously;
\item
The FM interactions can be additionally stabilized in the all-electron ${\rm Cr} \, 3d$+$X \, p$ model. However, the exchange interactions in this case depend on the strength of the effective Stoner coupling $\mathcal{I}_{X}$ on the ligand atoms and the proper definition of these Stoner parameters is still not completely resolved problem. As soon as the xc field is corrected to satisfy the sum rules for the given magnetization, the approximate $\hat{b}$ scheme works reasonably well. The main issue in this context is even \emph{not} the differences between the schemes $\hat{b}$ and $M$, but how to properly define $\mathcal{I}_{X}$ within each scheme. As an alternative solution, we have proposed the ${\rm L}0$ method, which makes the related to $\mathcal{I}_{X}$ contributions inactive. This method can be used for FM insulators and half-metallic materials;
\item
Another open question is how to properly merge the correlated Cr $3d$ bands and ligand $X$ $p$ bands. The exchange interactions can crucially depend on details of such merging. The method considered in \Sref{sec:CrX3_L_5o} is probably only the first step in this direction.
\end{itemize}

\section{\label{sec:HM} Half-metallic ferromagnets}
\par The half-metallicity is basically the peculiar type of the electronic structure where one spin channel is metallic while another one is semiconducting~\cite{deGroot}. Most of such materials are ferro- or ferrimagnets. The fully compensated half-metallic ferrimagnets, where 100\% spin polarization of the conduction electrons coexists with zero net magnetization, have been also proposed~\cite{vanLeuken}. In this section, we further explore abilities of the linear response theories for the analysis of interatomic exchange interactions in this type of materials with the emphasis on the Coulomb correlations and the ligand states. We consider two such examples: the canonical CrO$_2$~\cite{Schwarz} and more recent Co$_3$Sn$_2$S$_2$, which has attracted a great deal of attention due to the large anomalous Hall effect and other intriguing properties~\cite{Liu,Liu2,Minami,Jiao,Yanagi,PRB2022b}.

\par Although the electronic structure is metallic, the response tensor $\hat{\mathcal{R}}^{ \uparrow \downarrow}$ involves the transitions between occupied and empty states with opposite projections of spins. For the half-metallic compounds, these transitions will be gapped. In such situation, fine details of the electronic structure near the Fermi level do not play a primary role in the behavior of interatomic exchange interactions, as it could be expected, for instance, in the RKKY theory for the regular metals~\cite{Roth,BrunoChappert}.

\par We will start our analysis with LSDA (GGA). It should be noted that the alternative point of view on the electronic structure of CrO$_2$ and other rutile oxides is based on the LDA$+$$U$ concept~\cite{Korotin,Maurya}. Nevertheless, the situation is disputable~\cite{Mazin}. Moreover, the dynamic correlations are known to play a very important role in the half-metallic materials~\cite{HMRewModPhys} and can substantially revise the picture based on the static LDA$+$$U$ approach~\cite{PRB2015}. We will illustrate this idea by considering the behavior of interatomic exchange interactions in the case of CrO$_2$.

\subsection{\label{sec:CrO2} CrO$_2$}
\par The half-metallic ferromagnetism is not very common in stoichiometric transition-metal oxides. Nevertheless, there are exceptions and CrO$_2$ is one of them, which is widely considered in various applications related to spintronics and magnetic recording~\cite{Skomski}. One of the limitations of CrO$_2$ from this practical point of view is the relatively low $T_{\rm C} \sim 390$ K~\cite{Skomski}.
\noindent
\begin{figure}[t]
\begin{center}
\includegraphics[width=4.5cm]{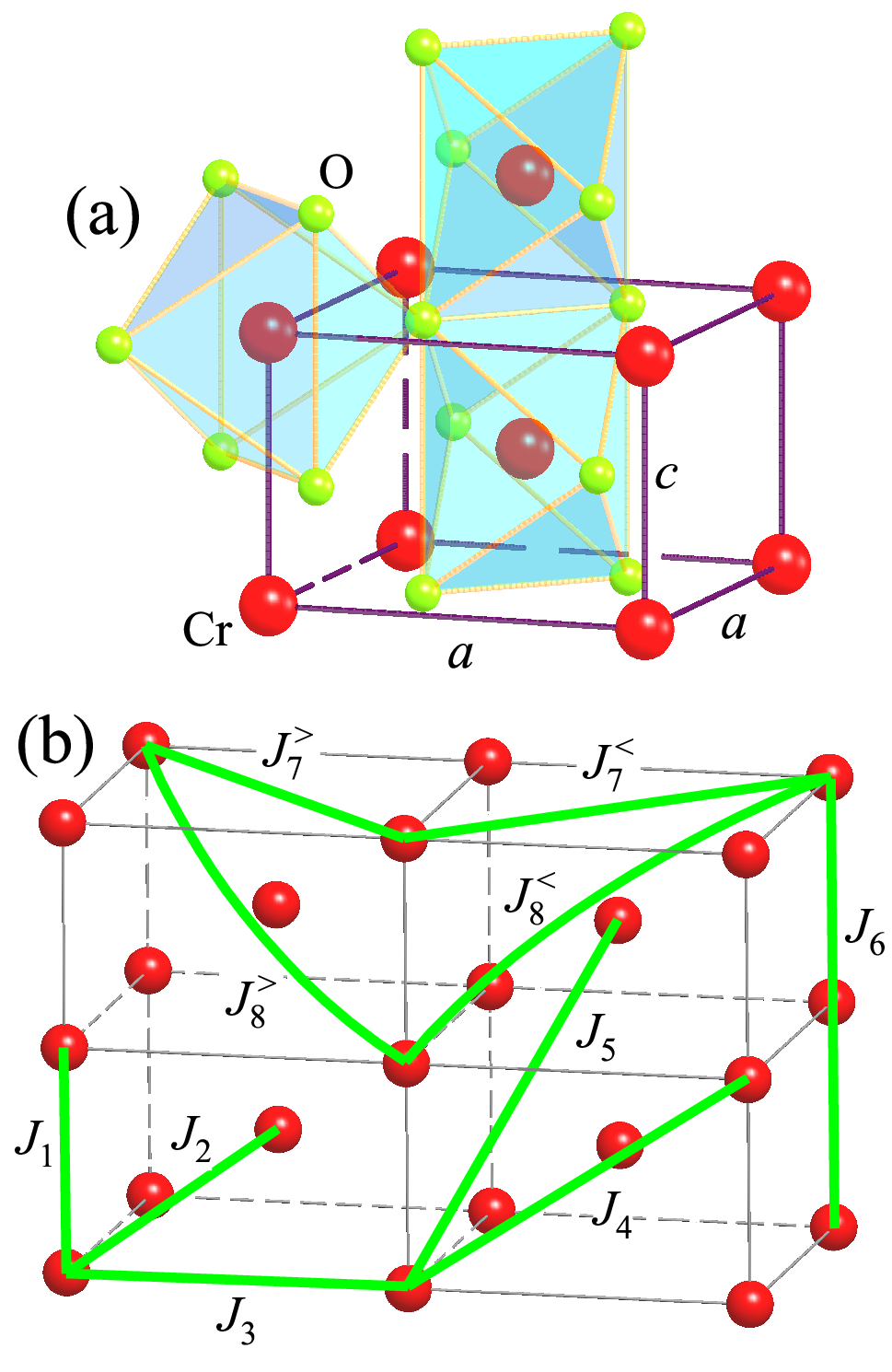}\hspace*{0.0cm}
\includegraphics[width=6.0cm]{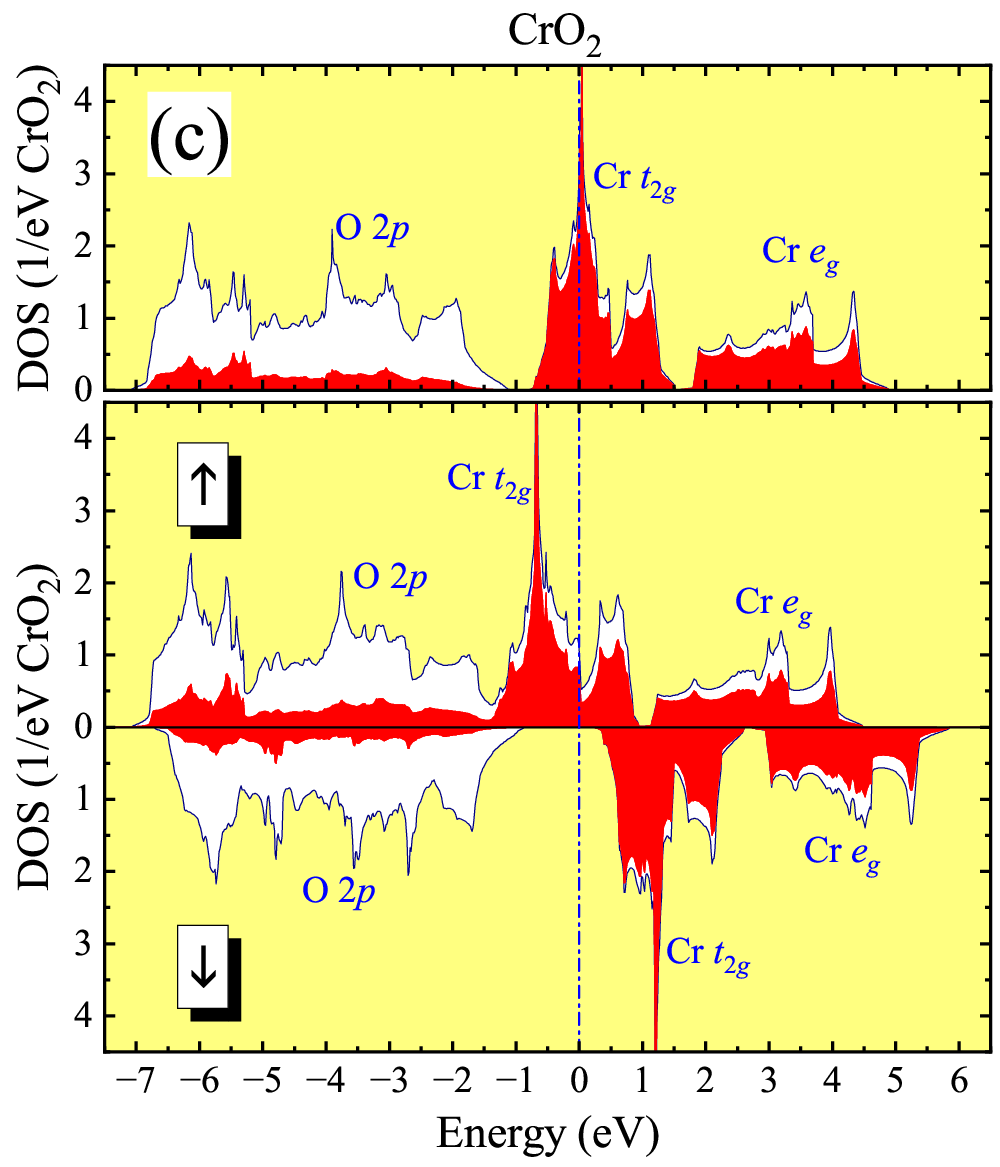}
\end{center}
\caption{(a) Fragment of the crystal structure of CrO$_2$, illustrating the arrangement of the CrO$_6$ octahedra. (b) The lattice of Cr atoms with the notations of exchange interactions. (c) Densities of states (DOS) for CrO$_2$ in LDA (top) and LSDA for the ferromagnetic state (bottom). Shaded areas show partial contributions of the Cr $3d$ states. The Fermi level is at zero energy.}
\label{fig.CrO2_basic}
\end{figure}

\par CrO$_2$ crystallizes in the rutile structure (the space group $P4_{2}/mnm$, \Fref{fig.CrO2_basic}a)~\cite{CrO2exp}. The exchange  interactions remain sizable up to at least the 8th coordination sphere. Moreover, since the rutile structure is nonsymmorphic, there are two types of interactions, $J_{7}$ and $J_{8}$, which are denoted by superscripts $>$ and $<$, as explained in \Fref{fig.CrO2_basic}b. The calculations are performed for the experimental parameters of the crystal structure~\cite{CrO2exp} on the mesh of the $10$$\times$$10$$\times$$16$ points both for $\boldsymbol{k}$ and $\boldsymbol{q}$.

\par
\begin{figure}[t]
\begin{center}
\includegraphics[width=4.5cm]{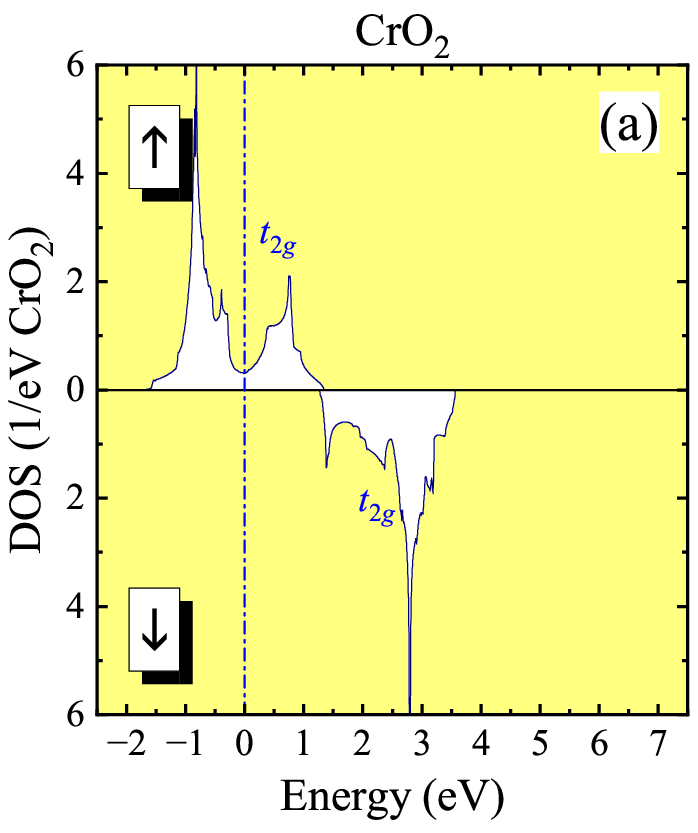}
\includegraphics[width=4.5cm]{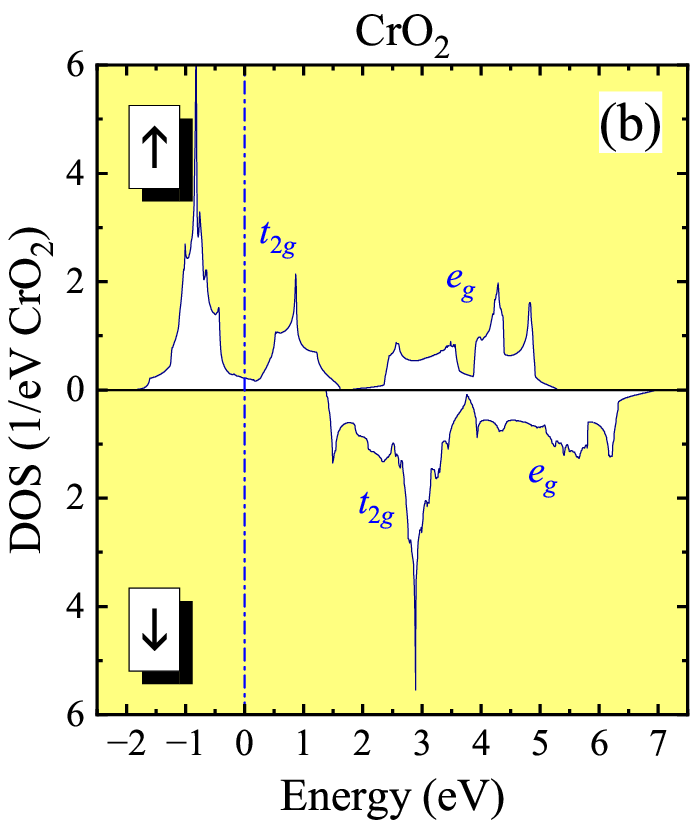}
\end{center}
\caption{Densities of states (DOS) for CrO$_2$ in the FM state: (a) Hartree-Fock approximation for the 3-orbital model and (b) the same for the 5-orbital model. The Fermi level is at zero energy.}
\label{fig.CrO2_35o}
\end{figure}

 The LDA band structure is featured by three separated bands: ${\rm O} \, 2p$ in the occupied part, ${\rm Cr} \, t_{2g}$ near the Fermi level, and ${\rm Cr} \, e_{g}$ in the unoccupied part. Therefore, one can consider two types of correlated models: the 3-orbital model for the Cr $t_{2g}$ bands and 5-orbital model both for Cr $t_{2g}$ and Cr $e_{g}$ bands. In the former case, the on-site Coulomb and exchange interactions can be described in terms of two Kanamori parameters~\cite{Kanamori1963}: the intraorbital Coulomb interaction $\mathcal{U}$ and the exchange interaction $\mathcal{J}$, which can be evaluated within cRPA as $2.84$ eV and $0.70$ eV, respectively~\cite{PRB2015}. The third parameter of interorbital Coulomb interaction can be obtained from these two as $\mathcal{U}' = \mathcal{U} - 2\mathcal{J}$~\cite{Kanamori1963}. The parameters of 5-orbital model are specified by $U=1.98$ eV, $J=0.94$ eV, and $B=0.09$ eV (see \ref{sec:Hubbard}). The corresponding densities of states, in the Hartree-Fock approximation, are shown in~\Fref{fig.CrO2_35o}.
\noindent
\noindent Alternatively, one can consider the all-electron ${\rm Cr} \, 3d$+${\rm O} \, 2p$ model in LSDA. The half-metallic character of the electronic structure is evident from the LSDA density of states (\Fref{fig.CrO2_basic}c). The Coulomb interactions additionally split the $\uparrow$-spin $t_{2g}$ band, placing the Fermi energy inside a pseudogap,\footnote{In the rutile structure, three $t_{2g}$ orbitals on the Cr sites belong to three different irreducible one-dimensional representations of the point group $mmm = D_{2h}$, meaning that the local (site-diagonal) part of the model Hamiltonian will be diagonal with respect to the $t_{2g}$ orbital indices. In such situation, the on-site Coulomb repulsion will tend to occupy two $t_{2g}$ levels with lower crystal-field energies and empty the remaining one. On the other hand, there is a strong hopping operating between ``occupied'' and ``empty'' orbitals of the different Cr sites, which leads to the formation of the pseudogap instead of the real gap. Further details can be found in ref.~\cite{PRB2015}.} and shift the $\downarrow$-spin states to the higher energy region, away from the Fermi level.

\par The parameters of exchange interactions are summarized in \Tref{tab.CrO2_J}.
\noindent
\begin{table}
\caption{\label{tab.CrO2_J} Parameters of exchange interactions in CrO$_2$ (in meV), as obtained in correlated 3- and 5-orbital models (respectively, 3o and 5o), and in LSDA for the all-electron ${\rm Cr} \, 3d$+${\rm O} \, 2p$ model using the schemes $\hat{b}$ and $M$ (corresponding to the infinitesimal rotations of the xc fields and local spin moments, respectively). In the L0 scheme, xc field was redefined to enforce $\mathcal{I}_{\rm O} = 0$.  The notations of exchange parameters are explained in \Fref{fig.CrO2_basic}b. $D^{xx}=D^{yy}$ and $D^{zz}$ are nonvanishing elements of the spin-stiffness tensor (in meV$\cdot$\AA$^{2}$). $T_{\rm C}$ is the Curie temperature in RPA (in K).}
\begin{indented}
\lineup
\item[]\begin{tabular}{@{}crrrrrrr}
\br
            & \centre{2}{3o}             & \centre{2}{5o}                & \centre{3}{LSDA}                       \cr
\ns
            & \crule{2}                  & \crule{2}                     & \crule{3}                              \cr
            & $\hat{b}$ \quad & $M$ \quad  & $\hat{b}$ \quad & $M$ \quad & $\hat{b}$ \quad & $M$ \quad & L0 \quad \cr
\mr
$J_{1}$     & $9.18$          & $15.76$    & $5.24$          & $17.47$   & $30.87$         & $36.86$   & $37.63$  \cr
$J_{2}$     & $10.94$         & $23.15$    & $11.47$         & $31.17$   & $21.39$         & $24.80$   & $25.32$  \cr
$J_{3}$     & $1.12$          & $1.55$     & $0.33$          & $0.44$    & $3.04$          & $2.54$    & $2.57$   \cr
$J_{4}$     & $0.84$          & $1.51$     & $0.71$          & $1.54$    & $1.49$          & $1.22$    & $1.20$   \cr
$J_{5}$     & $-0.34$         & $-0.38$    & $-0.12$         & $0.04$    & $-0.79$         & $-1.72$   & $-1.83$  \cr
$J_{6}$     & $-1.92$         & $-1.78$    & $-1.75$         & $-1.73$   & $-3.58$         & $-4.96$   & $-5.14$  \cr
$J_{7}^{>}$ & $-3.41$         & $-3.36$    & $-2.50$         & $-2.79$   & $-6.25$         & $-8.39$   & $-8.61$  \cr
$J_{7}^{<}$ & $-1.09$         & $-1.77$    & $-0.38$         & $-0.67$   & $-2.09$         & $-3.00$   & $-3.11$  \cr
$J_{8}^{>}$ & $-0.02$         & $0.19$     & $0.10$          & $0.62$    & $-1.50$         & $-1.98$   & $-2.08$  \cr
$J_{8}^{<}$ & $-0.52$         & $-0.64$    & $-0.39$         & $-0.30$   & $-0.53$         & $-0.66$   & $-0.73$  \cr
\mr
$D^{xx}$    &  $37$           & $258$      & $113$           & $532$     & $123$           & $113$     & $109$    \cr
$D^{zz}$    &  $11$           & $168$      & $39$            & $362$     & $161$           & $159$     & $152$    \cr
\mr
$T_{\rm C}$ &  $310$          & $1026$     & $430$           & $1563$    & $868$           & $880$     & $875$    \cr
\br
\end{tabular}
\end{indented}
\end{table}
\noindent First, there is a large difference in the parameters obtained in the schemes $\hat{b}$ and $M$, especially in the Hartree-Fock approximations for correlated 3- and 5-orbital models, where the more accurate method $M$ strengthens the FM interactions practically in all the bonds. The situation is more modest in LSDA for the all-electron ${\rm Cr} \, 3d$+${\rm O} \, 2p$ model: when going from the method $\hat{b}$ to the method $M$, the nearest FM interactions $J_{1}$ and $J_{2}$ increase only partially, while other longer-range interactions tend to decrease and become more antiferromagnetic.

\par In LSDA, the Cr and O atoms are antipolarized due to the joint effect of the hybridization and the intraatomic spin splitting, similar to Cr$X_3$. The corresponding spin magnetic moments are $M_{\rm Cr}=2.15$ $\mu_{\rm B}$ and $M_{\rm O}=-$$0.08$ $\mu_{\rm B}$. The effective Stoner parameters were calculated using the definitions III and IV, which should be used in the combination with the methods $\hat{b}$ and $M$, respectively (see \Sref{sec:CrX3_L_LSDA}). As expected, $\mathcal{I}_{\rm Cr} = 0.87$ eV practically does not depend on the definition. On the other hand, the definitions III and IV yield different values of $\mathcal{I}_{\rm O}$: $-$$0.78$ eV and $-$$0.48$ eV, respectively. Nevertheless, the parameters are pretty small and the exchange interactions in CrO$_2$ appears to be less sensitive to the ligand states (at least, in comparison with Cr$X_3$)~\cite{PRB2021}. For instance, the scheme L0, where we enforce $\mathcal{I}_{\rm O} = 0$, provides basically the same set of parameters as the $M$ scheme with $\mathcal{I}_{\rm O} = -$$0.48$ eV (see \Tref{tab.CrO2_J}).

\par
\begin{figure}[t]
\begin{center}
\includegraphics[width=8.0cm]{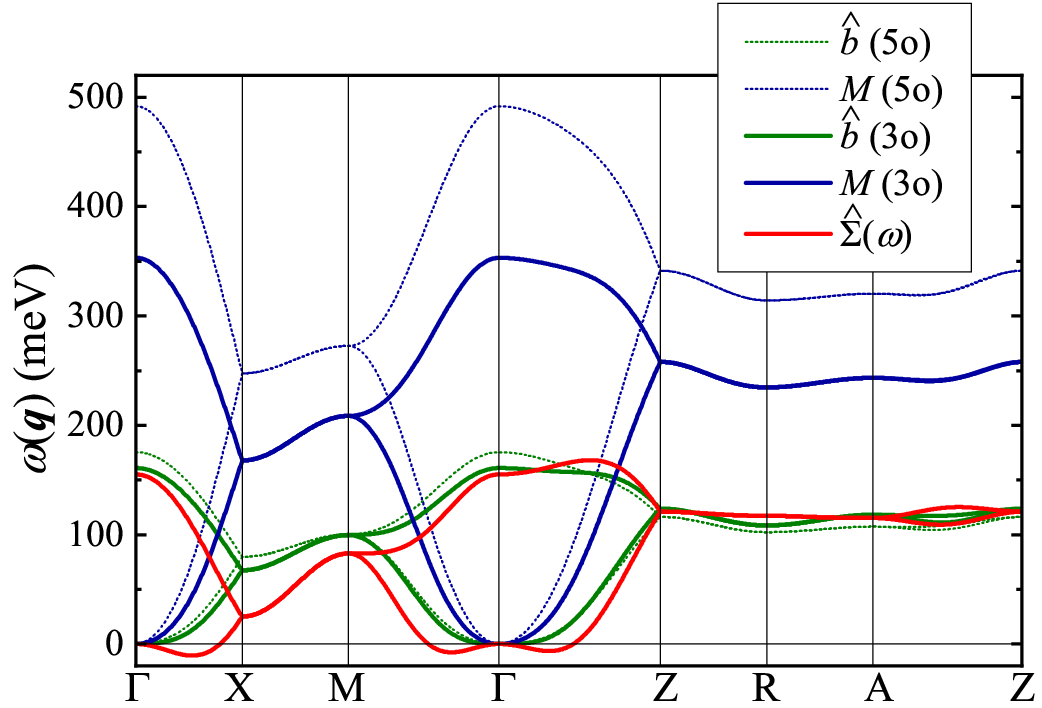}
\end{center}
\caption{Theoretical spin-wave dispersion for CrO$_2$ with the exchange parameters obtained in the Hartree-Fock approximation for correlated 3-orbital (3o) and 5-orbital (5o) models, by rotating either the xc field ($\hat{b}$) or the local magnetic moments ($M$). $\hat{\Sigma}(\omega)$ denotes results of ref.~\cite{PRB2015} (the extension of the scheme $\hat{b}$ for correlated 3-orbital model, where the scalar xc field was replaced by the frequency-dependent self-energy evaluated within the dynamical mean-field theory). The notations of the high-symmetry points of the Brillouin zone are taken from ref.~\cite{BradleyCracknell}.}
\label{fig.CrO2_SW}
\end{figure}

 The spin-wave dispersions calculated with different sets of parameters for the 3- and 5-orbital models is plotted in \Fref{fig.CrO2_SW} (see \ref{sec:RPA} for details).
\noindent
\noindent
The nonvanishing $xx$ and $zz$ elements of the spin-wave stiffness tensor $\hat{D} = [D^{\alpha \beta}]$,
\noindent
\begin{displaymath}
\omega_{A}(\boldsymbol{q}) \approx \sum_{\alpha, \beta} D^{\alpha \beta} q_{\alpha} q_{\beta},
\end{displaymath}
\noindent describing the dispersion of the acoustic ($A$) spin-wave branch near the $\Gamma$ point, are summarized in \Tref{tab.CrO2_J}. The experimental estimates for the averaged spin-wave stiffness in CrO$_2$ vary from $70$ to $150$ meV$\cdot$\AA$^{2}$~\cite{SimsCrO2,BeliayevCrO2}. The Curie temperature is about $390$ K~\cite{Skomski}. Then, the LSDA values of $T_{\rm C}$ seem to be overestimated, though the situation is rather subtle. On the one hand, the theoretical values of the spin-wave stiffness are within the experimental scatter. On the other hand, the exchange parameters derived for ground state are not supposed to reproduce $T_{\rm C}$ for this metallic system and more sophisticated methods, considering the temperature dependence of the exchange interactions, may be needed~\cite{Katanin2023,Oguchi,Staunton}. Anyways, in the light of well-known limitations of LSDA for the transition-metal oxides~\cite{AZA}, such moderate disagreement would not be surprising.

\par Even more interesting situation is realized in correlated 3- and 5-orbital models. At first glance, the approximate scheme $\hat{b}$ provides a much better description for the spin-wave stiffness and $T_{\rm C}$ in comparison with the scheme $M$, which is supposed to be more accurate. However, this is true only in the Hartree-Fock approximation, which has serious limitations for the metallic systems. If one goes beyond the Hartree-Fock approximation, the situation can change dramatically. Particularly, as expected for the half-metallic systems~\cite{HMRewModPhys}, the dynamic correlations lead to a strong redistribution of the electronic states. These effects can be treated in the framework of dynamical mean-field theory (DMFT)~\cite{DMFT1}. The latter can be regarded as an extension of SDFT in which the KS potential is replaced by the frequency-dependent self-energy $\Sigma^{\uparrow,\downarrow}(\omega)$~\cite{DMFT2}. This method has been applied to CrO$_2$ in ref.~\cite{PRB2015}, using the same correlated 3-orbital model. The exchange interactions were evaluated basically in the $\hat{b}$ scheme using \Eref{eq:JLKAG}, by considering the infinitesimal rotations of the frequency-dependent xc field $\Delta \Sigma(\omega) = \Sigma^{\uparrow}(\omega) - \Sigma^{\downarrow}(\omega)$~\cite{KL2000}. The dynamic correlations substantially reduce $\Delta \Sigma(\omega)$, resulting in much stronger AFM interactions $J_{7}$ and $J_{8}$, which overcome the effect of FM interactions $J_{1}$ and $J_{2}$, making the FM state unstable. The corresponding spin-wave dispersion is also plotted in \Fref{fig.CrO2_SW}, where the negative frequencies $\omega(\boldsymbol{q})$ along the directions $\Gamma$-${\rm X}$, $\Gamma$-${\rm M}$, and $\Gamma$-${\rm Z}$ mean that the theoretical ground state should be in one of these points. Therefore, it was concluded in ref.~\cite{PRB2015} that in order to reproduce the FM ground state of CrO$_2$, it is essential to extend the 3-orbital model, by adding new ingredients such as the Cr $e_{g}$ and O $2p$ bands. For instance, the spin-wave stiffness is systematically larger in the 5-orbital model (\Tref{tab.CrO2_J}), which makes the FM state more stable. Nevertheless, the present analysis also suggests that the problem may be not in the 3-orbital model itself, but in additional approximations underlying the scheme $\hat{b}$ for the exchange interactions. The method $M$ systematically improves the stability of the FM ground state in CrO$_2$. This tendency may be overestimated in the Hartree-Fock approximation. However, a more rigorous treatment of correlation effects in the framework of DMFT, in the combination with the method $M$, could possibly bring the situation to a better agreement with the experimental data.

\par The DM interactions can take place between atoms in different Cr sublattices. The strongest ones are realized in the 2nd coordination sphere (in the combination with $J_{2}$, as shown in \Fref{fig.CrO2_basic}b). If $\boldsymbol{\epsilon} = \frac{1}{\sqrt{2a^2+c^2}}(\pm a, \pm a, \pm c)$ are the directions of such bonds, the corresponding to them DM vectors have the following form $\boldsymbol{d} = d \epsilon^{z} [\boldsymbol{\epsilon} \times \boldsymbol{n}^{z}]$ (i.e., the vectors lie in the $xy$ plane, are perpendicular to the bonds, and have opposite signs for the bonds in the directions $\pm z$). The parameter $d$ can be estimated as $0.85$ and $1.38$ meV, thus yielding $\boldsymbol{d}=(\pm 0.23, \pm 0.23, 0)$ and  $(\pm 0.37, \pm 0.37, 0)$ meV  in the Hartree-Fock approximation for correlated 5-orbital model and all-electron LSDA, respectively (employing the scheme $M$ in both cases). The interesting point is that the DM interactions are expected to be strong (and comparable with the ones in CrI$_3$) even though CrO$_2$ does not contain heavy elements. This is basically the consequence of two factors: (i) the partial filling of the $t_{2g}$ states, opening the room for unquenched orbital magnetization; (ii) half-metallic character of the electronic structure, giving rise to the spin-current mechanism of the DM interactions in the $\uparrow$-spin $t_{2g}$ band~\cite{Katsnelson_DM,Kikuchi}. Nevertheless, $\boldsymbol{d}$ is much smaller than the isotropic interaction $J_{2}$ operating in the same bonds.

\subsection{\label{sec:Co3S2S2} Co$_3$Sn$_2$S$_2$}
\par
\begin{figure}[t]
\begin{center}
\hspace*{1.5cm}\includegraphics[width=2.68cm]{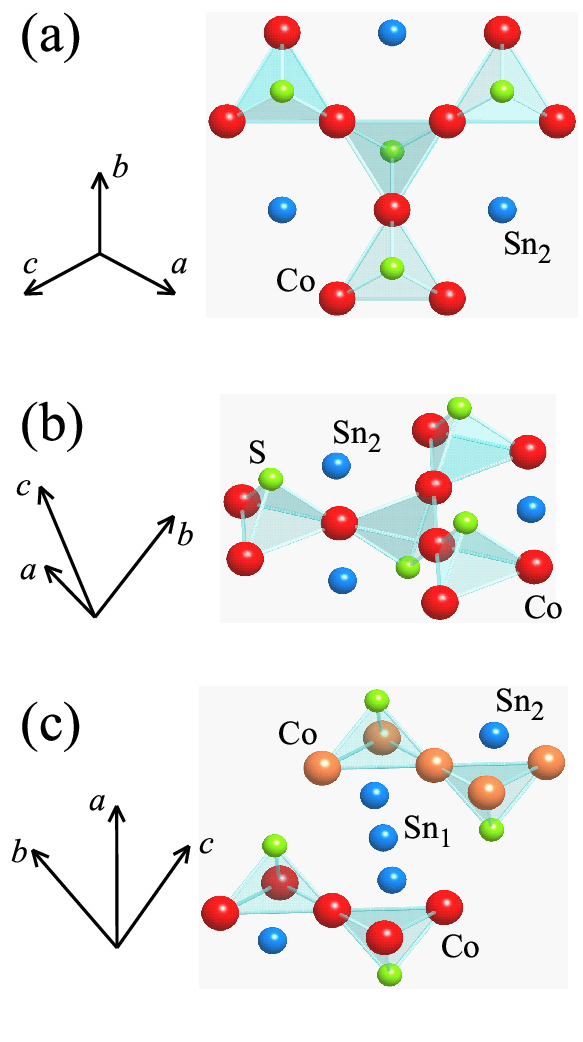}\hspace*{0.0cm}
\includegraphics[width=5.1cm]{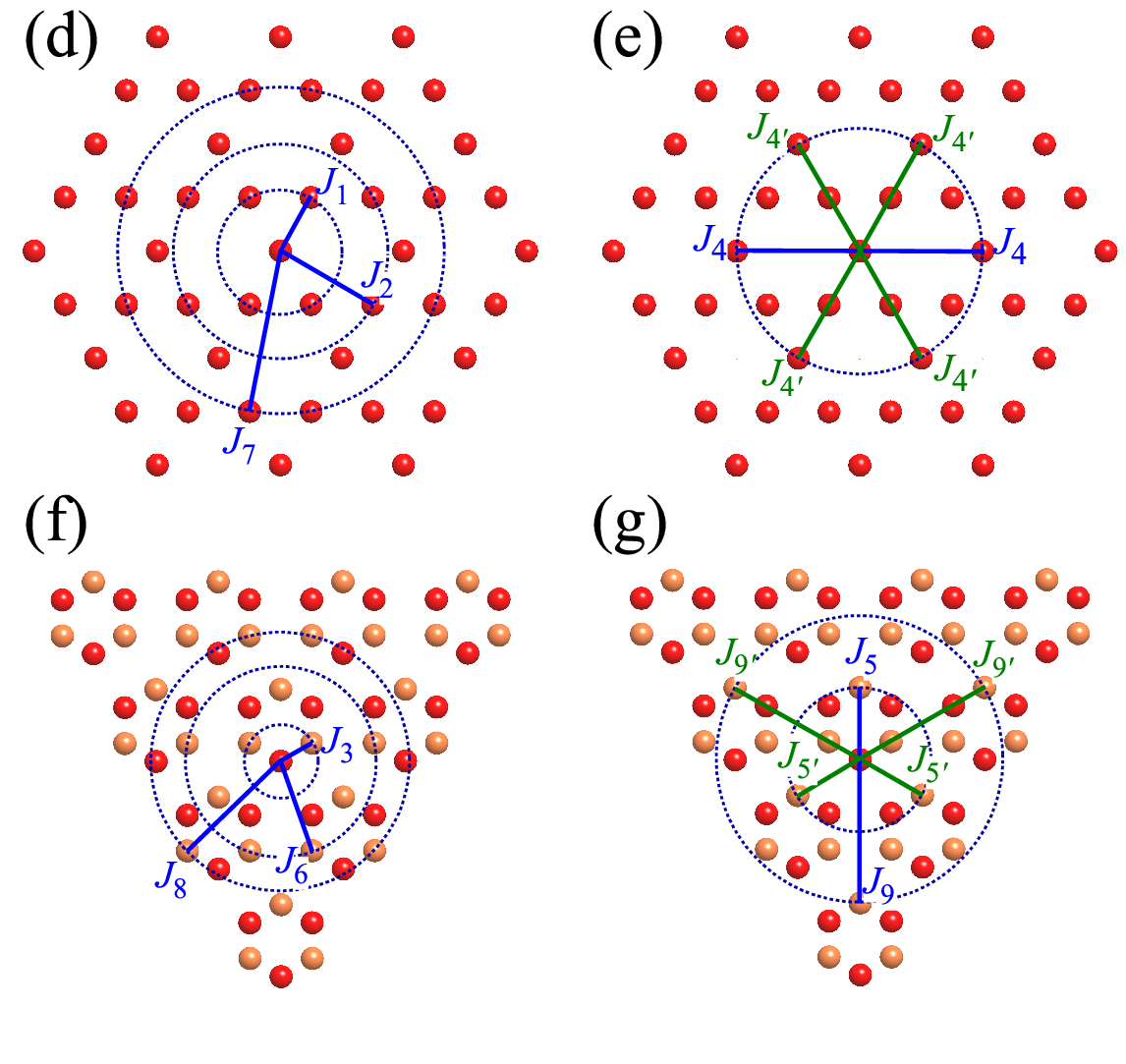}\hspace*{0.0cm}
\includegraphics[width=6.0cm]{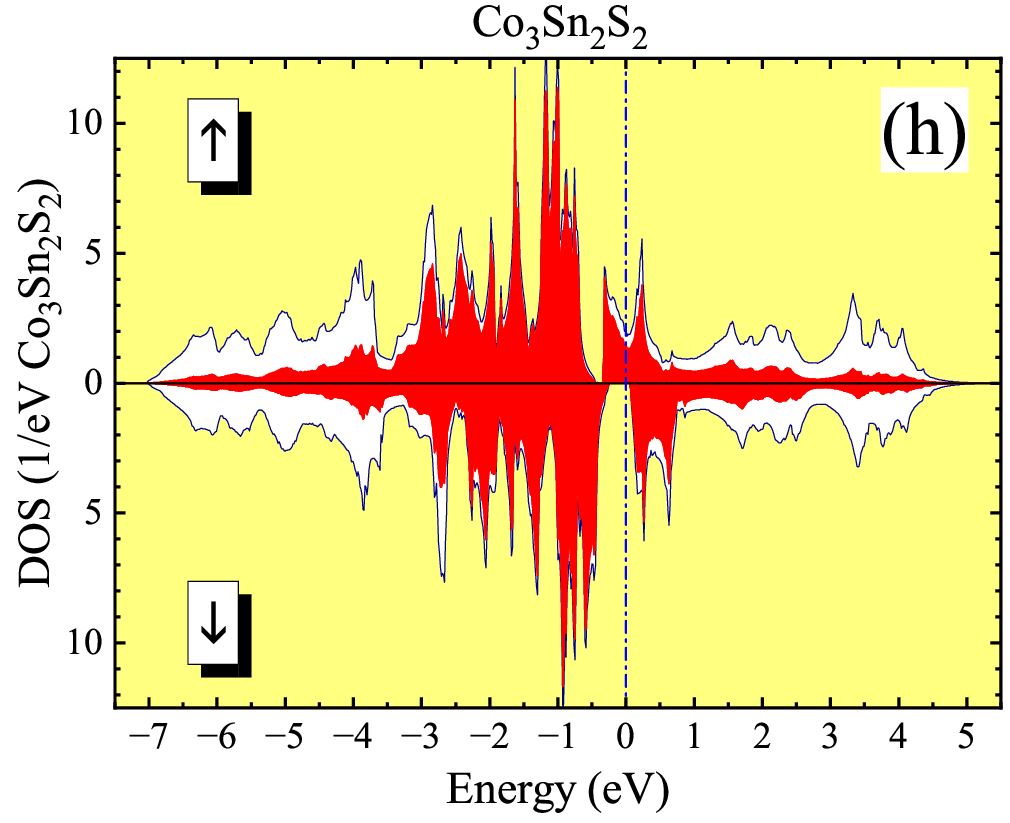}
\end{center}
\caption{(a)-(c) Fragments of the crystal structure of Co$_3$Sn$_2$S$_2$. (a) Top view on the kagome Cr layer surrounded by Sn and S atoms. (b) Side view on the same layer. (c) Relative arrangement of adjacent kagome layers. (d)-(g) Main exchange interactions. (d),(e) Interactions operating in the kagome plane. (f),(g) Interactions between adjacent planes (top view). The Co atoms located in different planes are denoted
by different colors, the same as in (c). The coordination spheres of Co atoms around the origin are denoted by dotted circles. (d),(f) Interactions, which are the same for all the bonds in the given coordination sphere.
(e),(g) Interactions, which are characterized by two different values for two types of the inequivalent bonds in the same coordination sphere. The behavior of $J_{k}$ around two other Co sites in the primitive cell are obtained by the threefold rotations. (h) GGA Density of states (DOS) for Co$_3$Sn$_2$S$_2$ in the ferromagnetic state. Shaded areas show partial contributions of the Co $3d$ states. The Fermi level is at zero energy.}
\label{fig.Co3Sn2S2_basic}
\end{figure}

 Co$_3$Sn$_2$S$_2$ is a ferromagnet in which small spontaneous magnetization (about $0.3$ $\mu_{\rm B}$ per Co atom) coexists with relatively high $T_{\rm C}=177$ K~\cite{Co3Sn2S2structure}. It crystallizes in the rhombohedral $R\overline{3}m$ structure, hosting the kagome lattice of Co ions (\Fref{fig.Co3Sn2S2_basic}a-c)~\cite{Co3Sn2S2structure}. The S atoms sit on the top of the Co$_3$ triangles, forming with them the network of alternating tetrahedra. There are two inequivalent Sn sites located in (Sn$_{1}$) and between (Sn$_{2}$) the kagome planes. Recently, Co$_3$Sn$_2$S$_2$ has attracted a lot of attention as a magnetic Weyl semimetal whose nontrivial topology of the electronic states gives rise to a large anomalous Hall effect~\cite{Liu,Liu2,Minami}.
\noindent
\noindent Furthermore, Co$_3$Sn$_2$S$_2$ appears to be half-metallic, as was demonstrated in experimental~\cite{Jiao} and theoretical~\cite{Yanagi,PRB2022b} studies.

\par We have chosen Co$_3$Sn$_2$S$_2$ to illustrate the complexities one can face in the analysis of interatomic exchange interactions for this type of itinerant electron systems. At the present stage, we have in our disposition only all-electron ${\rm Co} \, 3d$+${\rm Sn} \, 5p$+${\rm S} \, 3p$ model, which was constructed in ref.~\cite{PRB2022b} within GGA~\cite{GGA}, by employing the Vienna {\it ab initio} simulation package (VASP)~\cite{VASP} for the electronic structure calculation and the maximal localization technique for the Wannier functions~\cite{wannier90}. The corresponding density of states is shown in \Fref{fig.Co3Sn2S2_basic}h: the Fermi level crosses the $\uparrow$-spin band and falls inside the gap for the $\downarrow$-spin channel. The magnetic moments are $M_{\rm Co} = 0.33$, $M_{\rm Sn_{1}} = -0.03$, $M_{\rm Sn_{2}} = -0.04$, and $M_{\rm S} = 0.04$ $\mu_{\rm B}$, so that the total moment per one unit cell is $1$ $\mu_{\rm B}$. It is believed to be still an open question how good is GGA for Co$_3$Sn$_2$S$_2$. However, the attempts to construct a more compact model, which would also include the Coulomb correlations, were so far unsuccessful. Formally, the electronic structure in \Fref{fig.Co3Sn2S2_basic}h can be divided into fully occupied 19 bands and remaining 8 bands, which are separated by a band gap. However, the construction of the Wannier basis for the upper eight-band manifold is not straightforward, as these Wannier functions will probably reside not on single atomic sites~\cite{SavrasovPRB2021}. A compact model for Co$_3$Sn$_2$S$_2$ was constructed in ref.~\cite{OzawaNomura}, but using empirical arguments.

\par Co$_3$Sn$_2$S$_2$ provides an interesting example of itinerant magnetism. According to the GGA calculations~\cite{PRB2022b}, finite rotations of spins away from the FM ground state (for instance, by forcing three Co spins to form the umbrella texture) eventually leads to the collapse of magnetization, so that the system falls in the nonmagnetic state.\footnote{In the umbrella texture, three Co spins are rotated from the FM axis $z$ by the same angle such that their projections in the $xy$ plane form the $120^{\circ}$ texture.} A similar behavior was found in fcc Ni~\cite{Turzhevskii,Singer} and several Ru-based oxides~\cite{Streltsov,Schnelle2021}. Moreover, the size of the magnetic moments is expected to shrink due the temperature disorder, affecting both electronic structure and interatomic exchange interactions~\cite{PRB2022b}. Therefore, the bilinear Heisenberg model cannot be defined in the global sense (for instance, for describing simultaneously low-temperature spin-wave dispersion and $T_{\rm C}$). Nevertheless, it can still be defined locally, for the analysis of local stability of the FM state with respect to the infinitesimal rotations of spins. The exchange interactions spread at least up to 9th coordination sphere around each Co site, as explained in \Fref{fig.Co3Sn2S2_basic}d-g. Moreover, some of these interactions, formally belonging to the same coordination sphere, can be inequivalent (as $J_{4}$ and $J_{4}'$ in the plane, $J_{5}$ and $J_{5}'$, and $J_{9}$ and $J_{9}'$ between the planes). The calculations have been performed on the mesh of the $20\times20\times20$ points both for $\boldsymbol{k}$ and $\boldsymbol{q}$.

\par The effective Stoner parameters, calculated by using different definitions are summarized in \Tref{tab.Co3Sn2S2_Stoner}. As was discussed in \Sref{sec:CrX3_L}, the definitions II and IV should be used in combination with the spherically averaged scheme $M$, while the definitions I and III are more suitable for the matrix schemes $\hat{b}$ and $\hat{m}$.
\noindent
\begin{table}
\caption{\label{tab.Co3Sn2S2_Stoner} The effective Stoner parameters (in eV) obtained using definitions I-IV (as explained in \Sref{sec:CrX3_L_LSDA}). Sn$_{1}$ and Sn$_{2}$ are located, respectively, in and between kagome planes.}
\begin{indented}
\lineup
\item[]\begin{tabular}{@{}ccccc}
\br
definition & $\mathcal{I}_{\rm Co}$ & $\mathcal{I}_{\rm Sn_{1}}$ & $\mathcal{I}_{\rm Sn_{2}}$ & $\mathcal{I}_{\rm S}$ \cr
\mr
I          &  $1.30$                &  $\phantom{-}0.17$         & $-0.10$                    & $1.22$                \cr
II         &  $0.98$                &  $-0.06$                   & $-0.20$                    & $0.88$                \cr
III        &  $1.55$                &  $-0.65$                   & $-0.65$                    & $2.84$                \cr
IV         &  $1.45$                &  $-3.44$                   & $-2.53$                    & $3.85$                \cr
\br
\end{tabular}
\end{indented}
\end{table}
\noindent One can clearly see that all the parameters in Co$_3$Sn$_2$S$_2$ strongly depend on the definition. This holds even for $\mathcal{I}_{\rm Co}$, which can easily change by about 50\%. This is clearly in contrast with other considered compounds, where the value of $\mathcal{I}$ on the magnetic transition-metal site practically did not depend on the definition. Nevertheless, the exchange interactions \eref{eq:jTT1} between the Co sites do not explicitly depend on the choice of $\mathcal{I}_{\rm Co}$. The uncertainty with the choice of the parameters $\mathcal{I}_{\rm Sn}$ and $\mathcal{I}_{\rm S}$, which strongly depend on the way how they are defined, posses a more serious problem. The parameters $\mathcal{I}_{\rm Sn}$ are mainly negative (except $\mathcal{I}_{\rm Sn_{1}}$ in the case I) and tend to destabilize the FM interactions. On the other hand, positive $\mathcal{I}_{\rm S}$ strengthens the ferromagnetism. However, depending on the definition, $\mathcal{I}_{\rm S}$ can change by factor four, and $\mathcal{I}_{\rm Sn}$ changes by an order of magnitude. This dependence has a strong impact on the exchange interactions, which are summarized in \Tref{tab.Co3Sn2S2_J}.
\noindent
\begin{table}
\caption{\label{tab.Co3Sn2S2_J} Parameters of exchange interactions in Co$_3$Sn$_2$S$_2$ (in meV) calculated using the schemes $\hat{b}$, $\hat{m}$, and $M$ (the infinitesimal rotations of, respectively, the xc field, magnetization matrix, and local spin moments). In the scheme $\hat{b}$, the xc field was either associated with the site-diagonal part of the TB Hamiltonian (H) or derived from the sum rule (sr). The roman number in the parentheses stands for the set of the parameters $\mathcal{I}_{\rm Sn}$ and $\mathcal{I}_{\rm S}$ from \Tref{tab.Co3Sn2S2_Stoner}, which was used in the calculations of $J_{k}$. In the L0 scheme, xc field was redefined to ensure $\mathcal{I} = 0$ on the sites Sn and S.  The notations of $J_{k}$ are explained in \Fref{fig.Co3Sn2S2_basic}d-g. $D^{xx}=D^{yy}$ and $D^{zz}$ are nonvanishing elements of the spin-stiffness tensor (in meV$\cdot$\AA$^{2}$).}
\begin{indented}
\lineup
\item[]\begin{tabular}{@{}ccccccc}
\br
            & $\hat{b}$ (H,I)   & $\hat{b}$ (sr,III) & $\hat{m}$ (III)   & $M$ (II)          & $M$ (IV)          & L0                \cr
\mr
$J_{1}$     & $\phantom{-}1.60$ & $\phantom{-}2.54$  & $\phantom{-}7.71$ & $-0.08$           & $-1.51$           & $-0.09$           \cr
$J_{2}$     & $\phantom{-}0.05$ & $\phantom{-}0.08$  & $-0.14$           & $-0.06$           & $-0.25$           & $-0.06$           \cr
$J_{3}$     & $\phantom{-}0.09$ & $\phantom{-}0.14$  & $\phantom{-}0.22$ & $-0.06$           & $-1.11$           & $-0.07$           \cr
$J_{4}$     & $\phantom{-}0.18$ & $\phantom{-}0.22$  & $-0.01$           & $\phantom{-}0.34$ & $\phantom{-}0.35$ & $\phantom{-}0.41$ \cr
$J_{4}'$    & $\phantom{-}0.54$ & $\phantom{-}0.77$  & $\phantom{-}0.77$ & $\phantom{-}0.61$ & $\phantom{-}0.52$ & $\phantom{-}0.73$ \cr
$J_{5}$     & $\phantom{-}0.42$ & $\phantom{-}0.58$  & $\phantom{-}0.88$ & $\phantom{-}0.83$ & $-0.23$           & $\phantom{-}0.98$ \cr
$J_{5}'$    & $\phantom{-}0.65$ & $\phantom{-}0.90$  & $\phantom{-}1.56$ & $\phantom{-}0.95$ & $\phantom{-}0.92$ & $\phantom{-}1.13$ \cr
$J_{6}$     & $\phantom{-}0.29$ & $\phantom{-}0.41$  & $\phantom{-}0.64$ & $\phantom{-}0.44$ & $\phantom{-}0.51$ & $\phantom{-}0.53$ \cr
$J_{7}$     & $\phantom{-}0.10$ & $\phantom{-}0.14$  & $\phantom{-}0.08$ & $\phantom{-}0.12$ & $\phantom{-}0.18$ & $\phantom{-}0.15$ \cr
$J_{8}$     & $\phantom{-}0.04$ & $\phantom{-}0.06$  & $-0.13$           & $\phantom{-}0.10$ & $\phantom{-}0.07$ & $\phantom{-}0.11$ \cr
$J_{9}$     & $-0.13$           & $-0.19$            & $-0.66$           & $-0.27$           & $-0.25$           & $-0.31$           \cr
$J_{9}'$    & $\phantom{-}0.07$ & $\phantom{-}0.10$  & $\phantom{-}0.03$ & $\phantom{-}0.02$ & $\phantom{-}0.16$ & $\phantom{-}0.03$ \cr
\mr
$D^{xx}$    & $\phantom{-}521$  & $\phantom{-}727$   & $\phantom{-}383$  & $\phantom{-}490$  & $\phantom{-}36$   & $\phantom{-}582$  \cr
$D^{zz}$    & $\phantom{-}246$  & $\phantom{-}317$   & $\phantom{-}73$   & $\phantom{-}310$  & $\phantom{-}48$   & $\phantom{-}374$  \cr
\br
\end{tabular}
\end{indented}
\end{table}

\par As was pointed out before, the attempts to estimate the Curie temperature entirely from $J_{k}$, defined via the infinitesimal rotations of spins near the FM ground state, are pretty much meaningless in the case of Co$_3$Sn$_2$S$_2$, where proper theories for $T_{\rm C}$ should consider also the longitudinal change of the magnetization~\cite{PRB2022b}. Nevertheless, these $J_{k}$ can be used to evaluate the spin-wave dispersion and the spin stiffness. The latter has been measured experimentally and the nonvanishing elements of the spin-stiffness tensor are $D^{xx}=D^{yy}=803 \pm 46$ and $D^{zz}=237 \pm 13$ meV$\cdot$\AA$^{2}$~\cite{Liu2021}, which can be used for comparison with theoretical data.

\par First, we note that the results of the scheme $\hat{b}$ strongly depend on the definition of the xc field: the enforcement of the sum rule for $\hat{b}$ strengthens the FM interactions, bringing the spin stiffness to a good agreement with the experimental data. In this case, the exchange parameters are given mainly by the bare interactions, corresponding to the first term in \Eref{eq:jTT}. The corrections caused by the ligand states are small and do not play a decisive role. An interesting situation is realized in the scheme $\hat{m}$, based on the rigid rotations of the magnetization matrix: as expected, the individual parameters are overestimated (see \Sref{sec:object}). Nevertheless, the interactions are long-ranged and many of them are antiferromagnetic, resulting in the relatively small spin stiffness. The exchange interactions in the scheme $M$ are very sensitive to $\mathcal{I}_{\rm Sn}$ and $\mathcal{I}_{\rm S}$. If one uses the definition II, where the xc field in the expression for $\mathcal{I}$ is taken from the site-diagonal part of the TB Hamiltonian, $\mathcal{I}_{\rm Sn}$ and $\mathcal{I}_{\rm S}$ are relatively small so that the exchange interactions are given basically by the first term of \Eref{eq:jTT1}. In this case, $J_{1}$-$J_{3}$ are weakly antiferromagnetic, while other interactions are ferromagnetic (except $J_{9}$, which is antiferromagnetic in all considered methods). As the result, the FM state is stable, though $D^{xx}$ is somewhat underestimated in comparison with the experiment.

\par The situation changes dramatically if one uses the definition IV, where the xc field is taken from the sum rule, which substantially strengthens both $\mathcal{I}_{\rm Sn}$ and $\mathcal{I}_{\rm S}$. Apparently, negative $\mathcal{I}_{\rm Sn}$ plays a dominant role and is responsible for the AFM character of interactions in some of the bonds: particularly, $J_{1}$-$J_{3}$ become prominently antiferromagnetic and $J_{5}$ changes from strongly ferromagnetic to antiferromagnetic. Then, the FM state become unstable and the spins tend to form the $120^\circ$ in the $xy$ plane (at least, on the mean-field level). The corresponding spin stiffness is strongly underestimated compared to the experimental data. This is clearly an artefact of the model analysis, which is probably caused by unrealistic estimate for $\mathcal{I}_{\rm Sn}$. At least, the conclusion does not seem to be consistent with the brute-force GGA calculations, where the $120^\circ$ alignment of spins in the $xy$ plane leads to the collapse of the magnetic state~\cite{PRB2022b}.

\par The L0 approach allows us to eliminate the dependence of $J_{k}$ on $\mathcal{I}_{\rm Sn}$ and $\mathcal{I}_{\rm S}$, assuming that all magnetic moments are induces by $B_{\rm Co}$, while $B_{\rm Sn}=B_{\rm S}=0$ (so as $\mathcal{I}_{\rm Sn}$ and $\mathcal{I}_{\rm S}$). This results only in a small change of magnetic moments in comparison to the plain GGA values reported above: $M_{\rm Co} = 0.36$, $M_{\rm Sn_{1}} = -0.05$, $M_{\rm Sn_{2}} = -0.07$, and $M_{\rm S} = 0.02$ $\mu_{\rm B}$ (thus, $M_{\rm tot}$ remains equal to $1$ $\mu_{\rm B}$). The exchange interactions in these case reminiscent the ones obtained in the scheme $M$ using the definition II for $\mathcal{I}_{\rm Sn}$ and $\mathcal{I}_{\rm S}$ with somewhat stronger tendency towards the ferromagnetism, which is reflected in larger values of $D^{xx}$ and $D^{zz}$.

\par
\begin{figure}[t]
\begin{center}
\includegraphics[width=7.0cm]{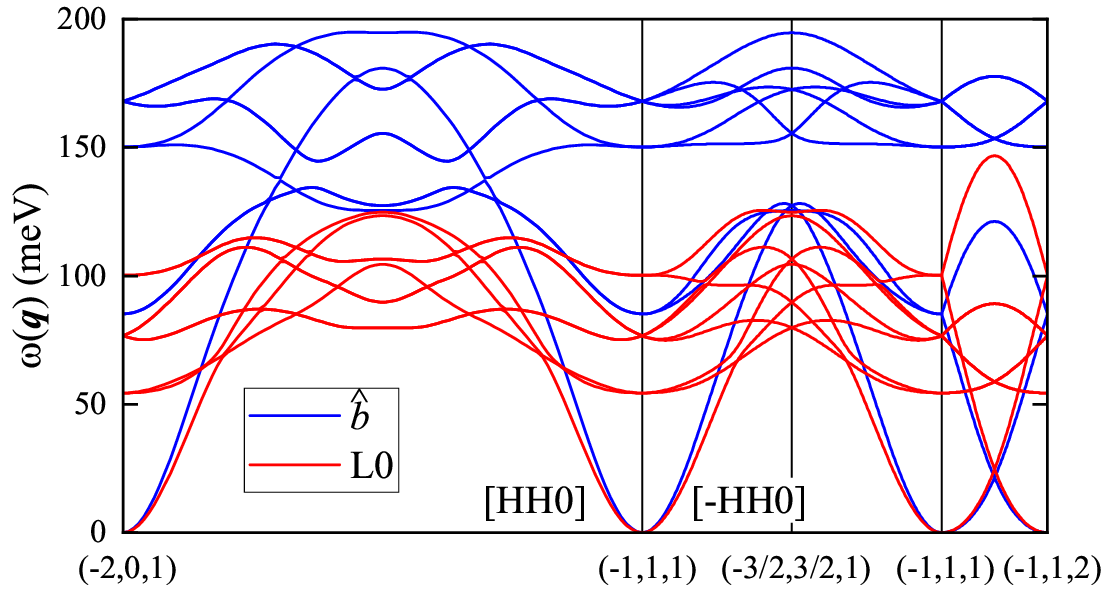}
\end{center}
\caption{Theoretical spin-wave dispersion for Co$_3$Sn$_2$S$_2$ with the exchange parameters derived in the schemes $\hat{b}$ and L0. The notations are taken from ref.~\cite{ZhangPRL} for the hexagonal unit cell.}
\label{fig.Co3Sn2S2_SW}
\end{figure}

 In principle, the schemes $\hat{b}$ and L0 both provide a reasonably good agreement with the experimental data for the spin-stiffness constants. Nevertheless, the behavior of exchange interactions is quite different. In the scheme $\hat{b}$, the strongest FM interaction is $J_{1}$. The longer-range interactions operating practically at the same distance in the 4th and 5th coordination spheres, in and between the kagome planes (see \Fref{fig.Co3Sn2S2_basic}e,g), are also sizable but smaller than $J_{1}$. In the scheme L0, $J_{1}$ is weakly antiferromagnetic, while the strongest FM interactions take place in the 4th and 5th coordination spheres. The corresponding spin-wave dispersions are plotted in \Fref{fig.Co3Sn2S2_SW}.
\noindent
\noindent In the long wavelength limit $\boldsymbol{q} \to 0$, two methods provide very similar description, while the main difference is in the position of optical branches in the higher-energy region. The results of recent inelastic neutron scattering data (which were nevertheless limited by the behavior of the acoustic branch in the vicinity of $\boldsymbol{q} = 0$) where interpreted in terms of four interactions (in our notations): $J_{2}=-0.32$, $J_{3}=0.64$, $\bar{J}_{4}=1.66$, and $\bar{J}_{5}= 0.09$ meV~\cite{ZhangPRL}, where $\bar{J}_{k}=\frac{1}{3}(J_{k} + 2J_{k}')$ stands for the averaged exchange interactions in the 4th and 5th coordination spheres.\footnote{In order to be consistent with our definition of the spin model, \Eref{eq:Hspin}, the exchange parameters reported in ref.~\cite{ZhangPRL} were additionally multiplied by $2S^2=\frac{1}{18}$.} There is certain similarity with the picture provided by the L0 scheme: at least $J_{1}$ is negligibly small, $J_{2}$ is antiferromagnetic, and the main FM interactions occurs in the next coordination spheres. Nevertheless, there are also the differences. Especially, experimental $\bar{J}_{5}$ appears to be smaller than $\bar{J}_{4}$, while theoretical parameters are at least comparable (see \Tref{tab.Co3Sn2S2_J}). According to the theoretical analysis, the values of inequivalent parameters $J_{k}$ and $J_{k}'$ can be very different and this difference was not considered in the fitting of the experimental spin-wave spectrum. On the other hand, it is not clear whether it can resolve the discrepancy between theoretical and experimental data. It is also unclear whether the experimental data available only in the vicinity of $\boldsymbol{q} = 0$ are enough to make a decisive conclusion about the complexity of the exchange interactions in Co$_3$Sn$_2$S$_2$.

\par
\begin{figure}[t]
\begin{center}
\includegraphics[width=3.0cm]{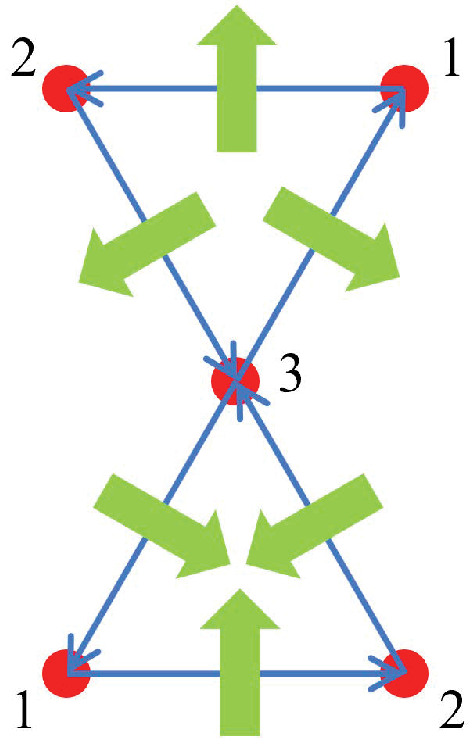}
\end{center}
\caption{Dzyaloshinskii-Moriya interactions between nearest neighbors in Co$_3$Sn$_2$S$_2$. The Co atoms belonging to one of the three sublattices are denoted by numbers. Blue vectors show the directions of the bonds. Green vectors are projections of the DM vectors onto the kagome plane for each such bond. The $d^{z}$ component is perpendicular to this plane and has the same direction for all such bonds.}
\label{fig.Co3Sn2S2_DM}
\end{figure}

 The Co$_3$Sn$_2$S$_2$ structure has three Co sublattices (which are transformed to each other by the threefold rotations). The spacial inversion transforms each Co atom to the same sublattice. Thus, there can be no DM interactions within the sublattices. Nevertheless, the DM interactions between the sublattices are permitted by the $R\overline{3}m$ space group. The strongest ones take place between nearest neighbors in the kagome plane. They are explained in \Fref{fig.Co3Sn2S2_DM}.
\noindent
\noindent Using the convention, where the directions of the bonds ($\boldsymbol{\epsilon}$) in the triangles can be transformed to each other by the threefold rotations, the DM vectors can be presented in the form: $\boldsymbol{d} = d^{xy}[\boldsymbol{\epsilon} \times \boldsymbol{n}^{z}] + d^{z}\boldsymbol{n}^{z}$. If the FM moment is parallel to $z$ (the experimental situation), $d^{z}$ does not contribute to the energy change, while $d^{xy}$ is responsible for the canting of spins and tends to form the umbrella texture. If $\boldsymbol{e}_{k} = (\theta \sin \frac{\pi k}{3}, \theta \cos \frac{\pi k}{3}, 1-\frac{\theta^2}{2})$ are the directions of spins in the triangle ($k$$= 1$, $2$, or $3$, as explained in \Fref{fig.Co3Sn2S2_DM}), the energy gain due to the DM interaction is $\delta E_{\rm DM} = -\sqrt{3}d^{xy} \theta$ (per one Co site). The corresponding energy loss due to the isotropic exchange is $\delta E_{\rm H} = \frac{3}{4} J_{1,23} \theta^2$, where $J_{1,23}$ is the effective interaction of the atom in the 1st sublattice with the atoms of the 2nd and 3rd sublattices in all the coordination spheres. The values of the parameters obtained in the scheme L0 are $J_{1,23}=6.04$ meV, $d^{xy}=0.20$ meV, and $d^{z}=-0.29$ meV.\footnote{The DM parameters are not sensitive to the definition and very similar values are obtained in the scheme $M$, taking $\mathcal{I}_{\rm Sn}$ and $\mathcal{I}_{\rm S}$ from the set IV. However, the same procedure would give us $J_{1,23} = -4.42$ meV, meaning that the FM state is unstable, as was already explained in the main text.} Thus, the angle $\theta = \frac{2d^{xy}}{\sqrt{3}J_{1,23}}$ can be estimates as $2^\circ$, which perfectly agrees with $\theta \sim 2^\circ$ obtained in the brute-force GGA calculations with the SO coupling~\cite{PRB2022b}.

\section{\label{sec:manganites} Orthorhombic perovskite manganites}
\par
\begin{figure}[t]
\begin{center}
\includegraphics[width=6.0cm]{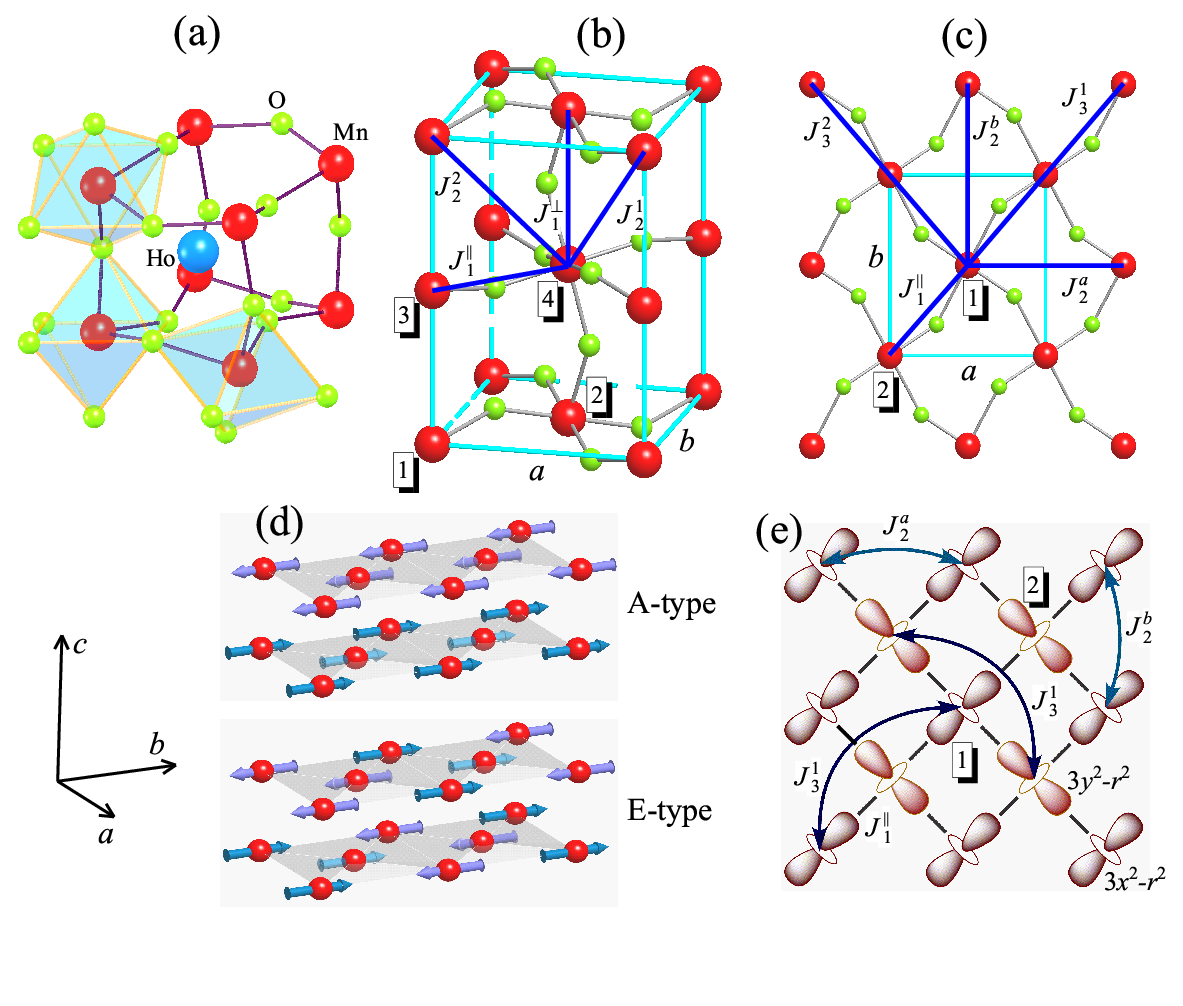}\hspace*{0.0cm}
\includegraphics[width=4.5cm]{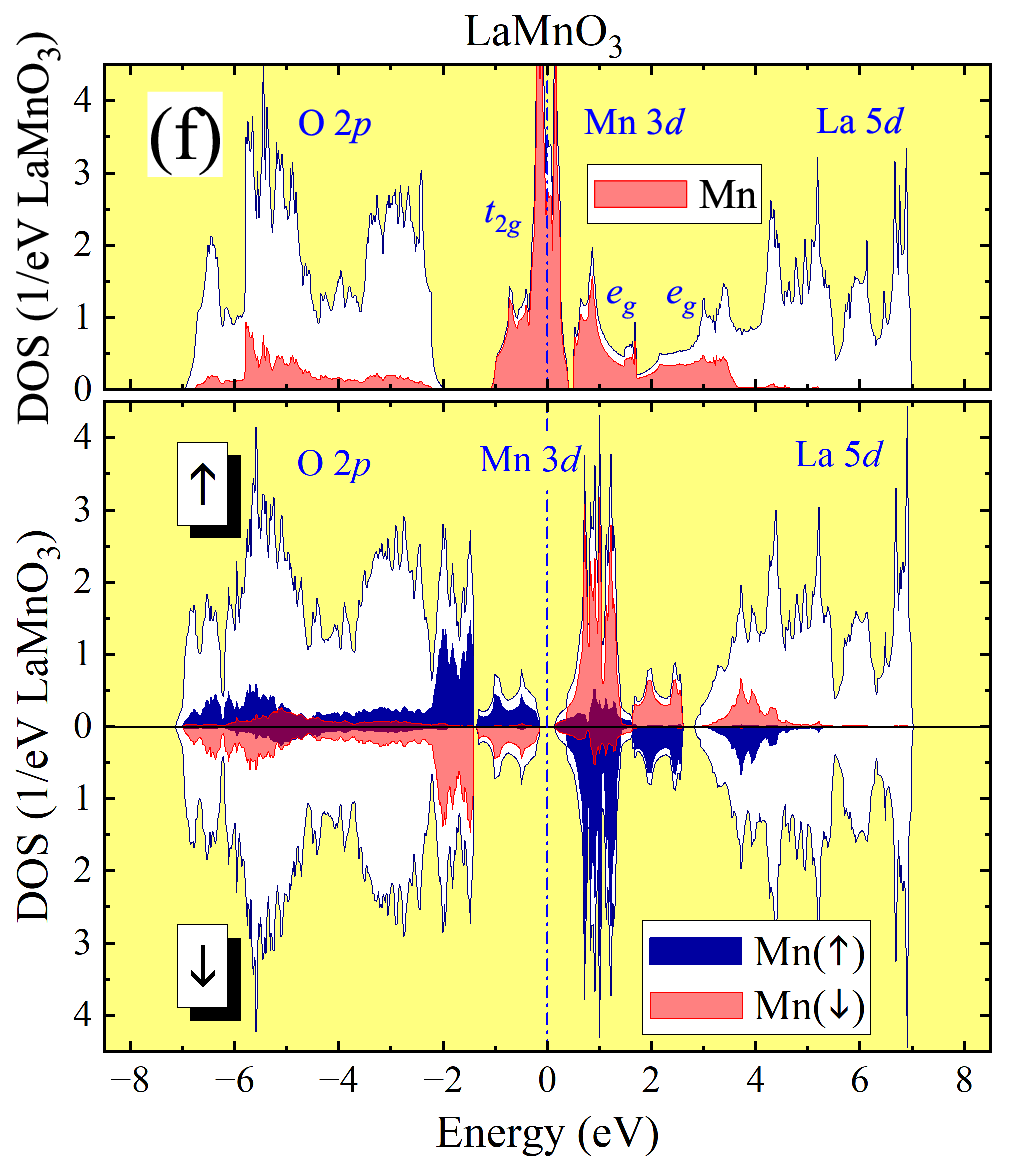}
\includegraphics[width=4.5cm]{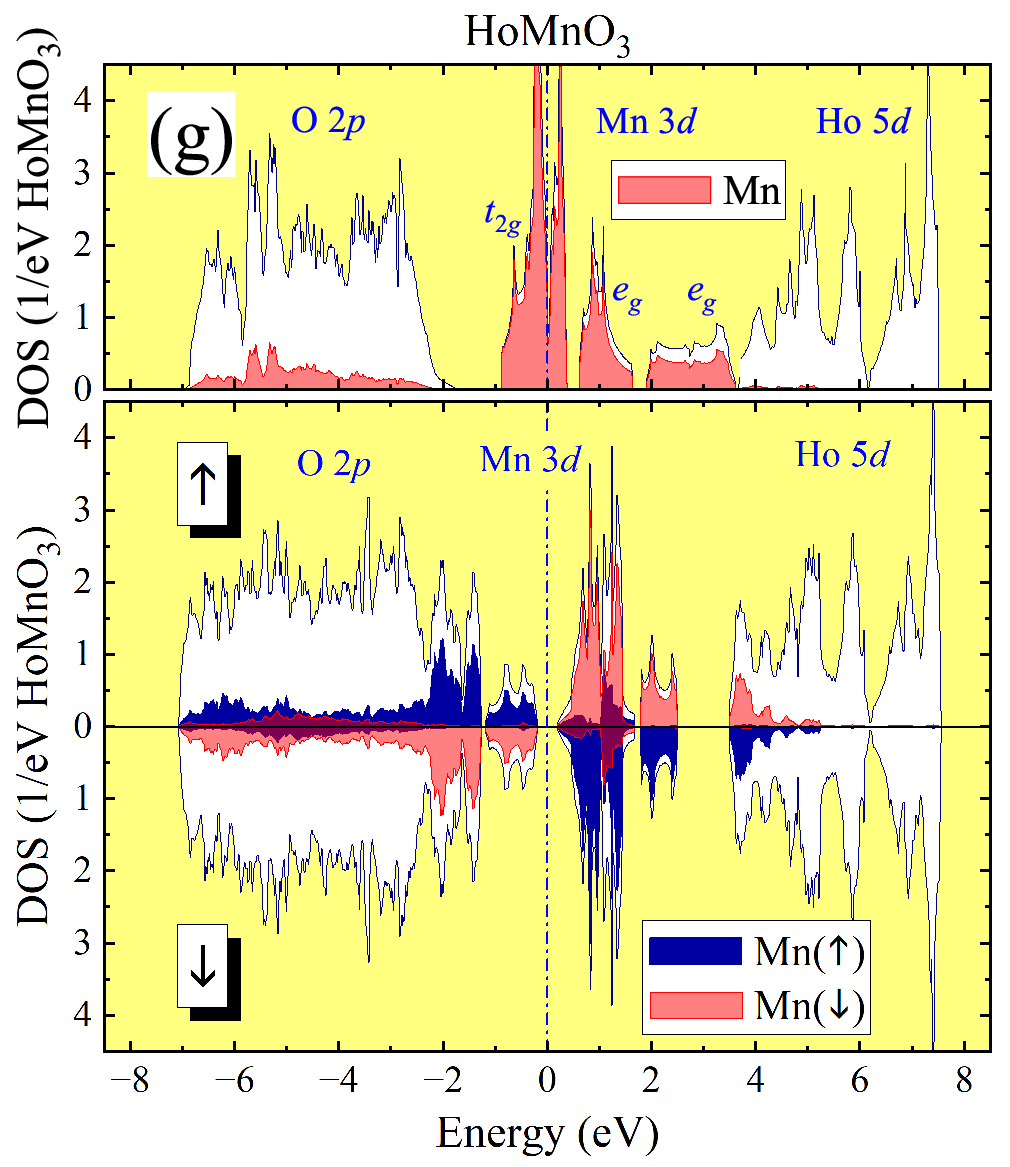}
\end{center}
\caption{(a) Fragment of distorted perovskite structure, illustrating the arrangement of the MnO$_6$ octahedra. (b) Main exchange interactions in the orthorhombic cell. (c) Main exchange interactions in the orthorhombic $ab$ plane. (d) AFM alignment of the types A and E. (e) Schematic view on the orbital ordering underlaying the behavior of exchange interactions in the $ab$ plane. Atoms forming four Mn sublattices are denoted by numbers. (f),(g) Densities of states (DOS) for LaMnO$_3$ and HoMnO$_3$ in LDA (top) and LSDA for the A-type AFM state (bottom). Shaded areas show partial contributions of the Mn $3d$ states (from two magnetic sublattices with the spins $\uparrow$ and $\downarrow$ in the case of A-type AFM order). The Fermi level is at zero energy (the middle of the band gap in the insulating phase).}
\label{fig.AMnO3_basic}
\end{figure}

 Perovskite manganites $A$MnO$_3$ (where $A$ is the trivalent rare earth or alkaline earth element) have attracted a great deal of attention. Most of them crystallize in the orthorhombic $Pbnm$ structure (\Fref{fig.AMnO3_basic}a).
\noindent
\noindent The magnetic Mn$^{3+}$ ions in the octahedral environment accommodate four electrons, three of which occupy the $t_{2g}$ states, while the remaining one resides in the doubly degenerate manifold of the $e_g$ states. Therefore, the system is subjected to the Jahn-Teller distortion, resulting in the peculiar orbital ordering (the alternation of occupied $e_g$ orbitals, as schematically shown in \Fref{fig.AMnO3_basic}e).

\par LaMnO$_3$ is the parent material of colossal magnetoresistive oxides~\cite{TokuraCMR} and a popular testbed system for studying abilities of first-principle electronic structure calculations, especially in reproducing the insulating behavior, the cooperative Jahn-Teller distortion, and associated with it A-type AFM ordering, where the FM coupling within the orthorhombic $ab$ plane coexists with weakly AFM coupling between the planes (see \Fref{fig.AMnO3_basic}d)~\cite{PRL96,Sawada,HeFranchini}. The orthorhombic distortion systematically increases with the decrease of the size of the $A^{3+}$ ions in the direction La$\to$Ho. At certain point it makes the A-type AFM phase unstable in the $ab$ plane. For instance, the ground state of TbMnO$_3$ is believed to be a spin spiral with the propagation vector $\boldsymbol{q} \approx (0,\frac{1}{4},0)$~\cite{Kimura_TbMnO3}, whereas HoMnO$_3$ forms the twofold periodic E-type AFM structure corresponding to $\boldsymbol{q} = (0,\frac{1}{2},0)$~\cite{HoMnO3exp,Ishiwata}. These magnetic superstructures breaks the inversion symmetry, giving rise to rich multiferroic activity~\cite{CheongMostovoy,KhomskiiMF,TokuraSeki}.

\par In this section we consider LaMnO$_3$ and HoMnO$_3$ as two characteristic example and discuss abilities of the linear-response based techniques for describing the change of the exchange interactions, which would eventually lead to the change of the magnetic structure from A to E. We use the experimental crystal-structure parameters reported in~\cite{LaMnO3exp} and~\cite{HoMnO3exp} for LaMnO$_3$ and HoMnO$_3$, respectively. The main parameters are summarized in \Tref{tab.AMnO3exp}.
\noindent
\begin{table}
\caption{\label{tab.AMnO3exp} Selected parameters of the orthorhombic structure for LaMnO$_3$~\cite{LaMnO3exp} and HoMnO$_3$~\cite{HoMnO3exp}: the orthorhombic lattice parameters $a$, $b$, and $c$; the Mn-O bondlengths ($d_{\rm Mn\mbox{-}O}$) in the $ab$ plane (the first two values) and along $c$ (the third value); and the Mn-O-Mn angles ($\angle \,{\rm Mn\mbox{-}O\mbox{-}Mn}$) in the $ab$ plane (first value) and between the planes (second value).}
\begin{indented}
\lineup
\item[]\begin{tabular}{@{}cccccc}
\br
compound   & $a$ (\AA) & $b$ (\AA) & $c$ (\AA) & $d_{\rm Mn\mbox{-}O}$ (\AA) & $\angle \,{\rm Mn\mbox{-}O\mbox{-}Mn}$ ($^{\circ}$) \cr
\mr
LaMnO$_3$  & $5.532$   & $5.742$   & $7.668$   & $1.906$, $2.118$, $1.959$   & $154$, $157$                                        \cr
HoMnO$_3$  & $5.257$   & $5.835$   & $7.361$   & $1.905$, $2.222$, $1.943$   & $144$, $142$                                        \cr
\br
\end{tabular}
\end{indented}
\end{table}
\noindent Particularly, the MnO$_6$ octahedra have two long Mn-O bonds in the $ab$ plane and four short ones, which is the signature of the Jahn-Teller distortion~\cite{KanamoriJT}. The Jahn-Teller distortion practically does not change when going from LaMnO$_3$ to HoMnO$_3$. On the other hand, the Mn-O-Mn angles change significantly with much stronger deviation from the ideal cubic value of $180^{\circ}$ in the case of HoMnO$_3$. This change is accompanied by some shrinking of the orthorhombic lattice along $a$ and $c$.

\par The orthorhombic distortion has a profound effect of the electronic structure in LDA/LSDA (\Fref{fig.AMnO3_basic}f,g), splitting the Mn $e_{g}$ band and opening the band gap in the A-type AFM phase. The latter is the consequence of the Jahn-Teller distortion \emph{and} quasi-two-dimensional character of the A-type AFM order~\cite{GorkovKresin}. The occupied Mn $e_{g}$ band is located around $-1$ eV and well separated from other bands. The corresponding distribution of the $e_g$ electron density has the form of the orbital ordering, which is schematically shown in \Fref{fig.AMnO3_basic}e.

\par The exchange interactions in LaMnO$_3$ and HoMnO$_3$ are rather complex (see \Fref{fig.AMnO3_basic}b,c)~\cite{JPSJ2009}. Particularly, in addition to the nn interactions in and between the $ab$ planes ($J_{1}^{\parallel}$ and $J_{1}^{\perp}$, respectively), there are several important longer-range interactions, such as: (i) the next-nn interactions between the planes, $J_{2}^{1}$ and $J_{2}^{2}$; (ii) the 2nd neighbor interactions in the plane, $J_{2}^{a}$ and $J_{2}^{b}$, operating along orthorhombic axes $a$ and $b$, respectively; and (iii) the 3rd neighbor interactions in the plane, $J_{3}^{1}$ and $J_{3}^{2}$. These interactions obey certain symmetry properties. For instance, considering Mn site 1 in \Fref{fig.AMnO3_basic}b as the reference point, $J_{3}^{1}$ and $J_{3}^{2}$ will operate in the bonds $\pm(a,a,0)$ and $\pm(a,-a,0)$, respectively. The parameters $J_{3}^{1}$ and $J_{3}^{2}$ around Mn sites 2, 3, and 4 are obtained by the $180^{\circ}$ rotations of these bonds about $a$, $b$, and $c$ combined with the lattice shifts by $(\frac{a}{2},\frac{b}{2},0)$, $(0,0,\frac{c}{2})$, and $(\frac{a}{2}, \frac{b}{2}, \frac{c}{2})$, respectively. The interactions $J_{1}^{\perp}$, $J_{2}^{1}$ and $J_{2}^{2}$ control the AFM coupling between the planes, while the formation of the long-periodic magnetic structures in the plane results from the interplay of $J_{1}^{\parallel}$, $J_{2}^{a}$, $J_{2}^{b}$, $J_{3}^{1}$ and $J_{3}^{2}$. The behavior of $J_{3}^{1}$ and $J_{3}^{2}$ is directly related to the orbital ordering: these 3rd neighbor interactions can be regarded as the super-superexchange ones, mediated by the states of intermediate Mn sites. If the lobes of the occupied $e_{g}$ orbitals are directed toward each other along the bond, as in the case of $J_{3}^{1}$, the interaction is strong. If the lobes are parallel to each other and perpendicular to the bond, as in the case of $J_{3}^{2}$, the interaction is weak. Moreover, two crystallographically different types of next-nn bonds between the planes result in slightly different values of the parameters $J_{2}^{1}$ and $J_{2}^{2}$~\cite{PRB2021}.

\par From the viewpoint of the electronic structure, one can construct two types of model: the correlated 5-orbital model, where the one-electron part is taken from LDA and combined with the screened Coulomb interactions obtained in cRPA (as explained in \ref{sec:Hubbard}), and the all electron model within LSDA, which includes Mn $3d$ as well as O $2p$ and $A$ $5d$ bands. The calculations are performed for the A-type AFM phase on the mesh of the $8$$\times$$8$$\times$$6$ points, both for $\boldsymbol{k}$ and $\boldsymbol{q}$, unless it is specified otherwise.

\subsection{\label{sec:AMnO3_5o} Correlated 5-orbital model}
\par First, we investigate abilities of correlated 5-electron model: whether it can reproduce the main tendencies in the behavior of interatomic exchange interactions in LaMnO$_3$ and HoMnO$_3$, stabilizing the A-type AFM state in the former case and making it unstable with respect to the formation of the spin superstructures along the orthorhombic axis $b$ in the latter one. Another important question is how good is the approximate scheme $\hat{b}$, which is frequently used for the analysis of interatomic exchange interactions in these orthorhombic manganites~\cite{JPSJ2009,PRL96}, in comparison with the more accurate scheme $M$. The averaged on-site parameters of the Coulomb repulsion, intraatomic exchange interaction, and nonsphericity (see \ref{sec:Hubbard} for definitions) calculated in cRPA are, respectively, $2.15$ ($2.16$), $0.85$ ($0.85$), and $0.09$ ($0.09$) eV for LaMnO$_3$ (HoMnO$_3$)~\cite{JPSJ2009}. Thus, the Coulomb $U$$\sim$$2$ eV is not particularly strong.\footnote{This is qualitatively consistent with old estimates, based in the constrained LSDA~\cite{PRB96}. The key point is that Mn $e_{g}$ electrons are efficiently screened by other Mn $3d$ electrons from other bands.} This finding is very important as it naturally explains the existence of the long-range exchange interaction, $J_{2}^{b}$ and $J_{3}^{1}$, arising in higher orders of $\hat{H}/U$ beyond the conventional superexchange approximation~\cite{Anderson1959}, and responsible for the formation of the spin superstructures along $b$. On the other hand, the relatively small $U$ also means that the strong-coupling limit is hardly satisfied. Therefore, it is reasonable to expect that the approximate scheme $\hat{b}$ may experience serious limitations for these orthorhombic manganites.

\par
\begin{figure}[t]
\begin{center}
\includegraphics[width=5.0cm]{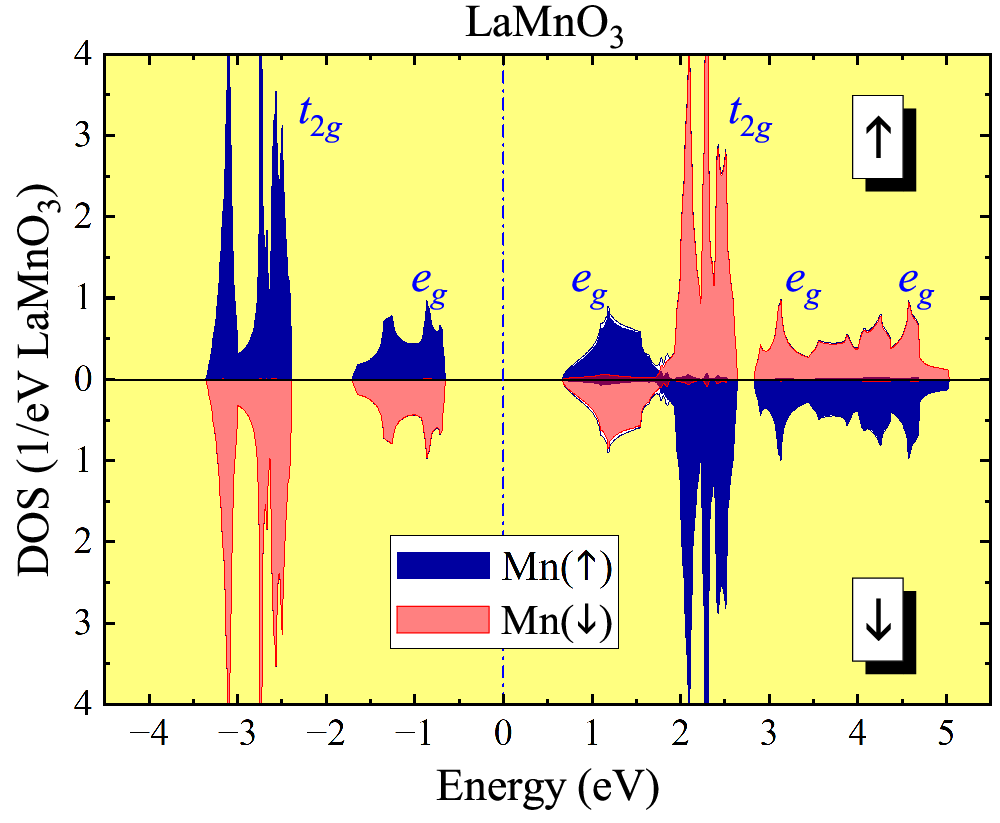}
\includegraphics[width=5.0cm]{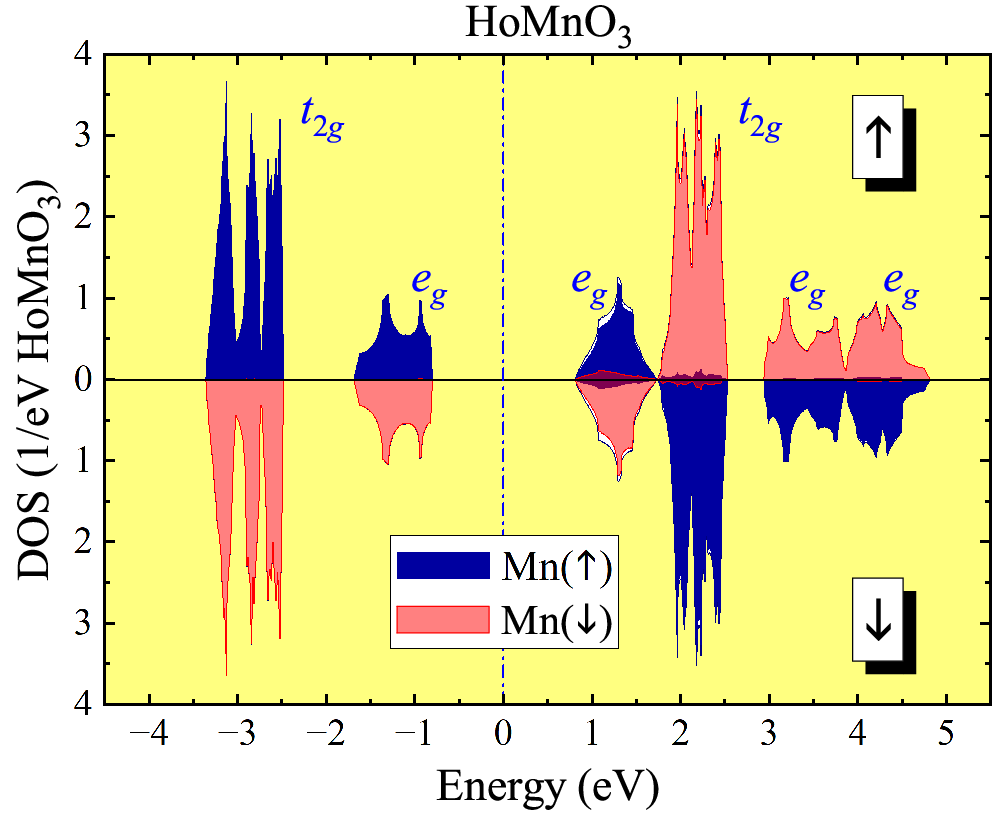}
\end{center}
\caption{Densities of states (DOS) for the A-type AFM order in LaMnO$_3$ and HoMnO$_3$ as obtained in the Hartree-Fock approximation for correlated 5-orbital model. Zero energy is in the middle of the band gap.}
\label{fig.AMnO3_5o}
\end{figure}

 The densities of states obtained in the Hartree-Fock approximation for the A-type AFM order are displayed in \Fref{fig.AMnO3_5o}. LaMnO$_3$ and HoMnO$_3$ have similar electronic structure. The only difference is slightly smaller $e_g$ bandwidth in the case of HoMnO$_3$.
\noindent
\noindent Nevertheless, the behavior exchange interactions appears to be very different. These interactions are summarized in Tables~\ref{tab.LaMnO3} and \ref{tab.HoMnO3} (lines 5o) for LaMnO$_3$ and HoMnO$_3$, respectively.
\noindent
\begin{table}
\caption{\label{tab.LaMnO3} Isotropic exchange interactions in LaMnO$_3$ (in meV) as obtained in the correlated 5-orbital (5o) model in comparison with LSDA for the all-electron ${\rm Mn} \, 3d$+${\rm O} \, 2p$+${\rm La} \, 5d$ model. The notations of parameters are explained in \Fref{fig.AMnO3_basic}b,c. The AFM interactions $J_{2}^{a}$ and $J_{3}^{2}$ are considerably weaker and not shown here. $T_{\rm N}$ is the N\'eel temperature evaluated in RPA (in K) and $\boldsymbol{q} = (0,q_{b},0)$ is the theoretical ground state propagation vector (in units of reciprocal lattice translations, where $\boldsymbol{q}=0$ corresponds to the A-type AFM state).}
\begin{indented}
\lineup
\item[]\begin{tabular}{@{}ccccccccc}
\br
method                       & $J_{1}^{\parallel}$ & $J_{1}^{\perp}$   & $J_{2}^{1}$ & $J_{2}^{2}$ & $J_{2}^{b}$ & $J_{3}^{1}$ & $T_{\rm N}$ & $q_{b}$  \cr
\mr
$\phantom{M}\hat{b}$ in 5o   & $\phantom{1}2.36$   & $-6.64$           & $-0.97$     & $-1.10$     & $-1.00$     & $-3.27$     & $80$        & $0.32$   \cr
$\phantom{\hat{b}}M$ in 5o   & $17.23$             & $\phantom{-}3.51$ & $-1.42$     & $-1.51$     & $-1.13$     & $-3.56$     & $186$       & $0$      \cr
$\phantom{M}\hat{b}$ in LSDA & $15.10$             & $\phantom{-}7.37$ & $-3.14$     & $-3.25$     & $-1.78$     & $-9.46$     & $131$       & $0$      \cr
$\phantom{\hat{b}}M$ in LSDA & $17.09$             & $\phantom{-}8.20$ & $-3.24$     & $-3.27$     & $-1.83$     & $-9.21$     & $185$       & $0$      \cr
\br
\end{tabular}
\end{indented}
\end{table}
\noindent
\begin{table}
\caption{\label{tab.HoMnO3} The same as \Tref{tab.LaMnO3} but for HoMnO$_3$. The LSDA results are obtained in the ${\rm Mn} \, 3d$+${\rm O} \, 2p$+${\rm Ho} \, 5d$ model.}
\begin{indented}
\lineup
\item[]\begin{tabular}{@{}ccccccccc}
\br
method                       & $J_{1}^{\parallel}$ & $J_{1}^{\perp}$   & $J_{2}^{1}$ & $J_{2}^{2}$ & $J_{2}^{b}$ & $J_{3}^{1}$ & $T_{\rm N}$ & $q_{b}$  \cr
\mr
$\phantom{M}\hat{b}$ in 5o   & $-6.53$             & $-7.58$           & $-0.54$     & $-0.72$     & $-1.44$     & $-2.22$     & $65$        & $0.22$   \cr
$\phantom{\hat{b}}M$ in 5o   & $\phantom{-}4.02$   & $-1.44$           & $-0.63$     & $-0.82$     & $-1.53$     & $-2.36$     & $53$        & $0.26$   \cr
$\phantom{M}\hat{b}$ in LSDA & $\phantom{-}2.23$   & $-1.64$           & $-1.72$     & $-1.85$     & $-2.57$     & $-6.90$     & $79$        & $0.39$   \cr
$\phantom{\hat{b}}M$ in LSDA & $\phantom{-}4.15$   & $-1.03$           & $-1.71$     & $-1.81$     & $-2.46$     & $-6.50$     & $79$        & $0.34$   \cr
\br
\end{tabular}
\end{indented}
\end{table}

\par The scheme $\hat{b}$ yields strong interlayer coupling $J_{\phantom{1}}^{\perp} = J_{1}^{\perp} + 2J_{2}^{1} + 2J_{2}^{2} = -$$10.78$ ($-$$10.1$) meV for LaMnO$_3$ (HoMnO$_3$), which is even stronger than the intralayer one $J_{1}^{\parallel} = 2.36$ ($-$$6.53$) meV. This behavior is inconsistent with the experimental neutron-scattering data for LaMnO$_3$, indicating that $J_{\phantom{1}}^{\perp} \sim -$$4.7$ meV is weaker than $J_{1}^{\parallel} = 6.6$ meV~\cite{LaMnO3Hirota,LaMnO3Moussa}.\footnote{In order to be consistent with our definition of the spin model, \Eref{eq:Hspin}, the experimental parameters have been multiplied by $2S^{2}=8$.} Furthermore, the longer-range AFM interactions $J_{2}^{b}$ and $J_{3}^{1}$ in LaMnO$_3$ are comparable or even stronger than $J_{1}^{\parallel}$, making the experimental A-type AFM structure unstable.

\par On the contrary, the scheme $M$ systematically improves the description of the interatomic exchange interactions. First, the interlayer coupling becomes considerably weaker: $J_{\phantom{1}}^{\perp}=-$$2.35$ ($-$$4.34$) meV for LaMnO$_3$ (HoMnO$_3$). Then, the itralayer interaction $J_{1}^{\parallel}$ in LaMnO$_3$ becomes strongly ferromagnetic, as expected for the ``antiferro'' orbital ordering~\cite{KugelKhomskii},\footnote{The alternation of the $3x^2$-$r^2$ and $3y^2$-$r^2$ orbitals in the $ab$ plane (see \Fref{fig.AMnO3_basic}e).} and overcomes the AFM long-range interactions $J_{2}^{b}$ and $J_{3}^{1}$. As the result, the experimental A-type AFM phase becomes stable. The theoretical $T_{\rm N}=186$ K, evaluated in RPA, is in fair agreement with the experimental value of $140$ K~\cite{LaMnO3Hirota,LaMnO3Moussa}. In HoMnO$_3$, $J_{1}^{\parallel}$ is significantly reduced due to the additional buckling of the Mn-O-Mn bonds to become comparable with $J_{2}^{b}$ and $J_{3}^{1}$. This makes the A-type AFM phase unstable. Regarding the direction of this instability, it is very important that $J_{2}^{a} = -$$0.35$ meV is much weaker than $J_{2}^{b} = -$$1.53$ meV. Therefore, the propagation vector for the expected theoretical ground state is parallel to $b$, in agreement with the experimental observation. This instability is further enhanced by the AFM interactions $J_{3}^{1}$. Nevertheless, in order to reproduce the experimental E phase with commensurate $\boldsymbol{q} = (0,\frac{1}{2},0)$, it is essential to consider other ingredients such as the exchange striction and single-ion anisotropy, which would lock the spin superstructure to the lattice~\cite{Picozzi,Okuyama,PRB2012}. In HoMnO$_3$, such lock-in transition occurs at $29$ K, while the N\'eel temperature is about $41$ K~\cite{HoMnO3exp,Ishiwata}, which is close to theoretical value of $T_{\rm N} = 53$ K. Other aspects of stability of the E-type AFM phase will be considered in \Sref{sec:AMnO3_E}.

\subsection{\label{sec:AMnO3_LSDA} All-electron model in LSDA}
\par The all-electron model in LSDA provides an alternative description for the exchange interactions. We start with the analysis of effective Stoner parameters. As was discussed in \Sref{sec:CrX3_L_LSDA}, amongst several possible definitions of the parameters $\mathcal{I}$, the most relevant seem to be III and IV, which should be considered in combination with the methods $\hat{b}$ and $M$, respectively. In the A-type AFM phase, the $A$ and apical O atoms, located between the antiferromagnetically coupled layers, are nonmagnetic. Therefore, we set $\mathcal{I}=0$ for them. The remaining parameters for Mn and planar O atoms are listed in \Tref{tab.AMnO3_Stoner}.
\noindent
\begin{table}[h]
\caption{\label{tab.AMnO3_Stoner} The effective Stoner parameters (in eV) for Mn and planar O atoms in LaMnO$_3$ and HoMnO$_3$, obtained using definitions III and IV (as explained in \Sref{sec:CrX3_L_LSDA}). The $A$ and apical O atoms are nonmagnetic in the A-type AFM phase and not considered here.}
\begin{indented}
\lineup
\item[]\begin{tabular}{@{}ccccc}
\br
           & \centre{2}{LaMnO$_3$}                          & \centre{2}{HoMnO$_3$}                          \cr
\ns
           & \crule{2}                                      & \crule{2}                                      \cr
definition & $\mathcal{I}_{\rm Mn}$ & $\mathcal{I}_{\rm O}$ & $\mathcal{I}_{\rm Mn}$ & $\mathcal{I}_{\rm O}$ \cr
\mr
III        & $0.98$                 &  $3.71$               & $0.97$                 & $4.20$                \cr
IV         & $0.98$                 &  $4.35$               & $0.97$                 & $4.23$                \cr
\br
\end{tabular}
\end{indented}
\end{table}
\noindent $\mathcal{I}_{\rm Mn}$ is practically identical for LaMnO$_3$ and HoMnO$_3$, and does not depend on the definition. $\mathcal{I}_{\rm O}$ appears to be more sensitive to the environment and the definition. Nevertheless, the change of $\mathcal{I}_{\rm O}$ is rather modest. For both definitions, $\mathcal{I}_{\rm O}$ is large and positive, that will additionally strengthens the FM interactions.

\par The exchange interactions are summarized in Tables~\ref{tab.LaMnO3} and \ref{tab.HoMnO3} (lines LSDA). Unlike for the correlated 5-orbital model, the methods $\hat{b}$ and $M$ in the all-electron LSDA provide a consistent description. On the one hand, there is an effect of the O $2p$ band, which is explicitly treated in this model. On the other hand, the underestimation of the FM interactions in the scheme $\hat{b}$ is partly compensated by $\mathcal{I}_{\rm O}$. Moreover, the correct definition of the xc field $\hat{b}$ using the sum rule is important. For instance, the use of $\hat{b}$ defined via site-diagonal elements of $\hat{H}^{\uparrow, \downarrow}$ underestimates the FM interactions and makes the experimental A-type AFM state unstable in LaMnO$_3$.

\par The exchange interactions are generally stronger than in correlated 5-orbital model (except $J_{1}^{\parallel}$), partly due to the fact that in these insulating materials the exchange interactions are expected to increase when the effective Coulomb repulsion decreases~\cite{Anderson1959}. Furthermore, the oxygen band can also contribute to the exchange interactions~\cite{Oguchi,ZaanenSawatzky}. Anyway, the total interlayer coupling in LaMnO$_3$, $J_{\phantom{1}}^{\perp} = -$$4.82$ ($-$$5.41$) in the scheme $M$ ($\hat{b}$), is consistent with the experimental data~\cite{LaMnO3Hirota,LaMnO3Moussa}. $J_{1}^{\parallel}$ is strongly ferromagnetic, which appears to be sufficient to overcome the strong AFM interactions $J_{2}^{b}$ and $J_{3}^{1}$, and stabilize the experimental A-type AFM phase. In HoMnO$_3$, this $J_{1}^{\parallel}$ is substantially weaker, making the A-type AFM phase unstable with respect to an incommensurate spin structure propagating along $b$. The theoretical $T_{\rm N}$ is in fair agreement with experimental data.

\subsection{\label{sec:AMnO3_DM} Dzyaloshinskii-Moriya interactions}
\par
\begin{figure}[t]
\begin{center}
\includegraphics[width=10.0cm]{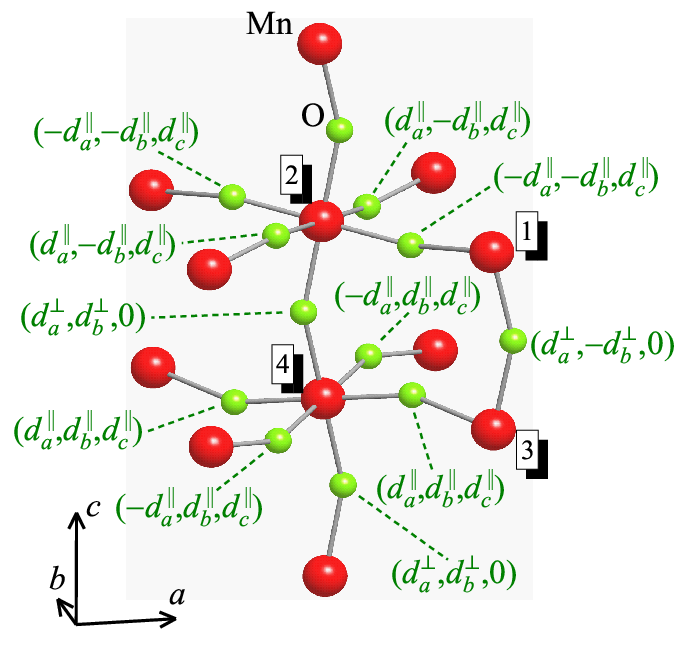}
\end{center}
\caption{Form of DM interactions operating between nearest neighbors in the orthorhombic structure, in and between the plane. The numbering of Mn sublattices is the same as in \Fref{fig.AMnO3_basic}. Each DM vector is attached to its Mn-O-Mn bond, starting from the atoms 2 or 4 for the in-plane interactions and atoms 3 or 4 for the out-of-plane interactions.}
\label{fig.AMnO3DM}
\end{figure}

 All nn DM interactions in the orthorhombic planes, $\boldsymbol{d}^{\, \parallel}_{ij}$, are transformed to each other by the symmetry operations of the space group $Pbnm$~\cite{PRL96}. The same holds for the interplane interactions $\boldsymbol{d}^{\perp}_{ij}$. Furthermore, since neighboring planes are connected by the mirror reflection, the $z$ ($c$) component of $\boldsymbol{d}^{\perp}_{ij}$ is equal to zero. Therefore, there are 5 parameters describing all nn DM interactions: $d^{\, \parallel}_{a}$, $d^{\, \parallel}_{b}$, and $d^{\, \parallel}_{c}$ for the in-plane interactions, and $d^{\perp}_{a}$ and $d^{\perp}_{b}$ for the out-of-plane interactions. The corresponding DM vectors, attached to neighboring Mn-O-Mn bonds, are shown in \Fref{fig.AMnO3DM}.
\noindent
\noindent The values of the parameters, obtained in the scheme $M$, are summarized in \Tref{tab.AMnO3DM}.
\noindent
\begin{table}[h]
\caption{\label{tab.AMnO3DM} Parameters of nearest-neighbor DM interactions (in meV) as obtained in the correlated 5-orbital (5o) model and all-electron LSDA using the scheme $M$. The results of ref.~\cite{PRL96}, employing mixed perturbation theory, are shown for comparison.}
\begin{indented}
\lineup
\item[]\begin{tabular}{@{}ccccccc}
\br
compound  & model        & $d^{\, \parallel}_{a}$ & $d^{\, \parallel}_{b}$ & $d^{\, \parallel}_{c}$ & $d^{\perp}_{a}$ & $d^{\perp}_{b}$ \cr
\mr
LaMnO$_3$ & \cite{PRL96} &  $0.44$                &  $0.33$                &  $0.53$                & $0.45$          & $0.71$          \cr
LaMnO$_3$ & LSDA         &  $0.46$                &  $0.35$                &  $0.63$                & $0.53$          & $0.60$          \cr
HoMnO$_3$ & LSDA         &  $0.54$                &  $0.38$                &  $0.73$                & $0.70$          & $0.47$          \cr
HoMnO$_3$ & 5o           &  $0.23$                &  $0.09$                &  $0.28$                & $0.20$          & $0.11$          \cr
\br
\end{tabular}
\end{indented}
\end{table}

\par For LaMnO$_3$ in LSDA we note a good agreement with the results of ref.~\cite{PRL96}, employing the mixed perturbation theory, where the rotation of the xc field on one Mn sites was combined with the SO coupling on another such site. This means that the $5d$ states of the heavy $A$ atoms, which are located far in the unoccupied part of the spectrum (see \Fref{fig.AMnO3_basic}), do not strongly contribute to the DM interactions. Partly, this may be due to the fact that the $A$ sites remain nonmagnetic in the A-type AFM phase.

\par The parameters obtained in the correlated 5-orbital model are generally smaller than in LSDA -- similar to what was found for the isotropic interactions. However, this is not surprising and can be again explained by the Coulomb $U$ in the denominator of the superexchange interactions~\cite{Anderson1959}.

\par The DM interactions in $A$MnO$_3$ mainly contribute to the spin canting. The magnetocrystalline anisotropy tends to align the spins parallel to the $b$ axis~\cite{Bozorth,Treves,LaMnO3Matsumoto}. In LaMnO$_3$, they order according to the A-type: $\boldsymbol{e}_{\mu} = (0,-1,0)$ for $\mu =1$ and $2$ and $\boldsymbol{e}_{\mu} = (0,1,0)$ for $\mu =3$ and $4$ in \Fref{fig.AMnO3DM}. This perfect AFM alignment will be further deformed by the DM interactions, which additionally rotate the spins along $a$ and $c$ by, respectively, $e_{a} = -d^{\, \parallel}_{c}/(J_{1}^{\parallel}-J_{2}^{1}-J_{2}^{2})$ and $e_{c} = d^{\, \perp}_{a}/(J_{1}^{\parallel}+2J_{2}^{1}+2J_{2}^{2})$~\cite{PRB2014}. The $a$  components will order according to the G-type,\footnote{The AFM coupling between all nearest neighbors, in and between the planes.} while $c$ components gives rise to the weak ferromagnetism~\cite{Bozorth,Treves}. Using the obtained parameters of exchange interactions, $|e_{a}|$ and $|e_{c}|$ can be estimated as $0.019$ and $0.110$, respectively. Taking into account that in all-electron LSDA $M=3.59$ $\mu_{\rm B}$, the weak FM moment can be estimated as $0.4$ $\mu_{\rm B}$ per Mn atom (being somewhat larger that the experimental $0.18$ $\mu_{\rm B}$, derived from magnetization measurements~\cite{Skumryev}).

\par An interesting question is whether the multiferroicity associated with the E-type AFM phase in HoMnO$_3$ can coexist with the weak ferromagnetism. In the E-type AFM structure, each of the $J_{2}^{1}$ and $J_{2}^{2}$ will contribute to two FM bonds and two AFM ones. Therefore, these contributions cancel each other and the spin canting along $c$ will be given by $e_{c} = \pm d^{\, \perp}_{a}/J_{1}^{\parallel}$ (for the sublattices with different spins in the $ab$ plane). Using the parameters for the $Pbnm$ structure of HoMnO$_3$, $|e_{c}|$ can be estimated as $0.169$ and $0.050$ in the all-electron LSDA and correlated 5-orbital model, respectively, where the difference mainly comes from $d^{\, \perp}_{a}$. Nevertheless, this component will repeat the pattern of the E-type AFM phase in each $ab$ plane and, therefore, there will be no weak ferromagnetism.

\subsection{\label{sec:AMnO3_E} Internal instability of the E phase}
\par An interesting aspect of interatomic exchange interactions defined via infinitesimal rotations of spins near some equilibrium magnetic states is that these exchange interactions depend on the magnetic state in which they are calculated, reflecting the dependence of the electronic structure on the magnetic state. If it happens, the bilinear spin model \eref{eq:Hspin} is ill-defined in the global sense, meaning, for instance, that the same set of the exchange parameters cannot be used for the analysis of the low-temperature spin-wave dispersion and the magnetic transition temperature, where the electronic structure is strongly modified by the spin disorder~\cite{Oguchi,Staunton}. Nevertheless, the model \eref{eq:Hspin} can be defined locally, near each magnetic equilibrium. Then, in principle, one can expect some kine of self-organization phenomenon, when the change of the electronic structure in certain magnetic state itself can stabilize this magnetic state, at least locally. This is what happens at least with some of the exchange interactions in the 50\% doped manganites, where the zigzag (CE-type) AFM alignment opens the band gap, induces the orbital ordering, and additionally stabilizes the FM coupling in the zigzag chains~\cite{PRL1999}. Can we expect a similar behavior in the E phase?

\par
\begin{figure}[t]
\begin{center}
\includegraphics[width=3.5cm]{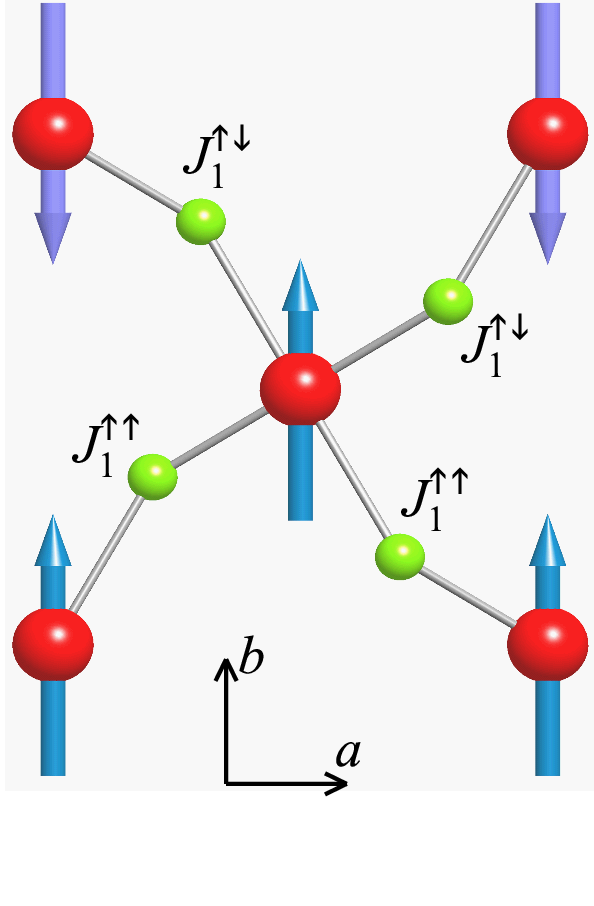}\hspace*{0.5cm} \includegraphics[width=7.0cm]{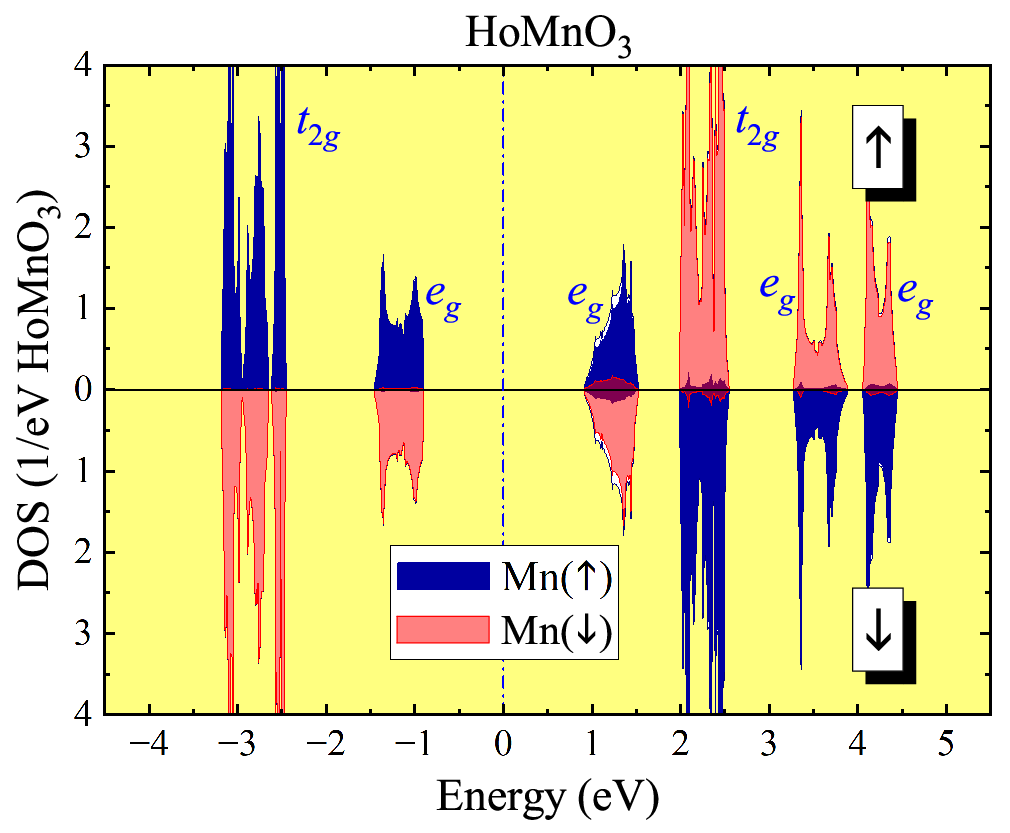}
\end{center}
\caption{(Left) Inequivalent nearest-neighbor interactions in the $ab$ plane of HoMnO$_3$ due to the E-type AFM order. (Right) Densities of states (DOS) for the E-type AFM state in HoMnO$_3$ as obtained in the Hartree-Fock approximation for the correlated 5-orbital model. Zero energy is in the middle of the band gap.}
\label{fig.HoMnO3E}
\end{figure}

 Intuitively, the reason for the formation of this noncentrosymmetric E-type AFM structure can be understood as follows: the AFM interactions $J_{2}^{b}$ and $J_{3}^{1}$ tend to align the corner spins, separated by the orthorhombic translations $a$ and $b$, antiferromagnetically, as shown in \Fref{fig.HoMnO3E}a. The central Mn atom is located in the inversion center. Then, due to the AFM alignment, the corner atoms should transform to each other by the symmetry operation $\hat{I}\hat{T}$ (where the spacial inversion $\hat{I}$ is combined with the time reversal $\hat{T}$), which is formally compatible with the $Pbnm$ space group. However, the same symmetry operation would make the central Mn atom nonmagnetic. This is an eligible scenario from the symmetry point of view, but would lead to gigantic energy loss, $\frac{1}{4}\mathcal{I}_{\rm Mn}M_{\rm Mn}^2\sim 4$ eV, caused by the violation of the first Hund's rule for the atoms, which are potentially expected to be in the high-spin state. Therefore, the more favorable scenario is to keep the Mn atom magnetic, but break the inversion symmetry, which is quite common in multiferroic materials, especially those with high-spin ions.

\par Then, the next question is whether the E-type AFM order is compatible with the change of the exchange interactions induced by this order.
\noindent

\par Below we consider results of correlated 5-orbital model (in the scheme $M$) for HoMnO$_3$. A qualitatively similar behavior has been found in LSDA for the all-electron model. The E-type AFM order results in the additional narrowing of the $t_{2g}$ and $e_{g}$ bands (\Fref{fig.HoMnO3E}), which is consistent with the decrease of the number of FM bonds around each Mn site (2 instead of 4 in the A-type AFM phase). The inversion symmetry breaking makes the nn interactions inequivalent, where $J^{\uparrow \uparrow}_{1}$ in the FM bond is generally different from $J^{\uparrow \downarrow}_{1}$ in the AFM one. In order to stabilize the E state, one would need at least $J^{\uparrow \uparrow}_{1} > J^{\uparrow \downarrow}_{1}$ (the FM coupling is stronger in the FM bond). However, we have found the opposite tendency: $J^{\uparrow \uparrow}_{1} = 2.99$ meV and $J^{\uparrow \downarrow}_{1} = 4.00$ meV, meaning that the E state corresponds to the energy \emph{maximum} and any small rotations of spins will slide the system away from this equilibrium towards a new magnetic state. Furthermore, there is an intrinsic mechanism inducing these rotations of spins. Indeed, the E-type AFM order, producing some changes in the electronic structure, can induce the electric polarization even in the centrosymmetric $Pbnm$ structure~\cite{Picozzi}. Then, it is reasonable to expect that the same changes in the electronic structure may lead to the appearance of new DM interactions, which would otherwise be forbidden in the $Pbnm$ structure. Particularly, we have found finite interactions $\boldsymbol{d}_{2}^{b} = (\pm 0.01,0,\pm 0.02)$ meV and $\boldsymbol{d}_{3}^{1} =(\pm 0.01,\pm 0.01,\pm 0.02)$ meV (being analogs of isotropic $J_{2}^{b}$ and $J_{3}^{1}$, respectively, where the $\pm$ signs depend on the origin and direction of the bond). Although these interactions are not particularly strong, they are sufficient to shift the system of spins away from the extremum (maximum) point, so that it will start to relax to a new magnetic equilibrium.

\par Thus, the E-type AFM state in HoMnO$_3$ is intrinsically unstable and cannot be stabilized by purely electronic mechanisms. For these purposes, it is essential to consider the exchange striction or the single-ion anisotropy (or both of them)~\cite{Picozzi,Okuyama,PRB2012}.

\subsection{\label{sec:AMnO3_summary} Brief summary}
\begin{itemize}
\item The minimal correlated 5-orbital model provides a consistent description for the interatomic exchange interactions in LaMnO$_3$ and HoMnO$_3$, explaining stability of the A-type AFM phase in the former material and the tendency towards formation of the spin superstructures along the orthorhombic axis $b$ in the case of HoMnO$_3$. The driving force behind this change of the magnetic structure is the buckling of the Mn-O-Mn bonds, which is stronger in HoMnO$_3$. The use of the scheme $M$ appears to be crucial within this 5-orbital model, while the approximate scheme $\hat{b}$ strongly underestimates the FM interactions and fails to explain the stability of the A-type AFM phase in LaMnO$_3$.
\item The all-electron model, explicitly treating the Mn $3d$, O $2p$, and $A$ $5d$ bands in LSDA provides an alternative description for the exchange interactions. Both 5-orbital model and all-electron LSDA capture the experimental situation pretty well. Why do we have such consistent description? Apparently, this is due to the cancellation of many contributions. For instance, the polarization of the oxygen states in LSDA strengthens the FM interactions~\cite{JPSJ2009}. In addition to them, there are FM superexchange interactions, which are controlled by the ratio of the on-site exchange and Coulomb repulsion, $J/U$, and expected to be  stronger for smaller $U$~\cite{KugelKhomskii}. However, these effects are compensated by stronger AFM interactions, also expected in the superexchange theory for smaller $U$~\cite{Anderson1959}. Furthermore, the AFM interactions are additionally stabilized by correlations effects beyond the Hartree-Fock approximation~\cite{JPSJ2009}.
\item Unlike in the correlated 5-orbital model, the schemes $\hat{b}$ and $M$ provide a consistent description within all-electron LSDA. Nevertheless, the special attention should be paid to the definition of the xc field and the choice of the parameters $\mathcal{I}_{\rm O}$. The incorrect choice of these parameters can worsens the description.
\item The inversion symmetry breaking, caused by the E-type AFM order in HoMnO$_3$, gives rise to the new DM interactions, operating across the inversion centers in the $Pbnm$ structure, which, in the combination with the magnetic-state dependence of the isotropic interactions, act against this E-type AFM state. Thus, the latter state can be stabilized only by extrinsic mechanisms such as the exchange striction and/or the single-ion anisotropy.
\end{itemize}

\section{\label{sec:other} Other developments}

\subsection{\label{sec:brealspace} Symmetric anisotropic exchange interactions}
\par The spin-spiral concept, which assumes that the perturbation of the magnetic ground state can be described in the form of an incommensurate spin spiral, is applicable only for calculations of isotropic and DM interactions~\cite{Sandratskii}. All these calculations are based on the generalized Bloch theorem, which allows us to deal with this spin-spiral periodicity~\cite{Sandratskii_review}. Unfortunately, this theorem is no longer applicable for calculations of the exchange anisotropy or any other symmetric anisotropic interaction, emerging in the second order of the SO coupling and involving spin diagonal as well as off-diagonal elements of this coupling. As was pointed out in \Sref{sec:SODM}, the DM interactions support the spin-spiral propagation, while the symmetric anisotropic interactions act against it. Nevertheless, one can still consider the perturbations caused by the infinitesimal rotations of the xc field and evaluate the $3$$\times$$3$ exchange tensor $\hat{J}_{ij}$ in the real space, \emph{separately for each magnetic bond}. Then, $\frac{1}{2}(\hat{J}_{ij}$$-$$\hat{J}_{ji})$, which has only three inequivalent matrix elements, can be related to the DM vector, while the traceless part of $\frac{1}{2}(\hat{J}_{ij}$$+$$\hat{J}_{ji})$ gives rise to the exchange anisotropy. This method was successfully implemented in several computational packages, which are actively used for calculations of isotropic as well anisotropic exchange interactions~\cite{TB2J,Kvashnin2020,Ebert,Ebert2023,Mahfouzi}. Nevertheless, one should remember that this method has limitations inherent to the scheme $\hat{b}$, which is an approximation. Unfortunately, only the scheme $\hat{b}$ can be easily reformulated in the real space so that the exchange interactions can be calculated separately for each bond. Due to the additional inversion of the response matrix in the scheme $M$, the exchange parameters in the bond will depend on the matrix elements of the response function in other bonds.

\subsection{\label{sec:DMFT} Dynamic electron correlations}
\par As was already pointed out in \Sref{sec:CrO2}, another interesting direction is to go beyond the conventional SDFT by incorporating the effects of dynamic electron correlations. The latter are typically evaluated in the framework of DMFT~\cite{DMFT1}, which provides a formal extension of the KS equations, where the local potential is replaced by also local, but frequency-dependent self-energy~\cite{DMFT2}. Then, the exchange interactions can be still evaluated using the scheme $\hat{b}$, but with the frequency-dependent xc field~\cite{KL2000}. This approach is especially important for metallic systems, where the static Hartree-Fock approximation is clearly insufficient and, as long as the Coulomb interactions are taken into account, they should be treated on a more rigorous footing. For instance, the dynamic correlations have a profound effect on the electronic structure of half-metallic compounds~\cite{HMRewModPhys}, which is reflected in the behavior of exchange interactions, as was demonstrated for CrO$_2$~\cite{PRB2015,JPCM2016}. However, as the most interesting applications of this method are beyond the strong-coupling limit, the scheme $\hat{b}$ can have serious limitations. Therefore, it can be important to reformulate the interatomic exchange interactions in terms of the inverse response function, as it is done in the scheme $M$. In this respect, \Eref{eq:techange} seems to be very general and could be a good starting point for such extension.

\par Another advantage of this DMFT based approach is that it provides a natural extension for the analysis of temperature dependence of the exchange interactions~\cite{Katanin2023}.

\section{\label{sec:Summary} Summary and conclusions}
\par The linear response theory becomes a powerful tool for calculations of interatomic exchange interactions in various substances. By treating infinitesimal rotations of spins as a perturbation, it allows us to present the total energy change caused by these rotations in the form of pairwise interactions. The main purpose of this topical review was to clarify basic principles of this technique and provide a transparent explanation to more recent developments and controversies related to its practical realization.

\par The first group of questions is related to the validity of the magnetic force theorem for the infinitesimal rotations of spins~\cite{Bruno2003,Antropov2004}. The magnetic force theorem is certainly valid and the total energy change underlying calculations of the exchange interactions can be replaced by the change of the single-particle energies, which seems to be a general fundamental property. However, it would be a mistake to make the equality between the magnetic force theorem and \Eref{eq:JLKAG}. Such attempts are obviously misleading as the correct use of the magnetic force theorem for the exchange interactions should also include contributions of the external magnetic field, which is needed to control the direction of the magnetization. \Eref{eq:JLKAG} does not take into account such contribution. This is an \emph{approximation} leading to the \emph{linear} dependence of the exchange interactions on the response tensor, which can be justified only in the long-wavelength and strong-coupling limits. In fact, the exact expression \eref{eq:techange} for the energy change is extremely simple. We only need to know how to find the external magnetic field and this can be done by using the linear response theory. The total energy change (and the interatomic exchange interactions) in this case will be proportional to the \emph{inverse} response tensor. The procedure is applicable for calculations of the isotropic exchange as well as the DM interactions. In the latter case, it requires the additional constraining field in order to compensate rotations caused by the DM interactions and we have shown how this field can be found using the linear response theory.

\par Besides fundamental aspects of the magnetic force theorem, there is a number of more practically oriented questions. The first one is which object is more suitable for the description of infinitesimal rotations of spins. In this respect, we have considered rigid rotations of the magnetization matrix and local magnetic moments (the schemes $\hat{m}$ and $M$, respectively). Since the local magnetic moment is nothing but the spherical part of the magnetization matrix and only this spherical part is subjected to the constraint conditions in the scheme $M$ (while other components of the magnetization matrix are allowed to relax) such perturbations are expected to cost less energy and, therefore, should be more suitable for the analysis of low-energy excitations.

\par Another important question is what to do with the ligand states. In most of the applications, these states play a role of effective medium participating in the electron transfer between magnetic transition-metal sites. The exchange interactions are typically calculated only between the transition-metal sites, while the contributions of the ligand sites, which can carry an appreciable portion of the magnetization due to the hybridization with transition-metal sites, are ignored. In this respect, we have proposed the downfolding method~\cite{PRB2021}, which allows us to eliminate the ligand states, by transferring their effect to the exchange interactions between the transition-metal sites. This can be achieved by employing the adiabaticity concept and assuming that the fast ligand states instantaneously follow the slow change of the magnetization on the transition-metal sites.

\par An important aspect of the downfolding method is that it can naturally incorporate the dependence of the exchange interactions on the strength of the Stoner coupling ${\cal I}_{\rm L}$ on the ligand sites, which is regarded to be the key ingredient of phenomenological GKA rules for the $90^{\circ}$-exchange and in certain cases primarily responsible for the FM character of this exchange. Nevertheless, the covalent mixing can easily make the effective coupling ${\cal I}_{\rm L}$ negative. Such situation is realized, for instance, in CrCl$_3$, CrI$_3$, and CrO$_2$, where the magnetic polarization of the ligand states is \emph{antiparallel} to the one of the transition-metal states. In such case, the negative ${\cal I}_{\rm L}$ acts against the FM coupling and the ferromagnetism is stabilized by other mechanisms involving the unoccupied Cr $e_g$ states.

\par For most of the applications, we have considered two pictures, which provide a supplementary to each other description for the exchange interactions. The first one is the correlated minimal model constructed for the magnetic, mainly transition-metal, bands near the Fermi level. The second one is based on the all-electron LSDA, which takes into consideration both magnetic transition-metal and ligand states. Once the Coulomb interactions are added to the TB model, they should be treated rigorously: if the static Hartree-Fock approximation is applicable for the analysis of exchange interactions in insulating systems with lifted orbital degeneracy, the situation in metallic systems can be very different. In certain cases, the use of the Hartree-Fock approximation can lead to a misleading answer, as was demonstrated for the half-metallic CrO$_2$~\cite{PRB2015}. However, this is not the only source of the error for the exchange interactions. Another problem is related to the fact that most of the real materials are far from the strong-coupling limit and the commonly used scheme $\hat{b}$, which leads to \Eref{eq:JLKAG} for the exchange interactions, is no longer justified. A more rigorous approach is to evaluate the exchange interactions via the inverse response (the scheme $M$), which is certainly more difficult from the computational point of view. However, such scheme tremendously improves the description of interatomic exchange interactions in the correlated model, as was demonstrated for insulating Cr$X_3$ ($X$$=$ Cl, I) and $A$MnO$_3$ ($A$$=$ La, Ho) in the framework of Hartree-Fock approximation. The application of correlated models for the exchange interactions in metallic systems should consider two aspects: (i) The method should be beyond the Hartree-Fock approximation. The commonly used alternative is DMFT~\cite{DMFT1,DMFT2}; (ii) The calculations of the exchange interactions themselves should be based on the inverse response, as in the scheme $M$.

\par The exchange interactions in all-electron LSDA (GGA) obey quite different principles. Certainly, this is a simple approximation derived in the limit of homogeneous electron gas. Nevertheless, it satisfies certain fundamental sum rules and on many occasions provides an insightful view on the properties of systems far beyond this limit~\cite{GL}. Thus, it can be regarded as an alternative view on the problem of exchange interactions, though not necessarily the perfect one. For the magnetic systems, LSDA and strong-coupling limit are basically incompatible with each other (or this limit can be strongly underestimated in LSDA). Therefore, there is absolutely no guarantee that the approximate scheme $\hat{b}$ can properly capture the behavior of interatomic exchange interactions and a more rigorous scheme $M$ looks more preferable. Nevertheless, the improvement expected in the scheme $M$ is somewhat diminished by the effects of other parameters controlling the properties of the exchange interactions in the all-electron case, particularly the choice of the xc field $\vec{b}$ and the effective Stoner coupling ${\cal I}_{\rm L}$ on the ligand sites. By taking the xc field from the sum rule, $\vec{b} = \hat{\mathcal{Q}}^{+}_{0} \vec{m}$, one can substantially improve the description in the framework of the method $\hat{b}$. However, the choice of the parameters ${\cal I}_{\rm L}$, which can be strongly depend on the definition, poses a more serious problem. We have considered several such definitions. In a number of cases, they provide a consistent description for the interatomic exchange interactions, but not always. It seems that there is still an ambiguity with the choice of ${\cal I}_{\rm L}$ and the isotropic exchange interactions depend on such ambiguity. On the other hand, the dependence of DM interactions on ${\cal I}_{\rm L}$ is considerably weaker. As an alternative approach, we have proposed the scheme L0, which assumes that the magnetic moments on the ligand sites are induced solely by the hybridization with the transition-metal sites, while all Stoner parameters ${\cal I}_{\rm L}$ can be set to zero. Such scheme is also formulated in terms of the linear response and most suitable for the analysis interatomic exchange interactions in FM insulators and half-metals.

\par The merging of correlated model for the transition-metal bands with LDA electronic structure for the ligand bands is another largely unresolved problem. Although such merging is expected to improve the description of the exchange interactions, by combining the effects of Coulomb correlations in transition-metal bands with the explicit treatment of the ligand states, the strategy suffers from many ambiguities, as was demonstrated for CrCl$_3$ and CrI$_3$. We are always forced to compromise between genuine physical effect and intrinsic error of such merging caused by additional approximations and assumptions, which are inevitable in this case. On the other hand, on many occasions such merging is not really needed as the physically meaningful picture for the exchange interactions can be obtained in correlated 5-orbital model constructed only for the magnetic $3d$ bands. As a supplementary step, one can always consider the all-electron LSDA (GGA), which gives an idea about the role played by the ligand states.

\par Amongst other interesting properties, CrI$_3$ has attracted a considerable attention due to the strong DM interaction induced by large SO coupling of the heavy I atoms~\cite{ChenPRX}. On the other hand, CrI$_3$ is a closed shell material, where orbital degrees of freedom are quenched by the large crystal-field splitting between occupied $t_{2g}$ and unoccupied $e_g$ states. Therefore, from the practical point of view, comparable or even stronger DM interactions can be expected in the open shell materials even without heavy $5p$ elements, as was demonstrated, for instance, for CrO$_2$, LaMnO$_3$, and HoMnO$_3$.

\ack I am grateful to Mikhail Katsnelson, Alexander Liechtenstein and Vladimir Antropov for valuable comments. The TB Hamiltonian for Co$_3$Sn$_2$S$_2$ in \Sref{sec:Co3S2S2} was constructed by Sergey Nikolaev~\cite{PRB2022b}. MANA is supported by World Premier International Research Center Initiative (WPI), MEXT, Japan.

\appendix
\section{\label{sec:Hubbard} Construction and solution of effective Hubbard-type models}
\par The Hubbard-type model,
\noindent
\begin{equation}
\hat{\cal{H}}  =  \sum_{ij} \sum_{\sigma} \sum_{ab}
H_{ia,jb}^{\sigma}
\hat{c}^{ \sigma \dagger}_{i a}
\hat{c}^{\sigma \phantom{\dagger}}_{j b} +
  \frac{1}{2}
\sum_{i}  \sum_{\sigma \sigma'} \sum_{abcd} U^{abcd}_{i} \,
\hat{c}^{\dagger \sigma}_{i a} \hat{c}^{\dagger \sigma'}_{i c}
\hat{c}^{\sigma \phantom{\dagger} }_{i b}
\hat{c}^{\sigma' \phantom{\dagger}}_{i d},
\label{eq:ManyBodyH}
\end{equation}
\noindent is constructed in the Wannier basis for some particular group bands (the so-called target bands)~\cite{review2008,ImadaMiyake,WannierRevModPhys}. The operator $\hat{c}^{\dagger \sigma}_{i a}$ ($\hat{c}^{\sigma \phantom{\dagger} }_{i a}$) stands for the creation (annihilation) of an electron with the spin $\sigma$ in the Wannier orbital $a$ on the site $i$. It is assumed that LDA (or LSDA) is the good starting point for the noninteracting one-electron part of the model so that the parameters $H_{ia,jb}^{\sigma}$ can be associated with the matrix elements of corresponding KS Hamiltonian in the Wannier basis. In LSDA, these parameters depend on spin. Furthermore, $\hat{H}^{\sigma} = [ H_{ia,jb}^{\sigma} ]$ can include spin-diagonal part of the SO coupling, which is needed to calculate DM interactions~\cite{Sandratskii,PRB2023}. In this case, $\hat{H}^{\sigma}$ also depends on spin, even in the non-magnetic LDA, where it holds: $\hat{H}^{\downarrow} = \hat{H}^{\uparrow *}$. For practical purposes we use mainly the LMTO method~\cite{LMTO1,LMTO2}. Then, the Wannier functions can be generated using the projector-operator technique~\cite{review2008,WannierRevModPhys}. The spin-diagonal part of the SO coupling is typically added to the site-diagonal part of the LMTO Hamiltonian as $\hat{H}_{ii}^{\uparrow, \downarrow} \to \hat{H}_{ii}^{\uparrow, \downarrow} \pm \frac{1}{2} \xi_{i} \hat{L}_{i}^{z}$. After that, the TB Hamiltonian is constructed in the Wannier basis for the target bands. Thus, even though the target bands are formally of the transition-metal $3d$ type, the corresponding Wannier functions and the TB Hamiltonian include some contributions of the SO coupling of the heavy ligand atoms (if any), which are admixed into the target bands via the hybridization effects.

\par The parameters of screened on-site Coulomb interactions, $\hat{U}=[U^{abcd}_{i}]$, are evaluated in cRPA starting with the LDA band structure~\cite{cRPA}. The basic idea of the constraint in this case is to get rid of nonphysical metallic screening in LDA emerging from the bands near the Fermi level. For the $d$ electrons, the matrix $\hat{U}$ can be fitted in terms of the on-site Coulomb repulsion $U = F^{0}$, the intraatomic exchange interaction $J=(F^{2}$$+$$F^{4})/14$, and ``nonsphericity'' $B=(9F^{2}$$-$$5F^{4})/441$ ($F^{0}$, $F^{2}$, and $F^{4}$ being radial Slater’s integrals), where $U$ is responsible for the charge stability of given electronic configuration, while $J$ and $B$ are responsible for the first and second Hund’s rules, respectively~\cite{review2008}. Strictly speaking, the parametrization in terms of $U$, $J$, and $B$ is valid only for the isolated atoms in the spherical environment~\cite{Slater}. It is used only for the explanatory purposes, while all numerical calculations are performed with the matrices of Coulomb interactions $\hat{U}$ extracted from cRPA without additional fitting. For the $t_{2g}$ electrons alone, it is convenient to use the Kanamori parametrization in terms of the intraorbital Coulomb repulsion $\mathcal{U}$ and the exchange interaction $\mathcal{J}$~\cite{Kanamori1963}. For the $d$ electrons, these parameters can be related to the above $U$ and $J$ as $\mathcal{U} \approx U +8J/7$ and $\mathcal{J} \approx 0.77 J$~\cite{review2008}. Moreover, for the $t_{2g}$ model, $\mathcal{U}$ can be reduced due to additional channels of screening by the $e_{g}$ electrons, which are typically considered in the $t_{2g}$ model, but not in the more general one constructed for all $d$ (i.e., $t_{2g}$+$e_{g}$) electrons~\cite{review2008}.

\par After the construction, the model is solved in the mean-field Hartree-Fock approximation, where the second term in \Eref{eq:ManyBodyH} is replaced by
\noindent
\begin{equation}
\sum_{i} \sum_{\sigma} \sum_{ab} {\cal V}_{i,ab}^{\sigma} \hat{c}^{ \sigma \dagger}_{i a} \hat{c}^{\sigma \phantom{\dagger}}_{i b}
\end{equation}
\noindent and the potential $\hat{\cal V}_{i}^{\sigma} = [{\cal V}_{i,ab}^{\sigma}]$ is found self-consistently. Then, the obtained electronic structure is used to calculate the exchange interactions. The xc field in this case is defined as $\hat{b}_{i} = \hat{\cal V}_{i}^{\uparrow} - \hat{\cal V}_{i}^{\downarrow}$. When the SO interaction is added to $\hat{H}^{\sigma}$, the potential $\hat{\cal V}_{i}^{\sigma}$ is recalculated self-consistently to include the effects or the SO coupling. After that the obtained electronic structure and $\hat{\cal V}_{i}^{\sigma}$ are used to calculate the DM interactions. The Hartree-Fock approximation is believed to be a good starting points for the analysis of magnetic properties of insulating materials, where the orbital degeneracy is lifted by the lattice distortions. Other details can be found in ref.~\cite{review2008}.

\section{\label{sec:RPA} Magnetic ground state and random-phase approximation for the magnetic transition temperature}
\par Here, we generalize the RPA expression for the critical transition temperature, $T_{\rm S}$, of compounds with multiple magnetic sublattices to the case, where the ground state is a noncollinear spin spiral with the propagation vector $\boldsymbol{q}$. Moreover, the sublattices are allowed to acquire the additional phases $\gamma_{\boldsymbol{q}, \, \mu}$, describing relative rotations of the magnetization in different sublattices relative to each other. In all other respects, we follow the derivation considered for the collinear case by Rusz \etal~\cite{TCRPA}.

\par Our first goal is to find the ground state, corresponding to the minimum of the energy
\noindent
\begin{equation}
E(\boldsymbol{q}) = -\frac{1}{2} \sum_{\mu \nu} c_{\boldsymbol{q}, \, \mu}^{*} J_{\boldsymbol{q}, \, \mu \nu}^{\phantom{*}} \, c_{\boldsymbol{q}, \, \nu}^{\phantom{*}},
\label{eqn.EMF}
\end{equation}
\noindent where $c_{\boldsymbol{q}, \, \mu} = e^{i \gamma_{\boldsymbol{q}, \, \mu}}$ and $J_{\boldsymbol{q}, \, \mu \nu}^{\phantom{*}}$ is the Fourier transform of $J_{ij}$ between the sublattices $\mu$ and $\nu$. \Eref{eqn.EMF} can be generalized to include the DM interactions $d_{ij}^{z}$ by replacing $J_{\boldsymbol{q}, \, \mu \nu}$ with $J_{\boldsymbol{q}, \, \mu \nu} -id_{\boldsymbol{q}, \, \mu \nu}^{z}$. For each $\boldsymbol{q}$, we minimize $E(\boldsymbol{q})$ with respect to $\gamma_{\boldsymbol{q}, \, \mu}$ using the gradient descent method. Then, we pick up $\boldsymbol{q}$ corresponding to the global minimum of $E(\boldsymbol{q})$ and for each $\boldsymbol{k}$ redefine $J_{\boldsymbol{k}, \, \mu \nu}$ as $J_{\boldsymbol{k}, \, \mu \nu} \rightarrow c_{\boldsymbol{q}, \, \mu}^{*}  J_{\boldsymbol{k}, \, \mu \nu}^{\phantom{*}} c_{\boldsymbol{q}, \, \nu}^{\phantom{*}}$ with the coefficient $c_{\boldsymbol{q}, \, \nu}$ determined for the ground state. This is nothing but the transformation to the new local coordinate frame corresponding to the energy minimum.

\par Then, the spin-wave energies can be associated with the eigenvalues of the positive-defined matrix $\hat{\Omega}_{\boldsymbol{k}} = \hat{S}^{-1} \hat{\mathbb{N}}_{\boldsymbol{k}}$, where
\noindent
\begin{equation}
\mathbb{N}_{\boldsymbol{k}, \, \mu \nu} =  \delta_{\mu \nu} \sum_{\nu'} J_{\boldsymbol{q}, \, \mu \nu'} -  J_{\boldsymbol{k}, \, \mu \nu},
\label{eqn.N}
\end{equation}
\noindent $\hat{S}=\hat{M}/2$, and $\hat{M}$ is the diagonal matrix of magnetic moments.\footnote{Here, we use the exchange parameters derived from the static response function and do not consider any renormalization effects~\cite{KL2004,Szczech}} In RPA, the corresponding magnetic transition temperature $T_{\rm S}^{\mu}$ for the sublattice $\mu$ is given by~\cite{TCRPA}:
\noindent
\begin{equation}
 T_{\rm S}^{\mu} = \frac{1 + 1/S_{\mu}}{3 k_{\rm B}} \left( \frac{1}{\Omega}_{\rm BZ} \int d \boldsymbol{k} \left[ \hat{\mathbb{N}}_{\boldsymbol{k}}^{-1} \right]_{\mu \mu} \right)^{-1}
\label{eqn.TN}
\end{equation}
\noindent ($\Omega_{\rm BZ}$ being the volume of the first Brillouin zone). For the inequivalent sublattices, this $T_{\rm S}^{\mu}$ depends on the ratio of the magnetic moments, $M_{\mu}/M_{\nu}$, which can also depend on the temperature. In order to find true transition temperature, these rations should be adjusted to make the same $T_{\rm S}^{\mu} = T_{\rm S}^{\nu} \equiv T_{\rm S}$ for all the sublattices~\cite{TCRPA}.

\par The spectral theorem is employed in order to calculate $\hat{\mathbb{N}}^{-1}_{\boldsymbol{k}}$. Then, a small imaginary part, $i\delta$ with $\delta = 0.01$ meV (corresponding to $\delta / k_{\rm B} \sim 0.1$ K) is added to the eigenvalues of $\hat{\mathbb{N}}_{\boldsymbol{k}}$ for $\boldsymbol{k}$ close to $\boldsymbol{q}$. Finally, the $\boldsymbol{k}$-space integration is replaced by the summation on a very dense mesh of $\boldsymbol{k}$-points (for instance, a typical mesh used for CrCl$_3$ and CrI$_3$ is $258$$\times$$258$$\times$$258$).

\end{document}